\documentclass[11pt]{article}
\usepackage[utf8]{inputenc}
\usepackage{amsmath}
\usepackage{tikz}
\usetikzlibrary{arrows.meta,positioning,calc}
\usepackage{placeins}
\usepackage{forest}
\usepackage{float}
\usepackage{xcolor}
\usepackage{authblk}
\usepackage{amssymb}
\usepackage[english]{babel}
\usepackage{hyperref}
\usepackage{apacite}
\usepackage{ragged2e}
\usepackage{enumerate}
\usepackage{enumitem}
\usepackage{authblk}
\usepackage{graphicx}
\usepackage{geometry}
\usepackage{pifont}
\usepackage{subcaption}
\usepackage{booktabs}
\usepackage{setspace}
\usepackage{xcolor}
\usepackage{multirow}
\usepackage{array}
\hypersetup{colorlinks=true,linkcolor=blue,anchorcolor=blue,citecolor=blue}
\usepackage{caption}
\newtheorem{example}{Example}
\newtheorem{definition}{Definition}
\newtheorem{remark}{Remark}
\newtheorem{theorem}{Theorem}
\newtheorem{lemma}{Lemma}
\newtheorem{proposition}{Proposition}

\newtheorem{assumption}{Assumption}
\newtheorem{corollary}{Corollary}
\newcommand{\e}{\mathbf{e}}

\renewcommand{\v}{\mathbf{v}}

\newcommand{\M}{\mathbf{M}}

\newcommand{\A}{\mathbf{A}}
\renewcommand{\i}{\mathbf{i}}
\newcommand{\R}{\mathbb{R}}
\newcommand{\1}{\mathbf{1}}

\newcommand{\F}{\mathcal F_S(\varepsilon_0,M_S,\rho_S)}

\newcommand{\cv}{\operatorname{cv}}

\renewcommand{\P}{\mathbb{P}}
\renewcommand{\r}{\mathbf{r}}

\newcommand{\E}{\mathbb{E}}

\title{Can We Ignore Data Dependence in Inference Implementation?\\ A Positive Answer for Regression Discontinuity Designs}
\author[1]{Zihao Yuan}
\affil[1]{University of Hamburg}
\date{}

\begin{document}
\maketitle

\begin{abstract}
In empirical economics, regression discontinuity design (RDD) inference is commonly based on asymptotic normality under an independence assumption, but the running variables or disturbances may exhibit unknown and complex dependence. We first rigorously identify a fundamental dilemma for this conventional practice. To address it, we propose a validity recovery procedure that redefines the roles of bias and stochastic calibration, thereby decoupling data dependence from stochastic calibration. Taking the independence-assumption-based confidence interval as our inferential anchor, our procedure restores its validity while preserving its asymptotic leading constant, even when the data are dependent. For fixed-design RDD based on local linear regression, our intervals achieve asymptotically exact nominal coverage and, compared with the anchor, exploit higher-order smoothness without enlarging the asymptotic variance constant, attaining the minimax-optimal shrinking rates up to a mild factor, such as $\sqrt{\log T})$. The procedure requires neither specific dependence frameworks (e.g., mixing or physical dependence) nor dependence-modeling techniques (e.g., long-run variance estimation). For random-design RDD, under more specific dependence assumptions, we derive the rate-explicit coverage error at each nominal level and characterize the trade-off between this rate and interval length. Moreover, modeling dependence becomes an optional enhancement that can further enable validity recovery and improve finite-sample coverage, while averaging over repeated calibrations stabilizes the randomized center and protects against undercoverage in finite samples without increasing interval length. In short, under mild dependence and moment conditions, our procedure makes the assumption of data independence an implementation default rather than a validity requirement. Simulations support the theoretical results.

\end{abstract}

\newpage
\section{Introduction}
\label{sec 1}
Many treatment assignment mechanisms operate through predetermined eligibility thresholds, creating opportunities for causal identification from observational data. In a sharp regression discontinuity design (RDD), treatment assignment changes deterministically when a running variable crosses a known cutoff, as in scholarship programs that award eligibility based on a minimum GPA. Provided that the conditional expectations of potential outcomes are continuous at the cutoff, the resulting discontinuity in the observed conditional mean identifies the average treatment effect at that point. This strategy has become an important tool in empirical economics and other social sciences, supported by an extensive literature on estimation and inference, particularly methods based on local linear regression. See, for example, \citeA{imbens2012optimal}, \citeA{lee2010regression}, \citeA{calonico2014robust}, \citeA{armstrong2020simple}, \citeA{noack2024bias}, and references therein.

Recall that the estimand of a sharp RDD is a linear functional of the regression function (see also \eqref{eq tau}); we denote it by $\theta(f)$, where $f\in \mathcal{F}$ and $\mathcal{F}$ is a function class. We denote the estimator as $\hat \theta_h$, where $h>0$ is the bandwidth. Recall that
\begin{align}
\label{eq decomposition}
    \frac{\hat \theta_h-\theta(f)}{\sqrt{\text{Var}(\hat \theta_h)}}=  \frac{\hat \theta_h-\E[\hat \theta_h]}{\sqrt{\text{Var}(\hat \theta_h)}}+  \frac{\E[\hat \theta_h]-\theta(f)}{\sqrt{\text{Var}(\hat \theta_h)}}=:Z+  \frac{B_h(f)}{\sqrt{\text{Var}(\hat \theta_h)}},
\end{align}
According to the literature reviewed above, a valid confidence interval often relies on the following two points: 
\begin{itemize}
    \item [(i)] proper correction or control of $\frac{B_h(f)}{\sqrt{\text{Var}(\hat \theta_h)}}$; 
    \item [(ii)] correct stochastic calibration (e.g., standard-normal approximation) of $Z$.
\end{itemize}

For point (i), recent progress on this topic falls into two streams: "bias-corrected inference" and "bias-aware inference". A prominent example of bias-corrected inference is robust bias correction
(RBC), proposed by \citeA{calonico2014robust} and further developed by
\citeA{calonico2018effect,calonico2022coverage}. The key idea of RBC is
not only to estimate the leading smoothing bias, but to incorporate this
estimated leading bias into the original local-polynomial estimator and
thereby form a new bias-corrected estimator, denoted here by
\(\hat\theta_{h,\mathrm{RBC}}\). Inference is then based on a suitable
asymptotic variance formula for \(\hat\theta_{h,\mathrm{RBC}}\), which
accounts for both the original estimation uncertainty and the additional
uncertainty introduced by estimating the leading bias. Under appropriate
regularity conditions, the remaining bias of
\(\hat\theta_{h,\mathrm{RBC}}\) is of higher order, so that
\[
    \frac{
    \mathbb E[\hat\theta_{h,\mathrm{RBC}}]-\theta(f)
    }{
    \sqrt{\operatorname{Var}(\hat\theta_{h,\mathrm{RBC}})}
    }
    \to 0 .
\]
Thus RBC addresses point (i) by transforming the estimator so that the
standardized residual bias is asymptotically negligible, while preserving
point (ii) through a variance formula appropriate for the RBC estimator. A second stream is bias-aware inference introduced in \citeA{armstrong2018optimal,armstrong2020simple}. This approach does not
attempt to make the standardized bias asymptotically negligible by
constructing a new bias-corrected estimator. Instead, it uses the
restrictions defining the function class \(\mathcal F_n\) to bound the
possible magnitude of the bias. If $ \bar B_h=\sup_{f\in\mathcal F_n}|B_h(f)|,$
then the bias-aware-type interval is calibrated using the property of the folded-normal distribution to incorporate the worst-case bias
\(\bar B_h/\sqrt{\operatorname{Var}(\hat\theta_{h})}\). A typical bias-aware interval takes the form $ \sqrt{\operatorname{Var}(\hat\theta_{h})}  \operatorname{cv}_{1-\alpha}\left(\frac{\bar B_h}{\sqrt{\operatorname{Var}(\hat\theta_{h})}}\right)$,
where \(\operatorname{cv}_{1-\alpha}(t)\) is the \((1-\alpha)\)-quantile of
\(|N(t,1)|\). Bias-aware inference therefore addresses point (i) by
controlling the standardized bias directly in the critical value, rather
than by constructing an estimator whose standardized bias vanishes.

The present paper focuses on point (ii) when the data dependence has been misspecified. In the usual inference implementation of sharp RD designs, “independent sample” is a widely employed assumption and a local linear or quadratic smoother is the benchmark choice for estimation and inference (see \citeA{gelman2019high} for more thorough discussions on this topic). Together with mild moment conditions and some simple algebra, the homoskedasticity and independence assumptions usually imply that 
\begin{align*}
    \sqrt{\text{Var}(\hat \theta_h)}= \frac{C_{K,p}\sigma}{\sqrt{T_{\text{eff}}}}(1+o(1)),\ \text{as}\ T\to\infty,
\end{align*}
where $T_{\text{eff}}$ is the effective sample size. $C_{K,p}>0$ is a fixed constant determined by the equivalent kernel of the local polynomial smoother and its order $p$. $\sigma^2$ is the variance of individual disturbances. Therefore, such an “independence-assumption-based oracle interval” can be expressed as 
\begin{align} 
\label{eq CI oracle} 
\mathbf{CI}_\mathrm{oracle}^\mathrm{ind}(\alpha):= \Big[\hat \theta_h-B_h(f)\pm C_{K,p}\sigma z_{1-\frac{\alpha}{2}}T_{\text{eff}}^{-\frac{1}{2}}\Big].
\end{align}
However, independence is an assumption usually too strong in real-world applications. When the true data-generating process possesses complex dependence, even the oracle interval $\mathbf{CI}_\mathrm{oracle}^\mathrm{ind}(\alpha)$ itself may be asymptotically invalid for any given $\alpha\in (0,1)$. This failure is not merely due to poor standard-error estimation, because, even if the variance under the independence-based assumed DGP were known exactly, the misspecification of data dependence could still lead to
wrong normalizing scale for the centered estimator under the dependent true DGP.

Before stating our research goals, we make precise which intervals we regard as non-trivial. Throughout, an interval for $\theta(f)$ takes the form $[C_T\pm r_T]$, where the center $C_T$ may be randomized and $r_T>0$ is the radius.

\begin{definition}[Trivial and non-trivial intervals]
\label{def nontrivial}
An interval $[C_T\pm r_T]$ for $\theta(f)$ is \emph{trivial} if its pivot at the radius scale is degenerate, that is, $(C_T-\theta(f))/r_T\xrightarrow{\P}0$ as $T\to\infty$. It is \emph{non-trivial} if $(C_T-\theta(f))/r_T$ converges weakly to a law with a positive Lebesgue density on $\R$.
\end{definition}

A trivial interval obeys a law of large numbers at the radius scale: the limit law of its pivot is a point mass, whose quantile function is constant. The critical value then corresponds to no quantile level, so the nominal level loses its calibration meaning. The induced test rejects with probability tending to zero for every $\alpha$, which produces systematic non-significance and tilts decisions towards the null. Stretching a rate-optimal interval by a diverging factor produces such an interval; see Observation 3 and \eqref{eq trivial coverage} in Section \ref{sec 3.2}. A non-trivial interval obeys a central limit theorem at the radius scale. Its coverage stays strictly inside $(0,1)$ and is governed by $\alpha$. Conventional asymptotic-normality-based intervals, such as $\mathbf{CI}_\mathrm{oracle}^\mathrm{ind}(\alpha)$ under a correctly specified independent DGP, are non-trivial. Validity is a separate requirement: a non-trivial interval may still undercover.

Therefore, this paper pursues the following three research goals.
\begin{enumerate}[label=\textbf{G\arabic*}, ref=G\arabic*]
 \item \label{G1} \textbf{(Diagnosis)} Quantify how risky it is to ignore the underlying data dependence in the inference implementation of RDD, and how difficult it can be to model the unknown dependence instead.
 \item \label{G2} \textbf{(Validity recovery with exact, non-trivial coverage)} Starting from a feasible but possibly invalid confidence interval of the form $\mathbf{CI}_\mathrm{oracle}^\mathrm{ind}(\alpha)$, obtained by misspecifying dependent data as independent ones, construct an interval $\mathbf{CI}_\mathrm{recover}^\mathrm{ind}(\alpha)$ without modeling the underlying dependence directly, such that, for every $\alpha\in(0,1)$,
 \begin{itemize}
  \item[(i)] \textit{asymptotic exact coverage}: $\sup_{f\in\mathcal F}\big|\P\big(\theta(f)\in\mathbf{CI}_\mathrm{recover}^\mathrm{ind}(\alpha)\big)-(1-\alpha)\big|\to 0$ as $T\to\infty$;
  \item[(ii)] \textit{non-triviality}: $\mathbf{CI}_\mathrm{recover}^\mathrm{ind}(\alpha)$ is non-trivial in the sense of Definition \ref{def nontrivial}; 
 \end{itemize}
 Where possible, we further want $\mathbf{CI}_\mathrm{recover}^\mathrm{ind}(\alpha)$ to turn higher-order smoothness into a larger effective sample size than conventional local linear inference permits.
 \item \label{G3} \textbf{(Mild efficiency cost)} Achieve \ref{G2} at a mild cost of efficiency: the radius of $\mathbf{CI}_\mathrm{recover}^\mathrm{ind}(\alpha)$ attains the minimax-optimal shrinking rate up to a mild diverging factor, such as $\sqrt{\log T}$.
\end{enumerate}
Definition \ref{def nontrivial} fixes the eligible class for every procedure in this paper: an interval must be non-trivial, so that the nominal level $\alpha$ determines the error rate it implements. Goal \ref{G2} asks for more, namely asymptotic exact coverage within this class, and Sections \ref{sec 3} and \ref{sec 4} deliver it. The finite-sample enhancement in Section \ref{sec 5} remain within the eligible class. Similar to bias-aware inference introduced in \citeA{armstrong2018optimal}, the coverage in Section \ref{sec 5} may exceed $1-\alpha$, but it remains governed by the given $\alpha$.

Under very general assumptions, this paper achieves Goals \ref{G1}--\ref{G3}. Section \ref{sec 2} achieves \ref{G1}: Propositions \ref{prop underestimation}--\ref{prop bias corrected} show that ignoring dependence can cause persistent undercoverage even under exact Gaussianity and oracle bias control, while Example \ref{ex 2} and Proposition \ref{prop impossible} show that learning a long-run variance can be unstable, fundamentally difficult, and unnecessary. Sections \ref{sec 3} and \ref{sec 4} achieve \ref{G2}--\ref{G3} for fixed- and random-design RDD (Theorems \ref{th 1} and \ref{th 2}) through the validity-recovery mapping below.
\begin{figure}[H]
\centering
\begin{tikzpicture}[
    box/.style={
        draw,
        rounded corners,
        align=center,
        text width=5cm,
        minimum height=1.8cm,
        inner sep=4pt
    },
    >=latex
]

\node[box] (left) 
{feasible but possibly invalid\\
independence-assumption-based interval};

\node[box, right=3.5cm of left] (right) 
{feasible, valid and non-trivial interval\\
inheriting the form of\\
independence-assumption-based interval};

\draw[->, thick] 
(left) -- node[above, align=center] {validity recovery} (right);

\end{tikzpicture}
\end{figure}
Building on the exogenous Gaussian noise in data fission \cite{leiner2025data} and the folded-normal calibration in bias-aware inference \cite{armstrong2018optimal}, the recovered interval also preserves the leading constant of the independence-assumption-based interval and exploits higher-order smoothness while retaining the asymptotic variance constant of local linear regression.

Beyond \ref{G1}--\ref{G3}, the framework yields three further results. Dependence-aware studentization is an optional enhancement that improves coverage but is not needed for validity (Section \ref{sec 3.4}). For random-design RDD, explicit undercoverage and overcoverage bounds characterize the trade-off with interval length (Section \ref{sec 4}). Averaging over repetitive generations stabilizes the randomized center and protects against undercoverage in finite samples without enlarging the interval (Section \ref{sec 5}). Section \ref{sec 7} reports simulations under strong serial dependence.

%%%%%%%%%%%%%%%%%%%%%%%%%%%%%%%%%%%%%%%%%%%%%%%%%%%%%%%%%%%%%%%%%%%%%%%%%%%%%%%%%%%%%%%%%%%%%%%%%%%%%%%%%%%%%%%%%%%%%%%%%%%%%%%%%%%%%%%%%%%%%%%%%%%%%%%%%%%%%%%%%%%%%%%%%%%%%%%%%%%%%%%%%%%%%%%%%%%%%%%%%%%%%%%%%%%%%%%%%%%%%%%%%%%%%%%%%%%%%%%%%%%%%%%%%%%%%%%%%%%%%%%%%%%%%%%%%%%%%%%%%%%%%%%%%%%%%%%%%%%%%%%%%%%%%%%%%%%%%%%%%%%%%%%%%%%%%%%%%%%%%%%%%%%%%%%%%%%%%%%%%%%%%%%%%%%%%%%%%%%%%%%%%%%%%%%%%%%%%%%%%%%%%%%%%%%%%%%%%%%%%%%%%%%%%%%%%%%%%%%%%%%%%%%%%%%%%%%%%%%%%%%%%%%%%%%%%%%%%%%%%%%%%%%%%%%%%%%%%%%%%%%%%%%%%%%%%%%%%%%%%%%%%%%%%%%%%%%%%%%%%%%%%%%%%%%%%%%%%%%%%%%%%%%%%%%%%%%%%%%%%%%%%%%%%%%%%%%%%%%%%%%%%%%%%%%%%%%%%%%%%%%%%%%%%%

\section{Dilemma of Conventional Approach under Dependence}
\label{sec 2}

The RDD inference discussed in nearly all the aforementioned literature relies on one conventional approach with three ingredients: \textit{asymptotic normality, proper bias tackling, and correct studentization}. However, most of this literature focuses on the second point, and its results depend heavily on a data-independence assumption that may be far from the truth in real applications. Section \ref{sec 2} discusses the risks and difficulties of ignoring and modeling data dependence, respectively. Overall, these reflections present a binary dilemma for RDD inference based on the asymptotic normality of the original centered RDD estimator. That is, for some normalizer \(\sigma_n\) (e.g., $\sqrt{\text{Var}(\hat\theta_h)}$), we have
\begin{align}
\label{eq ideal CLT}
    \frac{\hat\theta_n-E[\hat\theta_n]}{\sigma_n}
\Rightarrow N(0,1).
\end{align}
Now assume that \eqref{eq ideal CLT} holds for every dependent data-generating process. Unfortunately, even under these ideal circumstances, feasible inference would still require choosing how to approximate or estimate $\sigma_n$, which would no doubt create a simple but difficult “binary-dilemma-type” situation for practitioners. The following Figure \ref{figure 1} illustrates this.

% In the document
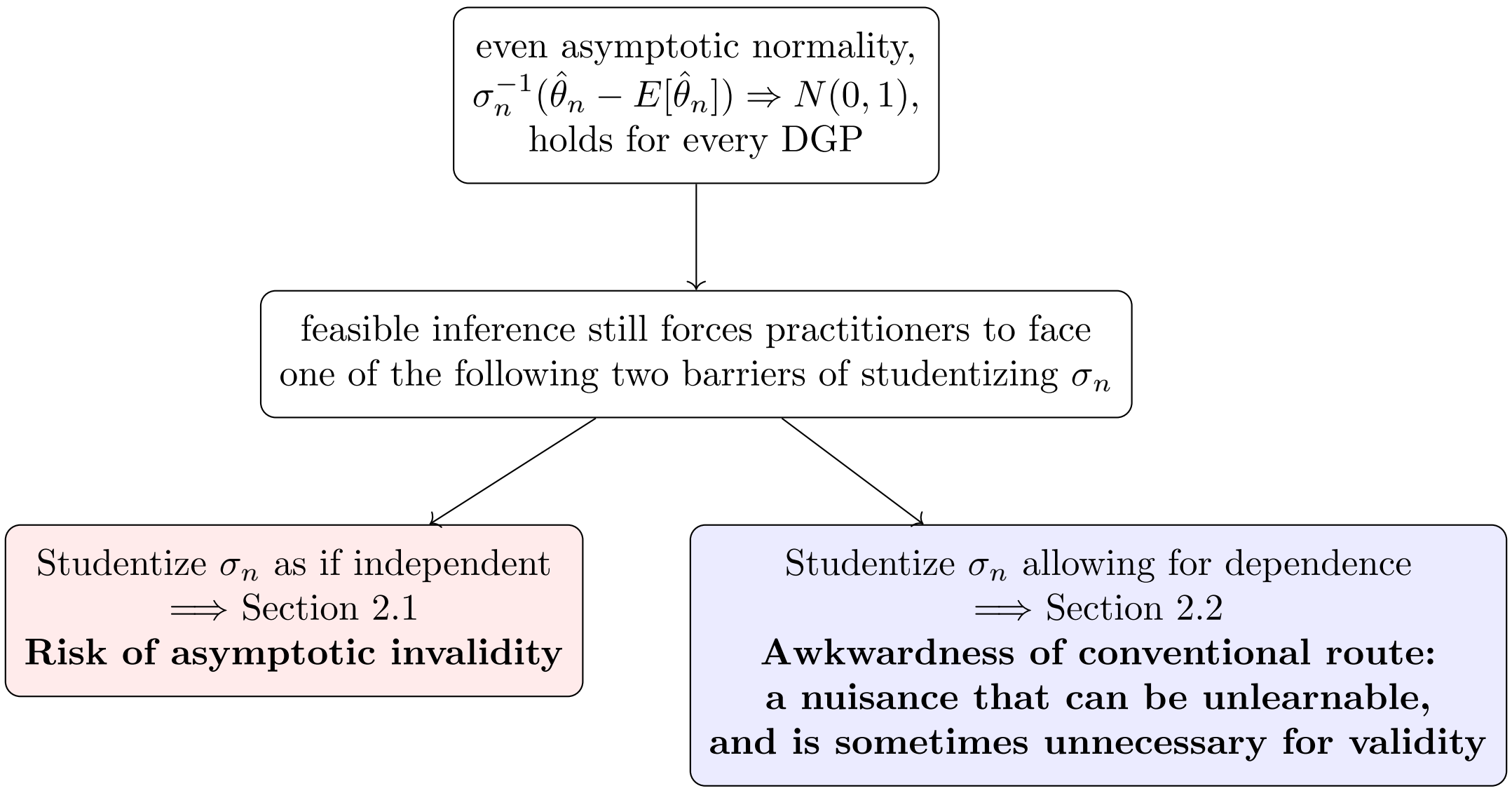
\begin{figure}[H]
\centering
\begin{forest}
for tree={
    draw,
    rounded corners,
    align=center,
    edge={->},
    l sep=10mm,
    s sep=10mm,
    inner sep=5pt,
    font=\small
}
[
{even asymptotic normality,\\
$\displaystyle
\sigma_n^{-1}(\hat\theta_n-E[\hat\theta_n])
\Rightarrow N(0,1),$\\
holds for every DGP}
    [
    {feasible inference still forces practitioners to face \\
   one of the following two barriers of studentizing $\sigma_n$}
        [
        {Studentize $\sigma_n$ as if independent\\
        $\Longrightarrow$ Section 2.1\\
        \textbf{Risk of asymptotic invalidity}},
        fill=red!8
        ]
        [
        {Studentize $\sigma_n$ allowing for dependence\\
        $\Longrightarrow$ Section 2.2\\
        \textbf{Awkwardness of conventional route:}\\
       \textbf{a nuisance that can be unlearnable,}\\
       \textbf{and is sometimes unnecessary for validity}},
        fill=blue!8
        ]
    ]
]
\end{forest}
\caption{Binary dilemma of conventional asymptotic-normality-based RDD inference.}
\label{figure 1}
\end{figure}

As Figure \ref{figure 1} shows, if the scale is calibrated under the independence working model, inference faces the invalidity risk studied in Section \ref{sec 2.1}. If the scale is calibrated to allow for dependence, practitioners typically follow the conventional route of modeling dependence through a long-run variance. Section \ref{sec 2.2} shows that this route leads to modeling and calibration difficulties, particularly when the dependence structure varies with the observation horizon, and, awkwardly, that the difficult nuisance-learning problem it creates is sometimes unnecessary for kernel-smoothing inference. 

\subsection{Invalidity caused by misspecified data dependence}
\label{sec 2.1}

Section \ref{sec 2.1} discusses the asymptotic invalidity risk of ignoring data dependence, particularly the dependence among running variables. We first use a simple random-design RDD model, Example \ref{ex 1}, to show that ignoring underlying data dependence may underestimate the asymptotic variance. Then, we develop two propositions showing that this underestimation immediately causes asymptotic invalidity. Neither bias awareness nor bias correction can fix this even if supplied with perfect asymptotic normality.

%In social science, there might be some cases where the data are nearly independent. A good example is locally interacted sequential social learning and informational cascades. Indeed, \citeA{banerjee1992simple} shows that, when later decision makers learn from choices made by their predecessors, rational behavior can generate herding in which individuals follow others rather than rely on their own private information. \citeA{bikhchandani1992theory} formalize the related notion of informational cascades and further note that similar mechanisms can arise even when decision makers observe only a few immediate predecessors. \citeA{anderson1997information} provide experimental evidence that such cascade behavior can indeed emerge in sequential decision environments. 

%We consider the following random-design RDD model whose running variables satisfy a minimal local interaction: individuals are linearly ordered, and the $i+1$-th individual can be influenced only by the immediately preceding $i$-th individual.
\begin{example} 
    \label{ex 1}
    Suppose the observations $\{(X_i,Y_i)\}_{i=1}^n$ admit the model
    \begin{align*}
        Y_i=f_-(X_i)\1[X_i\leq 0]+f_+(X_i)\1[X_i>0]+\varepsilon_i,
    \end{align*}
 where $f_-$ and $f_+$ have $S$-order $(S\geq 1)$ left and right derivatives at cutoff. $\{X_{i}\}_{i\geq 1}$ and $\{\varepsilon_{i}\}_{i\geq 1}$ are two mutually independent stochastic processes. $\{\varepsilon_{i}\}_{i\geq 1}$ is a mean-zero Gaussian process satisfying $\E[\varepsilon_{i}^2]=\sigma^2>0$ and $\text{Cov}(\varepsilon_{i},\varepsilon_{j})=\sigma^2(1+|i-j|)^{-\gamma}$, for some $\gamma>1$.  $\{X_{i}\}_{i\geq 0}$ admit
  \begin{align}
\label{eq markov}
    X_{i+1}=B_{i+1}X_i+(1-B_{i+1})Z_{i+1}.
\end{align}
Here $\{B_{i}\}_{i\geq 1}$ and $\{Z_{i}\}_{i\geq 1}$ are independent copies of $B_0$ and $Z_0$, where $X_0$ and $Z_0$ are uniformly distributed over $[-1,1]$ and $B_0\in\{0,1\}$ satisfies $\P(B_0=1)=\rho>0$. 
\end{example}
Data-generating process \eqref{eq markov} implies that individual $i+1$ repeats the $i$-th individual's reported value with positive probability and otherwise generates a new response. In fact, Example \ref{ex 1} is a minimal, stylized model that captures local interaction, which is widely studied in sequential social learning and informational cascades. See, for example, \citeA{banerjee1992simple},  \citeA{bikhchandani1992theory} and \citeA{anderson1997information}.

Suppose our target of interest is $\theta=\lim_{\varepsilon\to 0+}f_+(\varepsilon)-\lim_{\varepsilon\to 0-}f_-(\varepsilon)$ and our local linear estimator is generally formulated as $\tilde \theta_h=\sum_{i=1}^n(L_{ih,+}(0)-L_{ih,-}(0))Y_i$. Proposition \ref{prop underestimation} below quantifies the degree of asymptotic variance underestimation and the corresponding undercoverage when we ignore data dependence and studentize the estimator as if the data were independent.

\begin{proposition}[Variance underestimation and undercoverage]
\label{prop underestimation}
Consider Example \ref{ex 1} with $0<\rho<1$. Suppose $\tilde\theta_h$ is a local-linear RDD estimator defined as \eqref{eq local linear random} with kernel $K$ being a Lipschitz-continuous symmetric density with support $[-1,1]$. If $h\to0$ and $n^{-\frac{1}{2S+1}}/h\to 0$, we have
\begin{align}
\label{eq underestimation}
 \frac{\mathrm{Var}_{\mathrm{ind}}(\tilde\theta_h|X_1,...,X_n)}{\mathrm{Var}_{\mathrm{true}}(\tilde\theta_h\mid X_1,\ldots,X_n)}   :&=\frac{\sigma^2\sum_{i=1}^n(L_{ih,+}(0)-L_{ih,-}(0))^2}{\mathrm{Var}_{\mathrm{true}}(\tilde\theta_h\mid X_1,\ldots,X_n)}\notag\\
&\xrightarrow[n\to\infty]{\P}
\Big(1+2\sum_{k=1}^\infty\frac{\rho^k}{(1+k)^\gamma}\Big)^{-1}=:R(\rho,\gamma)\in (0,1),
\end{align}
where weight functions $L_{ih,\pm}(0)$ are introduced in \eqref{eq local linear random}. $\rho, \gamma$ are introduced in Example \ref{ex 1}. Consequently, provided that $\frac{\tilde\theta_h-\theta}{\sqrt{\mathrm{Var}_{\mathrm{true}}(\tilde\theta_h)}}\sim N(0,1)$ and $\frac{\tilde\theta_h-\theta}{\sqrt{\mathrm{Var}_{\mathrm{ind}}(\tilde\theta_h)}}$ is also normally distributed, 
\begin{align}
    \label{eq undercoverage}
   \lim_{n\to\infty} \P\Big(\theta\in \Big[\tilde\theta_h\pm z_{1-\alpha/2}\sqrt{\mathrm{Var}_{\mathrm{ind}}(\tilde\theta_h)}\Big]\Big)=2\Phi\Big(z_{1-\alpha/2}\sqrt{R(\rho,\gamma)}\Big)-1<1-\alpha,
\end{align}
where $R(\rho,\gamma)$ is introduced in \eqref{eq underestimation}.
\end{proposition}  
In general, \eqref{eq underestimation} and \eqref{eq undercoverage} in Proposition \ref{prop underestimation} reveal a pessimistic situation: perfect asymptotic normality and bias correction are still insufficient to address the risk of asymptotic invalidity caused by variance underestimation.

We now use the following two propositions to further show that this pessimistic result holds for nearly all pointwise inference frameworks depending on asymptotic normality and bias tackling. First, we introduce a mild assumption.
\begin{assumption}
    \label{as 1}
Suppose $\mathcal{F}_n$ is a function class allowed to depend on sample size $n$, and our target parameter of interest is $\theta(f)\in \mathbb{R}$ for some fixed $f\in \mathcal{F}_n$. Suppose $\hat \theta_n(f)$ is a well-defined estimator of $\theta(f)$ for each $f\in \mathcal{F}_n$ and $n\geq 1$. We then assume the following three points hold.
 \begin{itemize}
     \item [A1] We assume that $\sup_{f\in\mathcal{F}_n}|\E_f[\hat{\theta}_n(f)]-\theta(f)|=:\sup_{f\in \mathcal{F}_n}|b_n(f)|=:B_n$ is known, and for each $n\geq 1$, there exists $f_n^+\in \mathcal{F}_n$ such that $\E_{f_n^+}[\hat{\theta}_n(f_n^+)]-\theta(f_n^+)=B_n$.
     \item [A2] Suppose $\sigma_n$ is an oracle normalizer such that $Z_n:=\frac{\hat \theta_n(f)-\E[\hat \theta_n(f)]}{\sigma_n}$ has standard normal distribution for all $f\in \mathcal{F}_n$ and $n\geq 1$. 
     \item[A3] Let $\hat s_n$ be the estimator of $\sigma_n$ and we assume there exists a fixed constant $\beta>1$ such that $\lim_{n\to\infty}\P(|\frac{\sigma_n}{\hat s_n}|\geq \beta)=1.$
 \end{itemize}
\end{assumption}
A1-A2 in Assumption \ref{as 1} impose a perfect normal approximation and bias control for the estimator $\hat \theta_n(f)$, allowing us to isolate the variance underestimation described in A3.
\begin{proposition}
    \label{prop bias aware}
Under Assumption \ref{as 1}, for any fixed $\alpha\in (0, \frac{1}{2})$ and $\beta>1$, there exists a constant $\eta_{\alpha,\beta}>0$ depending only on $\alpha,\beta$ such that 
\begin{align}
\label{eq prop undercover eq1}
\limsup_{n\to\infty}\inf_{f\in \mathcal{F}_n}\P(\theta(f)\in [\hat\theta_n(f)\pm  \hat s_n\mathrm{cv}_{1-\alpha}(t_n)])<1-\alpha-\eta_{\alpha,\beta},
\end{align}
where $t_n=B_n/\hat s_n$ and $\text{cv}_{1-\alpha}(t)$ is the upper $\alpha$-quantile of folded normal distribution $|N(t,1)|$. 
\end{proposition}
\begin{proposition}
    \label{prop bias corrected}
Suppose that, for some fixed $f$, the exact estimation bias $b_n(f)$ mentioned in Assumption \ref{as 1} A1 is known. Then, together with A2 and A3 of Assumption \ref{as 1},  for any fixed $\alpha\in (0,0.5)$ and $\beta>1$, there exists an $\eta_{\alpha,\beta}'>0$ depending only on $\alpha,\beta$ such that
\begin{align*}
    \limsup_{n\to\infty}\P(\theta(f)\in [\hat \theta_n(f)-b_n(f)\pm \hat s_nz_{1-\frac{\alpha}{2}}])<1-\alpha-\eta_{\alpha,\beta}',
\end{align*}
where $z_{1-\alpha/2}$ indicates the $\alpha/2$-upper quantile of standard normal distribution.
\end{proposition}
Both Propositions \ref{prop bias aware} and \ref{prop bias corrected} show that neither bias-aware nor robust-bias-corrected inference introduced in \citeA{armstrong2018optimal,armstrong2020simple} and \citeA{calonico2018effect,calonico2022coverage} can overcome systematic variance underestimation, even under exact Gaussianity and perfect knowledge of the relevant bias. Therefore, proper bias treatment and distributional approximation cannot guarantee inferential validity when data dependence is ignored or misspecified. This leaves practitioners with the alternative of accounting for dependence in stochastic calibration, conventionally by modeling dependence through a long-run variance. Section \ref{sec 2.2} examines the difficulties and the awkwardness of this route.

\subsection{Difficulties and awkwardness in inference involving dependence}
\label{sec 2.2}

The previous section discussed the invalidity risk of ignoring dependence. However, in empirical economics, practitioners are usually reminded to model dependence when formulating the RDD model. A prominent example is the regression discontinuity in time (RDiT) model described as follows,
\begin{example} [rescaled RDiT]
    \label{ex 3}
   Provided that $t_0$ is the calendar time of cutoff, we define a rescaled regression discontinuity in time model as follows,
    \begin{align*}
        Y_{tT}=f_-(\frac{t-t_0}{T})\1[t\leq  t_0]+f_+(\frac{t-t_0}{T})\1[t>t_0]+\varepsilon_{tT},
    \end{align*}
    where, for any given $t_0-\frac{T}{2}\leq t\leq t_0+\frac{T}{2}$ and even $T\geq 2$, $\{\varepsilon_{tT}\}$ is some mean-zero time series with some unknown complex dependence. 
\end{example}
Even practitioners who choose to incorporate dependence in inference may still struggle with a range of difficulties, from choosing an appropriate dependence condition to finding a suitable modeling technique for correct studentization. We here focus on the following two difficulties.
\begin{itemize}
    \item [\textbf{O1}] Selecting a proper dependence condition that includes the dependence structure of the true data-generating process as a special case.
    \item [\textbf{O2}] Following the standard practice of modeling dependence through a long-run variance, practitioners face an unstable long-run variance and the impossibility of \textit{universally pointwise-consistent estimation} (see Definition \ref{def universal pointwise consistent}) under dynamic dependence (e.g., Example \ref{ex 2}).
\end{itemize}

Difficulty \textbf{O1} is relatively transparent in empirical research. After all, when we introduce a dependence condition into our statistical inference, we assume that this condition is general enough to incorporate the true data-generating process as a special case. One of the most widely used dependence conditions is the strong mixing condition (e.g., \citeA{doukhan2012mixing} and \citeA{bradley2005basic}). But \citeA{wu2005nonlinear} points out that this condition is usually hard to specify from data. Additionally, even some simple time series models do not obey this strong-mixing condition; see  \citeA{andrews1985nearly} for an example. Thus, many other important and general dependence conditions have been proposed to capture increasingly complex dependence. Good examples are physical dependence measures introduced in \citeA{wu2005nonlinear}, $\tau$-mixing condition exhibited in \citeA{dedecker2004coupling,dedecker2005new}, and mixingale condition demonstrated in \citeA{davidson1994stochastic}. In practice, this can be a serious difficulty because it pushes practitioners to choose one appropriate dependence condition from a long list, and some conditions are hard to specify from data.

For difficulty \textbf{O2}, Section \ref{sec 2.1} implies that correct studentization is crucial for the validity of a confidence interval calibrated by asymptotic standard normality. For the canonical rescaled RDiT model introduced in Example \ref{ex 3}, a long-run variance object naturally arises once serial dependence is taken seriously. When serial dependence is unknown, the standard inferential response is to model the dependence structure and estimate a long-run stochastic scale. Our approach also differs from self-normalization \citeA{shao2015self} and fixed-$b$ asymptotics (\citeA{kiefer2005new}), which avoid consistent long-run variance estimation but still rely on dependence-specific statistics and a functional central limit theorem. Here, the interval is implemented exactly as under independence, and its validity requires only summable covariances.

Moreover, under conventional weak-stationarity and short-range-dependence conditions, the leading variance of a local-polynomial estimator typically factors into a long-run variance of the disturbances\footnote{Given a real-valued process $\{X_t\}_{t\geq 1}$, its long-run variance is defined as $\lim_{T\to\infty}\Omega_T$, where $\Omega_T=\frac{1}{T}\text{Var}(\sum_{t=1}^TX_{t})$.} and a deterministic constant determined by the kernel, polynomial order, and local design. This further encourages practitioners to model the unknown dependence and estimate the long-run variance. In recent years, significant progress has been achieved in the area (e.g., \citeA{casini2024change}, \citeA{chan2024asymptotically} and \citeA{liu2026tail}), while in practice, things are usually much messier since the data-generating process may be totally non-stationary and the dependence law may change as the observation horizon expands. Similar phenomena have been widely discussed in economics. See, for example, \citeA{engle2002dynamic}, \citeA{primiceri2005time}, and \citeA{diebold2009measuring}.

We now use the following Example \ref{ex 2}, Propositions \ref{prop 1} and \ref{prop impossible} to show the aforementioned dynamic dependence can easily cause an awkward situation for practitioners intending to construct an interval by long-run variance\footnote{We say the process $\{X_{t}\}_{t\geq 1}$ has unstable long-run variance if $\liminf_{T\to\infty}\Omega_T<\limsup_{T\to\infty}\Omega_T$ holds strictly.} estimation. 

\begin{example}
    \label{ex 2}
    Suppose $\{g_{t}\}_{t\in \mathbb Z}$ is an independent and identically distributed Gaussian process such that $\E[g_{t}]=0$ and $\text{Var}(g_{t})=\sigma^2>0$. Suppose $\{I_m\}_{m\geq 1}$ is any given partition of the set of all positive integers and let  
    \begin{align*}
        a_t=\begin{cases}
   0,\ \ & \text{ if } t\in \bigcup_{m=1}^\infty I_{2m-1}; \\
   1,\ \  & \text{ if } t\in \bigcup_{m=1}^\infty I_{2m}.
   \end{cases}
    \end{align*}
    For any fixed $c>0$, define triangular array $\mathcal{E}_T(c):=\{\varepsilon_{tT}(c)\}_{t\leq T;T\geq 1}$ as follows,
    \begin{align*}
        \varepsilon_{tT}(c)=g_{t}+\frac{ca_T}{T}\sum_{j=1}^Tg_{t-j}, \ t=1,...,T,\ \  T\geq 1.
    \end{align*}
\end{example}
Some basic properties of Example \ref{ex 2} are as follows,
\begin{proposition}
    \label{prop 1}
    By letting $\Omega_T(c)=\frac{1}{T}\mathrm{Var}(\sum_{t=1}^T\varepsilon_{tT}(c))$, according to the definition of $\mathcal{E}_T(c)$ introduced in Example \ref{ex 2},  we obtain the following properties.
    \begin{itemize}
        \item [P1] For every given $T\geq 1$ and $c>0$, $\mathcal{E}_T(c)$ is a group of mean-zero Gaussian random variables such that $\mathrm{Cov}(\varepsilon_{tT}(c),\varepsilon_{t-k,T}(c))=0$ whenever $k> T$.
        \item [P2] For each $c>0$, $T\geq 1$ and $1\leq k\leq T$, we have 
        \begin{align*}
            \mathrm{Var}(\varepsilon_{tT}(c))=\sigma^2(1+\frac{c^2a_T}{T}),\ \mathrm{Cov}(\varepsilon_{tT}(c),\varepsilon_{t-k,T}(c))=\sigma^2[\frac{ca_T}{T}+\frac{c^2a_T(T-k)}{T^2}].
        \end{align*}
        \item [P3] For each $T\geq 1$ and $c>0$, $\Omega_T(c)=\sigma^2[1+a_T(\frac{c(T-1)}{T}+\frac{c^2}{T})+a_T\frac{c^2(T-1)(2T-1)}{3T^2}]$. Consequently, 
        \begin{align*}
         \liminf_{T\to\infty}\Omega_T(c)=\sigma^2,\qquad \limsup_{T\to\infty} \Omega_T(c)=\sigma^2(1+c+\frac{2c^2}{3}).
        \end{align*}
    \end{itemize}
\end{proposition}
 Recall that the serial dependence is usually unknown to practitioners. But model \eqref{model fixed-design} itself may have been sufficient to motivate practitioners to believe the existence of a long-run variance and estimate it. Unfortunately, together with Definition \ref{def universal pointwise consistent}, Proposition \ref{prop impossible} below shows that this nuisance-learning task can be fundamentally difficult.

\begin{definition}[Universal pointwise consistency (UPC)]
\label{def universal pointwise consistent}
Suppose $\{\mathcal E_T(c):T\ge1,c>0\}$ is the collection of triangular arrays introduced in Example \ref{ex 2}. For each $T\ge1$, let $\widehat\Omega_T:\mathbb R^T\to\mathbb R$ be a measurable estimator based on $T$ observations. We say that $\widehat{\Omega}_T$ has universal pointwise consistency (UPC) on this collection if, for every fixed $c>0$ and every $\eta>0$, its estimation error probability $\mathbb{P}(|\widehat{\Omega}_T-\Omega_T(c)|>\eta)$ converges to zero along every admissible triangular array, i.e., every triangular array generated by some partition $\{I_m\}_{m\geq 1}$ in Example \ref{ex 2}.
\end{definition}
Universal pointwise consistency (UPC) defined in Definition \ref{def universal pointwise consistent} is a natural requirement when inference is conducted over a family of triangular arrays. Intuitively, it requires an estimator to remain consistent along any admissible subsequence as the sample size diverges. This property holds in many standard parametric and nonparametric estimation procedures. For example, in RDiT model, local polynomial regressions can typically consistently estimate the target parameter along every triangular subarray satisfying some mild dependence conditions, like mixing or physical dependence measures. However, the following proposition shows that this property may be hard to achieve when we use data to capture long-run variance structures. 

\begin{proposition}[Impossibility of UPC]
\label{prop impossible}
    Based on the definition of $\mathcal{E}_T$ introduced in Example \ref{ex 2}, suppose $\hat\Omega_T$ is an estimator based on $T$-many observations. There exists no $\hat\Omega_T$ that has universal pointwise consistency $(\mathrm{UPC})$ on $\{\mathcal{E}_T(c):T\geq 1,c>0\}$. 
\end{proposition}
 Failure to achieve universal pointwise consistency (UPC) on $\{\mathcal{E}_T: T\geq 1,c>0\}$ has a substantive implication: under the maintained conditions introduced in Example \ref{ex 2} alone, consistency cannot be guaranteed automatically for every admissible DGP in this family of triangular arrays.

However, it is \textit{awkward} that this difficult nuisance-learning problem is sometimes unnecessary for kernel-smoothing inference. Under the very same data-generating process, Theorem \ref{th concentration fixed} and Proposition \ref{prop 1} P2 imply that the serial-covariance contribution to the variance of the local-polynomial smoother is of higher order for many standard bandwidth choices, so valid RDiT inference does not require estimating a long-run variance at all. Following the conventional route may therefore lead practitioners to solve a substantially harder problem than validity actually requires.

 In short, Section \ref{sec 2.2} reveals a vivid tension between “generality” and “learnability” in inference implementation under unknown and complex dependence.

\section{Dependence-Ignorance Validity Recovery for Fixed-Design RDD}
\label{sec 3}
For RDD inference with fixed-design running variables, this section develops an approach to recover the validity of a confidence interval built on the data-independence assumption when the data may actually be dependent.
\subsection{Model setups and assumptions}
\label{sec 3.1}
We now briefly review the basic settings of our RDD model with fixed-design running variables. Suppose that, for each given $T\geq 1$, our sample $\{Y_{tT}\}_{t=1}^T$ admits the following regression model,
\begin{align}
    \label{model fixed-design}
    Y_{tT}=f_{-}(x_{tT})\1[x_{tT}\leq 0]+f_{+}(x_{tT})\1[x_{tT}>0]+\varepsilon_{tT},\ \E[\varepsilon_{tT}]=0,\ \E[\varepsilon_{tT}^2]=\sigma^2,
\end{align}
and the target parameter of interest is 
\begin{align}
    \label{eq tau}
    \theta(f):=\lim_{\varepsilon\to 0+}f_{+}(\varepsilon)-\lim_{\varepsilon\to 0-}f_-(\varepsilon).
\end{align}
Example \ref{ex 3} is a special case of model \eqref{model fixed-design}, and we impose the following high-level assumption on the running variables.
\begin{assumption}[Fixed-design]
\label{as fixed-design}
Let $\{x_{1T},\ldots,x_{TT}\}\subset[-C_{\psi},C_{\psi}]$ be deterministic running variables, where $C_{\psi}>0$ is a known and fixed constant. Define $\Psi_T(x)=\frac{1}{T}\sum_{i=1}^T \mathbf{1}\{x_{iT}\le x\}$.
We assume there exists a non-negative and bounded function
$\Psi:\mathbb{R}\to\mathbb{R}_{+}$ such that
\[
0<\underline{c}_{\psi}\leq \frac{\Psi(y)-\Psi(x)}{y-x}\leq \bar c_{\psi}<\infty,
\qquad
\forall\ x,y\in [-C_{\psi},C_{\psi}],
\]
where $\underline{c}_{\psi}, \bar c_{\psi}>0$ are two fixed constants.  We also assume
\[
D_T:=\sup_{x\in[-C_{\psi},C_{\psi}]}|\Psi_T(x)-\Psi(x)|=o\left(\frac{\log T}{\sqrt{T}}\right),
\qquad T\to\infty.
\]
\end{assumption}
Assumption \ref{as fixed-design} incorporates the canonical rescaled RDiT model defined in Example \ref{ex 3}, in which we have $C_{\psi}=\frac{1}{2}$, $\Psi(x)=(x+\frac{1}{2})\1[|x|\leq \frac{1}{2}]+\1[x>\frac{1}{2}]$ and  $D_T=O(1/T)$. Similarly, Assumption \ref{as fixed-design} also incorporates  a more general setting, like $x_{tT}=G^{-1}(\frac{t-t_0}{T})$, where $G$ is an invertible positive-valued function (see \citeA{fan1992design,fan1993local} and \citeA{fan1992variable} for examples). $\Psi_T$ here is a counting device rather than an empirical distribution function, and thus $D_T$ is defined deterministically. The function $\Psi$ can be understood as a general limiting placement profile describing how the fixed-design running variables are asymptotically arranged. When $\Psi$ is a standard continuous distribution function defined on $[-C_{\psi},C_{\psi}]$, Assumption \ref{as fixed-design} implies some asymptotic random-like limiting placement of the fixed-design running variables. Some recent work on RDD also adopts a fixed-design assumption for the running variables. See, for example, \citeA{armstrong2018optimal,armstrong2020simple}.

Based on \citeA{gelman2019high}, we use the local linear smoother to estimate $\theta(f)$. In particular, it is defined as follows,
\begin{align}
    \label{eq local linear}
    \hat \theta_h&= \sum_{t=1}^T(l_{th}^+(0)-l^-_{th}(0))Y_{tT},\ \r(u)=(1,u)\in \R^2,\notag\\
    l_{th}^-(0)&=\frac{1}{Th}K(x_{tT}/h)\1[x_{tT}\in[-1, 0]] \Big(\e_0^{\top}S_{0,-}^{-1}\r(x_{tT}/h)\Big),\notag\\
    l_{th}^+(0)&=\frac{1}{Th}K(x_{tT}/h)\1[x_{tT}\in(0,1]] \Big(\e_0^{\top}S_{0,+}^{-1}\r(x_{tT}/h)\Big),\notag\\
    S_{0,-}&=\frac{1}{Th}\sum_{t=1}^TK(x_{tT}/h)\1[x_{tT}\in[-1,0]]\r(x_{tT}/h)\r^\top(x_{tT}/h),\notag\\
     S_{0,+}&=\frac{1}{Th}\sum_{t=1}^TK(x_{tT}/h)\1[x_{tT}\in(0,1]]\r(x_{tT}/h)\r^\top(x_{tT}/h),
\end{align}
where $K(u):[-1,1]\to[0,1]$ is a symmetric and Lipschitz continuous density function. Meanwhile, we choose the following residual-based estimator for the nuisance parameter $\sigma$,
 \begin{align}
     \label{eq variance estimation}
    & \hat \sigma^2:=\frac{1}{T}\sum_{t=1}^T(Y_{tT}-\hat f(x_{tT}))^2,\\ \hat f(x)=\hat f_{-}(x)\1[x\leq 0]&+\hat f_+(x)\1[x>0],\qquad \hat f_{\pm}(x)=\sum_{t=1}^Tl_{th}^{\pm}(x)Y_{tT}.\notag
 \end{align}
\eqref{eq variance estimation} is a standard residual-based variance estimator widely employed in nonparametric inference. \citeA{shen2020optimal} has shown that this is not an optimal choice, but we deliberately adopt the simplest residual-based estimator of the common marginal error variance, without attempting to account for cross-observation dependence. This choice reflects our intention to retain an elementary independence-based implementation rather than introduce dependence-aware variance estimation.

Additionally, we impose the following assumption of smoothness. 
\begin{assumption} [Smoothness]
\label{as smoothness}
For any integer \(S\geq 3\) and fixed $\varepsilon_0>0$, assume \(f\in\mathcal F_S(\varepsilon_0,M_S,\rho_S)\), where
\begin{align}
\label{eq funciton class}
\mathcal F_S(\varepsilon_0,M_S,\rho_S)
=\Big\{&f(x)=f_-(x)1[x\leq0]+f_+(x)1[x>0]:\\
&\left|
f_\pm(x+u)
-\sum_{s=0}^{S}
\frac{f_\pm^{(s)}(x)}{s!}u^s
\right|
\leq
M_S |u|^S\rho_S(|u|),\
\forall\,x\in[-\epsilon_0,\epsilon_0]\backslash\{0\},\ |u|\leq \varepsilon_0
\Big\}.\notag
\end{align}
Here \(M_S>0\) is a fixed constant and \(\rho_S:\mathbb R_+\to\mathbb R_+\) satisfies $\lim_{\delta\to 0+}\rho_S(\delta)\downarrow0$. Both \(M_S\) and \(\rho_S\) are independent of $f$. We only assume their existence, but their values are unknown.
\end{assumption}
 $S\geq 3$ means that, compared with the order of the local linear smoother, which is $1$, the regression functions $f_-$ and $f_+$ have higher-order smoothness at the cutoff. Besides being a standard assumption in nonparametric smoothing (e.g., \citeA{fan1993local} and \citeA{zhou2010simultaneous}), it is also widely employed in recent progress in robust nonparametric inference and its applications to causal inference, e.g., \citeA{calonico2019nprobust} and \citeA{calonico2022coverage}. Similar higher-order smoothness is also required for the theoretical guarantee and feasibility of coverage-error-optimal robust-bias-correction (CE-optimal RBC) inference introduced in \citeA{calonico2018effect}. In particular, CE-optimal RBC inference for a local polynomial smoother usually requires $p+2$ smoothness at the boundary point, where $p$ is the order of the local polynomial smoother. 

Additionally, we introduce the following assumption that describes misspecified data dependence.
\begin{assumption}[Misspecified Data Dependence]
\label{as misspecification 1}
We assume $\max_{t\leq T}\sqrt{\E[\varepsilon_{tT}^2]}=:E_2<\infty$, where $E_2$ is a fixed constant independent of $T\geq 1$. Moreover, we also assume that
\begin{itemize}
    \item [\rm{D1}] \rm{(Assumed Dependence)} \it{For the implementation of constructing a confidence interval of $\theta(f)$, we treat $\{\varepsilon_{tT}\}$ as mean-zero and independent random variables for each $T\geq 1$.} 
    \item [\rm{D2}] \rm{(True Dependence)} \it{For the actual dependence structure of $\{\varepsilon_{tT}\}$, provided that $Th\to\infty$ as $T\to\infty$, we assume}
    \begin{align*}
    &\sup_{t\leq T,T\geq 1} \sum_{k\neq t}|\mathrm{Cov}(\varepsilon_{tT},\varepsilon_{kT})|<\infty,\\
\limsup_{T\to\infty}\Omega_{Th}:&= \limsup_{T\to\infty}Th\mathrm{Var}(\sum^T_{t=1}(l_{th}^+(0)-l_{th}^{-}(0))\varepsilon_{tT})<\infty,
    \end{align*}
where $l_{th}^{+}(0)$ and $l_{th}^{-}(0)$ are introduced in \eqref{eq local linear}.
\item [\rm{D3}] \rm{(Non-degeneracy)} $\lim_{T\to\infty}\inf_{f\in \F}\P(\hat \sigma>\sigma_0)=1$ holds for some fixed $\sigma_0>0$.
\end{itemize}
\end{assumption}
The gap in dependence generality between D1 and D2 in Assumption \ref{as misspecification 1} describes the operational room for our validity-recovery procedure. D1 is only a working independence assumption used to implement the confidence interval, whereas D2 is the actual requirement imposed on the dependent data-generating process. The latter does not specify whether the disturbances are mixing, physically dependent, stationary, or generated by any particular dependence mechanism; it only controls the order of the variance of the local-linear RDD stochastic component and the summability of the covariance. Hence, the procedure allows the true dependence structure to be far richer than the independence model used for implementation. For instance, except for the widely discussed weakly dependent processes, Proposition \ref{prop 1} implies that Example \ref{ex 2} satisfies D2. 

D3 is only a nondegeneracy requirement. We do not require it to consistently estimate the standard deviation of the local-linear estimator under the true dependent DGP, nor does it require convergence to any deterministic long-run variance quantity. Indeed, the true dependence structure may remain misspecified in implementation. Instead, together with D1--D2, this condition guarantees that, uniformly over the function class $\F$, the coverage probability converges to the explicit level \(1-\alpha\). The validity procedure introduced in Section \ref{sec 3.3} does not attempt to repair misspecification by consistently learning the true dependence-adjusted variance.

\subsection{Observations and Motivations from Conventional Approach}
\label{sec 3.2}
This section outlines key observations from the conventional approach that strongly motivate our validity recovery procedure. Recall that, under the independence assumption, a natural way to construct a feasible version of $\mathbf{CI}_\mathrm{oracle}^\mathrm{ind}(\alpha)$ introduced in \eqref{eq CI oracle} is to simply plug in consistent estimators of bias $B_h$ and variance $\sigma C_{K,p}$, denoted as $\hat B_h$ and $\hat \sigma\hat C_{K,p}$, respectively, which yields the following independence-assumption-based and feasible confidence interval
\begin{align}
    \label{eq CI feasible p}
    &\qquad \mathbf{CI}_{p}^{\mathrm{ind}}(\alpha)=\Big[\hat \theta_h-\hat B_{h}\pm \hat \sigma\hat C_{K,p}z_{1-\alpha/2} T_{\text{eff}}^{-\frac{1}{2}}\Big],
\end{align}
Many bias-corrected RDD inference approaches fall into this regime, such as the aforementioned robust bias correction (RBC) and its recent improvement (see \citeA{cavaliere2026improvedinferencenonparametricregression}).

Based on the $\mathbf{CI}_p^{\mathrm{ind}}(\alpha)$ defined in \eqref{eq CI feasible p}, we highlight three key \textit{observations}.

\paragraph{Observation 1: A canonical asymptotic structure} Many modern nonparametric or semiparametric pointwise inference methods relying on asymptotic normality often admit the following canonical asymptotic structure: a nondegenerate Gaussian leading term and an asymptotically negligible remainder, namely,
\begin{align}
    \label{eq canonical}
    N(0,1)+o_\P(1) (\text{or}\ o(1)).
\end{align}
In addition to the RDD literature reviewed in Section \ref{sec 1}, other examples are undersmoothing (e.g., \citeA{li2007nonparametric}), robust bias correction (e.g., \citeA{calonico2022coverage}), and orthogonal machine learning for additive model (e.g., \citeA{bach2025estimation}). These methods often differ in how the remainder is controlled, but their first-order inferential goal is of this same form. Therefore, formula \eqref{eq canonical} points out a canonical structure we should achieve but never restricts us to the regime of asymptotic normality based on the original centered estimator.

\paragraph{Observation 2: source of Gaussianity} Under the independence assumption, given nuisance estimators $(\hat B_h,\hat \sigma \hat C_{K,p})$, the validity of $\mathbf{CI}_p^\mathrm{ind}(\alpha)$ for the local linear smoother often relies on the weak convergence  
\begin{align*}
    \frac{(\hat \theta_h-\theta(f)-\hat B_h)}{\hat \sigma\hat C_{K,p}/\sqrt{T_{\text{eff}}}}\xrightarrow{w} N(0,1).
\end{align*}
However, for every given integer $m_{T_{\text{eff}}}\geq 1$ and $(\hat B_h, \hat \sigma \hat C_{K,p} )$, by letting $G_{jT}\overset{\text{i.i.d}}{\sim} N(\hat B_h,(\hat \sigma\hat C_{K,p})^2)$, $1\leq j\leq m_{T_{\text{eff}}}$, we also have 
\begin{align}
\label{eq gaussian noise}
  \frac{m_{T_{\text{eff}}}^{-1/2}\sum_{j=1}^{m_{T_{\text{eff}}}}(G_{jT}-\hat B_h)}{\hat \sigma\hat C_{K,p}}\sim N(0,1),
\end{align}
where $m_{T_{\text{eff}}}$ is an integer that diverges as $T_{\text{eff}}\to\infty$. This gives us two hints: (i) Asymptotic normality does not monopolize the source of pivot normality; (ii) Slutsky's theorem may not be the unique way for \((\hat B_h,\hat \sigma\hat C_{K,p})\) to enter the inference implementation.

\paragraph{Observation 3: a simple but trivial solution} A simple but trivial solution to bypass the invalidity caused by dependence misspecification is to simply stretch the interval. Recall that,  under some given smoothness index $S>0$ and summability condition of the estimator's covariance, using an optimal bandwidth $h_{\text{opt}}=T^{-\frac{1}{2S+1}}$ often yields an estimator $\hat \theta_{h_{\text{opt}}}$ (e.g., a local polynomial estimator of order $S-1$) with both bias and standard deviation of order $T^{-\frac{S}{2S+1}}$. Then, since Assumption \ref{as misspecification 1} D2 usually implies a constant-level difference between the long-run variance and the individual one,  multiplying an arbitrary diverging factor $d_T$ leads to an honest but trivial confidence interval, i.e.,
\begin{align}
\label{eq trivial coverage}
    \P(\theta(f)\in [\hat \theta_{h_ \text{opt}}\pm \hat \sigma\hat C_{K,p}z_{1-\alpha/2}(d_TT^{-\frac{S}{2S+1}})])\to 1.
\end{align}
This shows that the dependence misspecification can be absorbed by some sort of rate difference, but the resulting interval is trivial in the sense of Definition \ref{def nontrivial} and is therefore excluded by \ref{G2}. Section~\ref{sec 5.3} shows that the AoRG interval introduced in Section~\ref{sec 5} approaches an interval of this type as the number of generations grows; this limit is likewise trivial in the sense of Definition \ref{def nontrivial} and is therefore excluded.

\paragraph{Motivation} Observation 1 reveals a destination-like asymptotic structure, $N(0,1)+o_{\P}(1)$ (or $o(1)$); Observation 2 reminds us of another path of attaining this structure; Formula \eqref{eq trivial coverage} in Observation 3 rules out trivial coverage inflation. Thus, in the upcoming Section \ref{sec 3.3}, based on Observation 2, we build a validity recovery procedure that has the asymptotic representation $N(0,1)+o_{\P}(1)$ (or $o(1)$) but bypasses the risk and difficulties mentioned in Section \ref{sec 2}. In particular, this procedure rules out trivial intervals in the sense of Definition \ref{def nontrivial}, such as \eqref{eq trivial coverage}, and obtains asymptotically exact $(1-\alpha)$-coverage for each fixed $\alpha\in (0,1)$.

%It is often true that combining asymptotic normality and Slutsky's theorem, and the independence assumption implies, for every fixed $\alpha\in (0,1)$,
%\begin{align}
%\label{eq asymptotic coverage}
    %\P(\theta(f)\in \mathbf{CI}^{\mathrm{ind}}(\alpha))=\P\left(
%\left|
%\frac{\hat\theta_h-\theta(f)-\hat B_{h}}
%{\hat \sigma\hat C_{K,p}/\sqrt{Th}}
%\right|
%\le z_{1-\alpha/2}
%\right)\xrightarrow[n\to\infty]{} 1-\alpha.
%\end{align}
%However, Example \ref{ex 1} and Proposition \ref{prop bias corrected} have revealed that, when the data-generating process has complex dependence, \eqref{eq asymptotic coverage} may fail, particularly when $\hat \sigma\hat C_{K,p}$ underestimates the true asymptotic variance. Thus, it seems unrealistic to wish for dependence-ignorance implementation and asymptotic validity at the same time.m.

%The validity-recovery procedure developed in the incoming Section \ref{sec 3.3} exploits precisely \ref{R1}-\ref{R3}. Rather than requiring the original dependent data to generate the desired Gaussian calibration, we construct an auxiliary Gaussian component and embed the feasible bias and variance ingredients directly into
%this artificial randomness. In this way, the familiar Gaussian distribution remains the inferential anchor, while the source of the pivotal randomness and the way in which nuisance estimates enter the construction are fundamentally altered. The following sections formalize this alternative route and show how it can be used to recover the validity of independence assumption-based RDD inference under misspecified data dependence

\subsection{Validity Recovery}
\label{sec 3.3}
Motivated by Sections \ref{sec 2} and \ref{sec 3.2}, we seek a validity-recovery procedure that, while ignoring data dependence in implementation, achieves Goals \ref{G2}--\ref{G3} and, in addition, preserves the asymptotic leading constant of the independence-assumption-based benchmark. We now show that, for some given widely-used $(\hat B_h, \hat \sigma\hat C_{K,p})$, these requirements can be met.

First, we inherit the standard settings of $(\hat B_h,\hat\sigma \hat C_{K,p})$ widely used in RDD inference under the data-independence assumption. For each given $T\geq 1$, local linear estimator $\hat\theta_h$ has the exact smoothing bias
\begin{align*}
B_h
:=&\sum_{t=1}^{T}l^+_{th}(0)
\left[f_+(x_{tT})-f_+(0+)\right]
-\sum_{t=1}^{T}l^-_{th}(0)
\left[f_-(x_{tT})-f_-(0-)\right].
\end{align*}
We can follow \citeA{calonico2018effect,calonico2022coverage} and do the bias correction by estimating higher-order derivatives of $f_{\pm}$. Note that $B_h$ is a well-defined functional of the original regression function and, under Assumption \ref{as smoothness}, can be estimated by the plug-in method. Motivated by \citeA{cavaliere2026improvedinferencenonparametricregression}, we introduce bias and variance proxies
\begin{align}
      \label{eq bias correction}
      \hat B_{h}
&:=\sum_{t=1}^{T}l^+_{th}(0)
\left[\hat f_{+,b}(x_{tT})-\hat f_{+,b}(0+)\right]-\sum_{t=1}^{T}l^-_{th}(0)
\left[\hat f_{-,b}(x_{tT})-\hat f_{-,b}(0-)\right],\\
&\hat f_{+,b}(x)=\sum_{j=1}^Tw_{jb}(x;0,1)Y_{jT},\qquad \hat f_{-,b}(x)=\sum_{j=1}^Tw_{jb}(x;-1,0)Y_{jT},\notag
\end{align}
 where $b>0$ is a pilot bandwidth. The weight function $w_{jb}(x;a,b)$ is introduced in Appendix \ref{app concentration: fixed} with $q=S$.  

Additionally,  since we have ignored the underlying dependence (see Assumption \ref{as misspecification 1}) and assumed homoscedasticity (see \eqref{model fixed-design}), $\text{Var}_{\text{ind}}(\hat \theta_h)=\sigma^2\sum_{t=1}^T(l_{th}^+(0)-l_{th}^-(0))^2.$ Together with $\sigma$ estimator \eqref{eq variance estimation}, a widely employed plug-in-type studentizer is
\begin{align}
    \label{eq variance proxy fixed}
    \hat \sigma\hat C_{K,1}=\Big\{\Big[\frac{1}{T}\sum_{t=1}^T(Y_{tT}-\hat f(x_{tT}))^2\Big]\Big[Th\sum_{t=1}^T(l^+_{th}(0)-l^-_{th}(0))^2\Big]\Big\}^{\frac{1}{2}}.
\end{align}
Importantly, this studentizer is intentionally based on the variance of the original local-linear estimator under the independence assumption, without accounting for either cross-observation covariances or the additional variability induced by bias correction. We retain this deliberately elementary construction as an input to our validity-recovery procedure.

Then, for the local linear RDD estimator, an independence-assumption-based and feasible confidence interval can be formulated as 
\begin{align}
    \label{eq CI feasible}
     \mathbf{CI}_{1}^{\mathrm{ind}}(\alpha)=\Big[\hat \theta_{h_0}-\hat B_{h_0}\pm \underbrace{\hat \sigma_{h_0}\hat C_{K,1,h_0}z_{1-\alpha/2}}_{\text{leading constant}} \underbrace{(Th_0)^{-\frac{1}{2}}}_{\text{anchor's own bandwidth}}\Big].
\end{align}
Here the subscript $h_0$ indicates that \eqref{eq local linear}, \eqref{eq bias correction}, \eqref{eq variance estimation} and \eqref{eq variance proxy fixed} are evaluated at the anchor's own bandwidths $(h_0,b_0)$, chosen as under independence, with $h_0\to0$ and $D_T/h_0\to0$ (e.g., $h_0\asymp T^{-1/(2s+1)}$ for any $s\geq1$; this implies $Th_0\to\infty$); without the subscript, these quantities are evaluated at the bandwidths $(h,b)$ of Theorem \ref{th 1}. Validity recovery inherits the form of \eqref{eq CI feasible}, not its bandwidths. In particular, the effective sample size $m_T$ of the recovered interval is not tied to the anchor's $Th_0$ and can exceed it: for the conventional choice $h_0\asymp T^{-1/5}$, $m_T\asymp T^{\frac{2S}{2S+1}}/\log T$ exceeds $Th_0\asymp T^{4/5}$ for every $S\geq3$ (Remark \ref{remark 1}).

Section \ref{sec 2.1} shows that \eqref{eq CI feasible} may be invalid. We thus design the following two-step validity-recovery procedure to restore validity without modeling dependence or changing the form of the baseline local linear estimator and of the studentizer.

Our procedure has two main steps: \textit{noise injection} and \textit{interval construction}.

\paragraph{Step 1. Noise injection} According to $(\hat B_h, \hat \sigma^2\hat C^2_{K,1})$ introduced above, for each given $T\geq 1$, we generate auxiliary samples $\{G_{iT}\}_{i=1}^{m_T}$ as follows,
\begin{align}
\label{eq noise}
   G_{iT}\overset{iid}{\sim} G_0,\ 1\leq i\leq m_T ,\ \text{and}\  G_0\sim N(\hat B_h, (\hat \sigma\hat C_{K,1})^2),
\end{align}
where $m_T\geq 1$ is an integer determined by $T,b,h$.
\paragraph{Step 2. Interval construction} Let $\bar G_T:=\frac{1}{m_T}\sum_{i=1}^{m_T}G_{iT}$. For each given $f\in \mathcal F_S$, our confidence interval of $\theta(f)$ is constructed as follows,
\begin{align*}
    \mathbf{CI}_\mathrm{recover}^\mathrm{ind}(\alpha):&=\Big [\hat \theta_h-\bar G_T\pm \Big\{\hat \sigma\hat C_{K,1}\mathrm{cv}_{1-\alpha}(1/\sqrt{l_T})\Big\}m_T^{-\frac{1}{2}}  \Big],
\end{align*}
where $l_T\to\infty$ is a diverging sequence specified in Theorem \ref{th 1}, so that $1/\sqrt{l_T}\to 0$. $\text{cv}_{1-\alpha}(t)$ is the $1-\alpha$ quantile of the folded normal distribution $|N(t,1)|$. \\

The preceding two steps define a randomized validity-recovery mapping from the possibly-invalid $\mathbf{CI}_1^\mathrm{ind}(\alpha)$ interval to a valid $\mathbf{CI}_\mathrm{recover}^\mathrm{ind}(\alpha)$. Let $\mathcal{G}_{m_T}:=\{G_{iT}\}_{i=1}^{m_T}.$ We write this mapping as
\begin{align}
\label{eq recovery mapping}
    \mathcal{R}\!\left(
\mathbf{CI}_1^{\mathrm{ind}}(\alpha);
\mathcal{G}_{m_T}
\right)
=
\mathbf{CI}_{\mathrm{recover}}^{\mathrm{ind}}(\alpha).
\end{align}
We refer to the mapping $\mathcal{R}$ as a \textit{validity-recovery} mapping because, as shown in Theorem \ref{th 1}, $\mathcal{R}$ takes the possibly invalid $\mathbf{CI}_1^{\mathrm{ind}}(\alpha)$ as its inferential input and, through the auxiliary Gaussian randomization $\mathcal{G}_{m_T}$, transforms it into a valid $\mathbf{CI}_{\mathrm{recover}}^{\mathrm{ind}}(\alpha)$. 

\begin{remark} [Why Gaussian noise?]
\label{remark Gaussian}
  \rm{The choice of Gaussian noise is not arbitrary. Our validity recovery should \textit{not} be read as a new inference principle, but as a procedure that recovers the validity of our inferential anchor, the independence-assumption-based interval $\mathbf{CI}_1^{\mathrm{ind}}(\alpha)$ defined in \eqref{eq CI feasible}. Thus, preserving its $\alpha$-quantile is mandatory. Gaussian injection achieves precisely this goal: it replaces the unreliable sampling-based calibration with an auxiliary Gaussian pivot, using independence-based bias and scale ingredients of the same form. }
\end{remark}    
\begin{remark}[Why these bias and variance estimators?]
\label{remark bias and variance}
   \rm{Note that one benefit of the bias estimator defined in \eqref{eq bias correction} is that it allows us to nearly exhaust the assumed smoothness $S\geq 3$ (see Remark \ref{remark 1}). The standard deviation estimator introduced in \eqref{eq variance proxy fixed} is also standard and widely employed in studentization under the data-independence assumption. 
    However, the choice of bias and variance estimators is not mandatory for implementing validity recovery. We simply inherit their definitions from our inference anchor, $\mathbf{CI}_1^{\mathrm{ind}}(\alpha)$. This means that the two-step validity recovery procedure above can be immediately applied to the cases where we use some other studentizers and bias estimators to construct $\mathbf{CI}_1^{\mathrm{ind}}(\alpha)$. See Corollary \ref{th DA} in Section \ref{sec 3.4} for an example. Over there, the validity recovery is implemented on a confidence interval relying on dependence-aware studentizer and pure undersmoothing-type bias control, in which the $\hat \theta_h$ is a local linear estimator with undersmoothing bandwidth and $\hat B_h=0$.}
\end{remark}

%Importantly, $\mathcal R$ preserves the independence-assumption-based studentization used in $\mathbf{CI}_{\mathrm{feasible}}^{\mathrm{ind}}(\alpha)$, without requiring long-run variance modeling, while the radius of $\mathbf{CI}_{\mathrm{recover}}^{\mathrm{ind}}(\alpha)$ has the same asymptotic leading constant as that of its independence-assumption-based benchmark.

\begin{theorem}
    \label{th 1}
  Based on the $\varepsilon_0$ introduced in Assumption \ref{as smoothness}, suppose $0<h< \varepsilon_0$ satisfies $h\to0$, $T^{-\frac{1}{2S+1}}/h=o(1)$ and $b=T^{-\frac{1}{2S+1}}$. Let $m_T=[l_T^{-1}T^{\frac{2S}{2S+1}}]+1$and $\bar G_T=\frac{1}{m_T}\sum_{i=1}^{m_T}G_{iT}$, where $G_{iT}$ is introduced in \eqref{eq noise}. Under Assumptions \ref{as fixed-design}-\ref{as misspecification 1}, for every fixed $\alpha\in (0,1)$ and each given $l_T\to\infty$ satisfying $l_T=o(T^{\frac{S}{2S+1}})$,
    \begin{align}
    \label{eq CI recover}
    {\mathbf{CI}_{\mathrm{recover}}^{\mathrm{ind}}(\alpha)}:&=\Big [\hat \theta_h-\bar G_T\pm \underbrace{\Big\{\hat \sigma\hat C_{K,1}\mathrm{cv}_{1-\alpha}(1/\sqrt{l_T})\Big\}}_{\mathrm{leading\  constant}}m_T^{-\frac{1}{2}} \Big]
    \end{align}
   has the following asymptotic properties:
   \begin{itemize}
       \item \rm{(\textbf{Uniformly exact coverage}) $\sup_{f\in \F}|\P(\theta(f)\in \mathbf{CI}_\mathrm{recover}^\mathrm{ind}(\alpha))-(1-\alpha)|\xrightarrow[T\to\infty]{}0$}
       \item (\textbf{Recovery of leading constant}) Suppose, in addition, that $\Psi$ in Assumption \ref{as fixed-design} is differentiable from the left and from the right at $0$, and that $f_-$ and $f_+$ are bounded and piecewise continuous on $[-C_\psi,0]$ and $(0,C_\psi]$, respectively. Then, for the anchor \eqref{eq CI feasible} implemented at any $h_0$ with $h_0\to0$ and $D_T/h_0\to0$ (e.g., $h_0\asymp T^{-1/(2s+1)}$ for any $s\geq1$), the leading constants of \eqref{eq CI feasible} and \eqref{eq CI recover} satisfy
       \begin{align*}
           \frac{\mathrm{leading\ constant}( \mathbf{CI}_\mathrm{recover}^\mathrm{ind}(\alpha))}{\mathrm{leading\ constant}( \mathbf{CI}_1^\mathrm{ind}(\alpha))}\xrightarrow[T\to\infty]{\P}1.
       \end{align*}
       More specifically, for every fixed $\alpha\in (0,1)$, we have $\text{cv}_{1-\alpha}(1/\sqrt{l_T})\to z_{1-\alpha/2}$ as $T\to\infty$.
   \end{itemize}
   Moreover, $\mathbf{CI}_\mathrm{recover}^\mathrm{ind}(\alpha)$ is non-trivial in the sense of Definition \ref{def nontrivial}.
\end{theorem}
\begin{remark}[Undersmoothing factor $l_T$ and higher-order smoothness]
\label{remark 1}
  \rm{The sequence \(l_T\) is similar to the “undersmoothing factor” used in nonparametric smoothing. Both govern the extent to which inferential bias is rendered negligible relative to the stochastic calibration scale. Hence, $l_T$ governs the additional efficiency cost relative to the \(S\)-smooth minimax benchmark. To see this, as $l_T\to\infty$, $\mathrm{cv}_{1-\alpha}(1/\sqrt{l_T})=z_{1-\alpha/2}(1+\frac{1}{2l_T}+o(l_T^{-1}))$. Therefore, the radius of $\mathbf{CI}_{\mathrm{recover}}^{\mathrm{ind}}(\alpha)$ satisfies $\hat \sigma \hat C_{K,1}z_{1-\alpha/2}(1+\frac{1}{2l_T}+o(l_T^{-1}))\sqrt{l_T}\,T^{-\frac{S}{2S+1}}$. If the requirement $l_T\to\infty$ were ignored, the asymptotic leading term of the radius would favor a constant-order choice of $l_T$. However, validity recovery requires $l_T$ to diverge. Hence, choose $l_T$ to diverge as slowly as possible. We here choose \(l_T=\log T\).  When $l_T=\log T$, the radius is $$\hat\sigma \hat C_{K,1}z_{1-\alpha/2}\sqrt{\log T}\,T^{-\frac{S}{2S+1}}\{1+o(1)\},$$ which is minimax-optimal up to a logarithmic factor.\footnote{Here minimax-optimality refers to the shrinking rate: apart from the data-dependent scale $\hat\sigma\hat C_{K,1}$, the radius shrinks at the rate $\sqrt{\log T}\,T^{-\frac{S}{2S+1}}$, which matches the $S$-smooth minimax benchmark $T^{-\frac{S}{2S+1}}$ up to $\sqrt{\log T}$. This rate is also attained in the worst-case sense if $\mathcal F_S(\varepsilon_0,M_S,\rho_S)$ is further restricted to functions with $\sup_{x\in[-C_\psi,0]}|f_-(x)|+\sup_{x\in(0,C_\psi]}|f_+(x)|\leq M$ for a fixed constant $M$. Indeed, the local linear weights $l^{\pm}_{th}(x)$ in \eqref{eq variance estimation} have $\ell_1$ norms bounded uniformly in $x$, by the argument leading to \eqref{3}--\eqref{4}. Hence the fitting bias in \eqref{eq variance estimation} is bounded by a multiple of $M$, and Assumption \ref{as misspecification 1} gives $\sup_f\E[\hat\sigma^2]<\infty$ over the restricted class. Since $\hat C_{K,1}$ depends only on the design and is bounded by \eqref{3}--\eqref{4}, the expected length of $\mathbf{CI}_{\mathrm{recover}}^{\mathrm{ind}}(\alpha)$ is $O(\sqrt{\log T}\,T^{-\frac{S}{2S+1}})$ uniformly over the restricted class.} This means that, when smoothness order $S$ is much larger than the order of the local linear smoother, $1$, the shrinking rate of our  $\mathbf{CI}_{\mathrm{recover}}^{\mathrm{ind}}(\alpha)$ is not restricted to the RMSE optimal rate $T^{-\frac{2}{5}}$. This is the gain anticipated in \ref{G2}: with $l_T=\log T$, the effective sample size $m_T$ is of order $T^{\frac{2S}{2S+1}}/\log T$, which exceeds the $T^{4/5}$ order of conventional local linear inference for every $S\geq 3$.}
\end{remark}
\begin{remark} [Why a randomized center?]
    \label{remark 2}
   \rm{The randomized center $\bar G_T$ reflects a fundamental distinction between conventional inference and our validity-recovery procedure. In conventional inference, the nondegenerate Gaussian pivot arises from the sampling fluctuation of the original baseline estimator. In contrast, our procedure replaces this source of pivotal randomness with the average of auxiliary Gaussian draws. As shown in \eqref{eq redefine}, the original data-dependent component becomes asymptotically negligible, while the average of auxiliary Gaussians supplies the nondegenerate pivot. Consequently, the recovered interval necessarily has a randomized center, \(\hat{\theta}_h-\bar{G}_T\). This randomization is therefore not incidental, but a direct consequence of changing the source of stochastic calibration. Variation of the confidence set across realizations of the auxiliary randomness is a common feature of procedures based on external randomization, such as randomized tests \cite{lehmann2022testing}, selective inference with a randomized response \cite{tian2018selective}, data fission \cite{leiner2025data} and data thinning \cite{neufeld2024data}; see also \citeA{ren2023derandomizing} on derandomization. The relevant question is therefore whether this variability can be controlled. Section~\ref{sec 5} shows that averaging over repetitive generation (AoRG) controls it and delivers further gains at the same time: it reduces the conditional variance of the randomized center by the factor $1/K$, enlarges the range of tolerated displacements and strengthens the protection against undercoverage, all without changing the interval radius (Theorem~\ref{thm:aorg-finite}).}
\end{remark}

 We now highlight four noteworthy points of Theorem \ref{th 1}.

\paragraph{Source of Gaussianity and non-trivial coverage}
To see this distinction, write
\begin{align}
\label{eq new decomposition}
    \hat\theta_h-\bar G_T-\theta(f)&=(\hat\theta_h-E[\hat\theta_h])+(B_h-\hat B_h)-(\bar G_T-\hat B_h)\notag \\
&:=\underbrace{S_{1T}+S_{2T}}_{\text{displacement (random bias)}}-(\bar G_T-\hat B_h).
\end{align}
Compared with \eqref{eq decomposition}, recall that a conventional inflation argument would make the radius asymptotically dominate the center's entire fluctuation, thereby driving coverage to one and eliminating \(\alpha\)-interpretability. Our construction instead makes only the data-dependent displacement \(S_{1T}+S_{2T}\) negligible relative to the scale of the auxiliary randomization. Conditional on the original sample set $\mathcal{D}_T$, the standardized auxiliary component remains exactly Gaussian,
\begin{align}
\label{eq redefine}
    \frac{\hat\theta_h-\bar G_T-\theta(f)}
{\hat\sigma\hat C_{K,1}/\sqrt{m_T}}
=
-
\underbrace{\frac{\bar G_T-\hat B_h}
{\hat\sigma\hat C_{K,1}/\sqrt{m_T}}}_{ N(0,1)|\mathcal{D}_T\atop \text{exogenous randomness}}
+
\underbrace{\frac{S_{1T}+S_{2T}}
{\hat\sigma\hat C_{K,1}/\sqrt{m_T}}}_{o_\P(1)},
\end{align}
This supplies a nondegenerate source of pivotal randomness even after the original sampling fluctuation has become asymptotically negligible. Under the nondegeneracy condition in Theorem \ref{th 1}, the standardized displacement vanishes while the standardized auxiliary Gaussian component does not, and the recovered interval therefore has the same canonical nondegenerate limit \([Z\pm z_{1-\alpha/2}]\), \(Z\sim N(0,1)\), as a correctly calibrated Gaussian oracle interval, rather than coverage converging trivially to one. This is the non-triviality stated in Theorem \ref{th 1}.

\paragraph{Redefining inferential roles} The differences between decompositions \eqref{eq decomposition} and \eqref{eq new decomposition} also redefine the roles of bias and stochastic calibration in RDD inference. First, it actually yields
a decoupling property of bandwidth selection. Note that the discussion above actually reveals that the main bandwidth is no longer required to balance smoothing bias against sampling variation to produce a Gaussian pivot: the centered estimator $S_{1T}$, bias estimation error $S_{2T}$, and Gaussian calibration ($\bar G_T-\hat B_h$) play separate roles. Under this circumstance, $S_{1T}+S_{2T}$ serves as a \textit{random bias} in the new calibration framework, while the \textit{exogenous stochastic calibration} (Gaussian noise) preserves the nontrivial asymptotic $1-\alpha$ coverage. In this sense, validity recovery is random-bias-aware inference: the exogenous Gaussian injection of data fission supplies the pivot, and the folded-normal calibration of bias-aware inference absorbs the random bias $S_{1T}+S_{2T}$, with $1/\sqrt{l_T}$ in the role of the standardized bias bound. We thus only need to choose the bandwidths, $h$ and $b$, that make $|S_{1T}+S_{2T}|$ as small as possible. In short, we present the following table.
\begin{table}[H]
\centering
\begin{tabular}{|p{0.43\textwidth}|p{0.43\textwidth}|}
\hline
\textbf{Conventional Approach} & \textbf{Validity Recovery} \\
\hline
Tuning and calibration are coupled.
&
Tuning and calibration are decoupled.
\\
\hline

Bandwidths must jointly manage bias and sampling variation to support a native Gaussian pivot.
&
Bandwidths only need to reduce the displacement (random bias) $S_{1T}+S_{2T}$ as much as possible
\\
\hline

Non-degenerate calibration randomness comes from the asymptotic Gaussianity of the original-data-based centered estimator.
&
Non-degenerate calibration randomness comes separately from the exogenous and auxiliary Gaussian noise.
\\
\hline
\end{tabular}
\end{table}

\paragraph{Decouple dependence from calibration} Recall that the Gaussian noise $G_{iT}$ actually opens a fresh conditional probability space, in which the $1-\alpha$ calibration of the radius has been done. The dependence of the raw data thus only matters in $S_{1T}+S_{2T}$. This immediately decouples the original data dependence and stochastic calibration. But, rather than a distributional approximation requirement, $o_{\P}(1)$ is actually a concentration requirement, in which dependence-adjusted studentization is not needed. This also explains why we introduce Assumption \ref{as misspecification 1}. As for the possibility that $\hat \sigma\hat C_{K,1}$ may be inconsistent with respect to the true standard deviation, we highlight two points: (i) This inconsistency is benign to asymptotic validity and correct coverage; (ii) This is the cost we have to pay if practitioners have chosen to ignore the data dependence.

\paragraph{Retain canonical structure without discarding information}
Equation \eqref{eq redefine} shows that our validity-recovery procedure retains the canonical structure $N(0,1)+o_\P(1)$ defined in \eqref{eq canonical} but changes the origin of its two components.
%In conventional inference, the nondegenerate Gaussian leading term is generated by the sampling fluctuation of the original estimator, while bias, nuisance-estimation errors, and other approximation errors are required to enter the remainder. In contrast, under validity recovery,
However, a noteworthy point is 
\begin{align*}
    \text{exogenous randomness}\neq\text{discard original-data information}.
\end{align*}
Because a Gaussian law is fully characterized by its mean and variance, we parameterize the conditional distribution by the same two original-data-dependent nuisance quantities, \((\widehat B_h,\widehat\sigma\hat C_{K,1})\), which is always required for feasible inference under the independence assumption. Therefore, when the inference implementation relies on the data-independence assumption, the original data still determine the location and scale of the exogenous Gaussian noise.

\subsection{Modeling dependence: an enhancement rather than a necessity}
\label{sec 3.4}
Section \ref{sec 3.3} shows that we can decouple data dependence from stochastic calibration by introducing exogenous Gaussian randomness. In particular, the stochastic fluctuation generated by the original dependent observations is absorbed into an asymptotically negligible displacement, while the auxiliary Gaussian component supplies the nondegenerate randomness required for asymptotically exact inference. Consequently, the validity of our recovered confidence interval does not rely on consistently estimating the true dependence-adjusted variance. Nevertheless, asymptotic validity does not eliminate potential finite-sample difficulties. When the independence-assumption-based studentizer substantially underestimates the true stochastic scale, the resulting constant-order discrepancy must be absorbed by the diverging factor \(l_T\). Although this discrepancy becomes asymptotically negligible under Theorem 1, a slowly diverging \(l_T\) may provide insufficient compensation at empirically relevant sample sizes, potentially resulting in noticeable undercoverage.

 This raises a natural question: \textit{Once dependence estimation is no longer necessary for validity, can a studentizer with a dependence-aware component, such as a long-run variance estimator, still improve the finite-sample performance of validity recovery?}

  This section suggests an optimistic answer. According to \eqref{eq variance proxy fixed}, the standard-deviation proxy used for the independence-assumption-based interval is $\hat\sigma\hat C_{K,1}$. Let $\widehat{\mathrm{sd}}_{\mathrm{DA}}$ be any given standard deviation estimator that may contain a dependence-aware (DA) component, such as a heteroskedasticity and autocorrelation consistent (HAC) estimator, and $\widehat{\mathrm{sd}}_{\mathrm{DA}}$ includes $\hat\sigma\hat C_{K,1}$ as a special case. Likewise, let $\check B_h$ be any given bias-correction quantity used by the anchor interval. It includes $\hat B_h$ in \eqref{eq bias correction} as a special case, and $\check B_h=0$ corresponds to pure undersmoothing.

 Let the dependence-aware version of the diverging factor mentioned in Remark \ref{remark 1} be denoted as $l_T^{\mathrm{DA}}$, and we also define
\begin{equation*}
 m_T^{\mathrm{DA}}=\left\lfloor \frac{T^{2S/(2S+1)}}{l_T^{\mathrm{DA}}}\right\rfloor+1,
\end{equation*}
where $m_T^{\mathrm{DA}}=o(T^{\frac{2S}{2S+1}})$. Similar to \eqref{eq noise}, the corresponding Gaussian generation becomes
\begin{align}
\label{eq noise DA}
    G_{iT}^{\mathrm{DA}}\mid\mathcal D_T\overset{\mathrm{iid}}{\sim}
 N\!\left(\check B_h,(\widehat{\mathrm{sd}}_{\mathrm{DA}})^2\right),
 \qquad i=1,\ldots,m_T^{\mathrm{DA}},
\end{align}
and the same interval construction gives
\begin{align}
\label{eq CI recover fixed DA}
    \mathbf{CI}_{\mathrm{recover}}^{\mathrm{DA}}(\alpha)
 =\Big[\widehat\theta_h-\overline G_T^{\mathrm{DA}} \pm
 \widehat{\mathrm{sd}}_{\mathrm{DA}}
 \operatorname{cv}_{1-\alpha}\left(1/\sqrt{l_T^{\mathrm{DA}}}\right)
 (m_T^{\mathrm{DA}})^{-\frac{1}{2}}\Big],\ \overline G_T^{\mathrm{DA}}=\frac{1}{m_T^{\mathrm{DA}}}\sum_{i=1}^{m_T^{\mathrm{DA}}}G_{iT}^{\mathrm{DA}}.
\end{align}
Of course, when dependence awareness is discarded and data-independence assumption is imposed, we have $\check B_h=\hat B_h$, $m_{T}^{\mathrm{DA}}=m_T$, $ l_T^{\mathrm{DA}}=l_T$, $ \overline{G}_T^{\mathrm{DA}}=\overline{G}_T$, and $\widehat{\mathrm{sd}}_{\mathrm{DA}}=\hat \sigma\hat C_{K,1}$, which makes $\eqref{eq CI recover fixed DA}$ equal to $\mathbf{CI}_{\mathrm{recover}}^{\mathrm{ind}}(\alpha)$.

Furthermore, we use the following theorem to show the unconditional coverage error under both dependence-aware and dependence-ignored settings.
%In parallel, the $\mathbf{CI}_{\mathrm{recover}}^{\mathrm{ind}}(\alpha)$ defined in \eqref{eq CI recover} can be rewritten as 
%\begin{align}
%\label{eq CI recover 2}
   % \mathbf{CI}_{\mathrm{recover}}^{\mathrm{ind}}(\alpha)
 %=\left[\widehat\theta_h-\overline G_T^{\mathrm{ind}}\ \pm\
%\widehat{\mathrm{sd}}_{\mathrm{ind}}
% \operatorname{cv}_{1-\alpha}\!\left(1/\sqrt{l_T^{\mathrm{ind}}}\right)
% (m_T^{\mathrm{ind}})^{-\frac{1}{2}}\right].
%\end{align}

\begin{theorem}[General unconditional coverage expansion]
    \label{prop DA}
    Fix $S\geq 1$, $\alpha\in (0,1)$ and $f\in \F$. Define $A_T=T^{\frac{S}{2S+1}}(\hat \theta_h-\check B_h-\theta(f))$. Suppose the terms $l_{T}^{\mathrm{DA}}$, $m_T^{\mathrm{DA}}$ and $\widehat{\mathrm{sd}}_{\mathrm{DA}}$ defined above satisfy the following conditions,
    \begin{align*}
&l_T^{\mathrm{DA}}\to\infty,\ 
m_T^{\mathrm{DA}}\to\infty,\
m_T^{\mathrm{DA}}=o\!\left(T^{2S/(2S+1)}\right),\\
\widehat{\mathrm{sd}}_{\mathrm{DA}}&>0\ \ \P\text{-a.s.},\ \ \ \ 
\sup_{T\geq 1}\sup_{f\in\F} \E(A_T/\widehat{\mathrm{sd}}_{\mathrm{DA}})^4<\infty.
    \end{align*}
Then, by letting $z_{1-\alpha/2}$ be the $1-\alpha/2$-quantile of the standard normal distribution and $\phi$ be the standard normal density, the unconditional coverage expansion is
\begin{align*}
&\P\!\left(\theta(f)\in\mathrm{CI}_{\mathrm{recover}}^{\mathrm{DA}}(\alpha)\right)=(1-\alpha)+\mathbf{CE}_{\mathrm{lead}}+\mathbf{CE}_{\mathrm{remainder}},\\
&\mathbf{CE}_{\mathrm{lead}}=\frac{z_{1-\alpha/2}\phi(z_{1-\alpha/2})}{l_T^{\mathrm{DA}}}
\Big[1-\frac{m_T^{\mathrm{DA}}l_T^{\mathrm{DA}}}{T^{2S/(2S+1)}}
 \E\!(A_T/\widehat{\mathrm{sd}}_{\mathrm{DA}})^2\Big],\\
 &\mathbf{CE}_{\mathrm{remainder}}=O\!\Big(
\frac{1}{(l_T^{\mathrm{DA}})^2}
+\frac{m_T^{\mathrm{DA}}}{T^{2S/(2S+1)}l_T^{\mathrm{DA}}}
+\frac{(m_T^{\mathrm{DA}})^2}{T^{4S/(2S+1)}}
\Big).
\end{align*}
\end{theorem}
Theorem \ref{prop DA} holds for arbitrary given $\hat \theta_h$, $\check B_h$, and integer $S\geq 1$. It shows that, under the 4th-moment condition only, the unconditional coverage error depends on the normalized second moment $\E[(A_T/\widehat{\mathrm{sd}}_{\mathrm{DA}})^2]$. Theorem \ref{th 2} later shows that stronger conditions on moments, dependence and selection of $l_T$ immediately imply much sharper control of undercoverage error. Recall that $\widehat{\mathrm{sd}}_{\mathrm{DA}}$ incorporates its independent version $\hat\sigma\hat C_{K,1}$ as a special case. When the asymptotic variance of the estimator $\hat\theta_h-\check B_h$ is significantly underestimated by its independence-assumption-based version, combining proper moment conditions and the dominated convergence theorem implies the possibility of $\lim_{T\to\infty}\E[(A_T/\widehat{\mathrm{sd}}_{\mathrm{DA}})^2]<\lim_{T\to\infty}\E[(A_T/\hat\sigma\hat C_{K,1})^2]$, which yields the following important corollary.

\begin{corollary}[Enhancement of coverage by modeling dependence]
    \label{corollary DA}
 Fix an estimator $\hat \theta_h-\check B_h$ and let $l_T=l_T^{\mathrm{DA}}$, $m_T=m_T^{\mathrm{DA}}$. Based on the $A_T$ defined in Theorem \ref{prop DA}, suppose
   \begin{align*}
\lim_{T\to\infty}\E[(A_T/\widehat{\mathrm{sd}}_{\mathrm{DA}})^2]=c_1\qquad \text{and}\qquad \lim_{T\to\infty}\E[(A_T/\hat\sigma\hat C_{K,1})^2]=c_2
   \end{align*}
 hold for some fixed constants $0\leq c_1<c_2<\infty$.  The following inequality holds for all $T\geq T_0$,
 \begin{align*}
\P(\theta(f)\in\mathrm{CI}_{\mathrm{recover}}^{\mathrm{DA}}(\alpha))-\P(\theta(f)\in \mathrm{CI}_{\mathrm{recover}}^{\mathrm{ind}}(\alpha))
>\frac{z_{1-\alpha/2}\phi(z_{1-\alpha/2})m_T^{\mathrm{DA}}}{2T^{2S/(2S+1)}}(c_2-c_1)>0
 \end{align*}
 where $T_0\geq 1$ is an integer independent of $T,b,h$. Consistency of $\widehat{\mathrm{sd}}_{\mathrm{DA}}$ and $\hat \sigma\hat C_{K,1}$ is not required.
\end{corollary}
Corollary \ref{corollary DA} highlights an important distinction between asymptotic validity and coverage enhancement. Although modeling data dependence is unnecessary for the asymptotic validity of our recovered confidence interval, it can still yield a strict improvement in unconditional coverage. Specifically, dependence-aware studentization can mitigate coverage loss from the original-data displacement by reducing its limiting normalized second moment, while preserving the exogenous Gaussian source of stochastic calibration. Importantly, this improvement does not require the dependence-aware studentizer \(\widehat{\mathrm{sd}}_{\mathrm{DA}}\) to consistently estimate the true dependence-adjusted standard deviation; the strict ordering of the limiting normalized second moments specified in Corollary \ref{corollary DA} is sufficient under the stated conditions. Thus, modeling dependence becomes an optional enhancement rather than a prerequisite for valid inference, and the resulting coverage improvement does not necessarily imply a reduction in interval length.

The preceding results characterize the potential coverage gains from dependence-aware studentization within the validity-recovery framework. We now illustrate how this framework can be combined with conventional undersmoothing and HAC-based variance estimation. As emphasized in Remark \ref{remark bias and variance}, the plug-in bias correction in \eqref{eq bias correction} is not essential to validity recovery. Instead of estimating and subtracting the smoothing bias, we may control it directly by choosing an undersmoothing bandwidth and setting \(\check B_h=0\). We can then use a HAC-based \(\widehat{\mathrm{sd}}_{\mathrm{DA}}\) in the Gaussian injection \eqref{eq noise DA}, while leaving the validity-recovery mechanism unchanged. The following corollary establishes the asymptotic validity of this simple undersmoothing–HAC-type validity recovery.

\begin{corollary}[HAC-type validity recovery with pure undersmoothing]
\label{th DA} Suppose Assumptions \ref{as fixed-design} and \ref{as misspecification 1} D1-D2 hold. Fix $S= 2$, $\alpha\in(0,1)$, and
$f\in\mathcal F_2(\varepsilon_0,M_2,\rho_2)$, i.e., the class \eqref{eq funciton class} with $S=2$. Consider the local linear estimator $\widehat\theta_h$ defined in \eqref{eq local linear} with $h=T^{-1/5}$ and $\check B_h=0.$ Choose $l_T^{\mathrm{DA}}\to\infty$ satisfying $l_T^{\mathrm{DA}}=o(T^{\frac{4}{5}})$ and let $ m_T^{\mathrm{DA}}
 =\left\lfloor\frac{T^{2S/(2S+1)}}{l_T^{\mathrm{DA}}}\right\rfloor+1$. Meanwhile, let $\widehat{\mathrm{sd}}_{\mathrm{DA}}$ be a HAC-based standard-deviation estimator that is positive for each $T\geq 1$ and satisfies $\sup_{T\geq 1}\E[(A_T/\widehat{\mathrm{sd}}_{\mathrm{DA}})^4]<\infty$, where $A_T=T^{\frac{2}{5}}(\hat\theta_h-\theta(f))$. Then the confidence interval defined in \eqref{eq CI recover fixed DA} satisfies
 \begin{align*}
     \P(\theta(f)\in \mathbf{CI}_{\mathrm{recover}}^{\mathrm{DA}}(\alpha))-(1-\alpha)= \frac{z_{1-\alpha/2}\phi(z_{1-\alpha/2})}{l_{T}^{\mathrm{DA}}}\Big[1-\E[(A_T/\widehat{\mathrm{sd}}_{\mathrm{DA}})^2]\Big]+O((l_{T}^{\mathrm{DA}})^{-2}+T^{-4/5})
 \end{align*}
\end{corollary}

Corollary \ref{th DA} highlights two important features of validity recovery.

First, unlike conventional inference based on asymptotic normality, our procedure permits the use of the “\(S=2\)”-optimal bandwidth \(h=T^{-1/5}\) without explicit bias correction. This is because the role traditionally played by undersmoothing is now transferred to the diverging calibration factor \(l_T^{\mathrm{DA}}\). In particular, increasing \(l_T^{\mathrm{DA}}\) reduces the order of the unconditional coverage error while increasing the interval radius. Corollary \ref{th DA} therefore makes the \textit{coverage–length trade-off} explicit, without requiring undersmoothing of the original local-linear estimator. Second, Corollary \ref{th DA} obtains its coverage guarantee under relatively mild dependence and moment conditions. If researchers are willing to impose stronger restrictions, more refined coverage-error guarantees may become available, particularly for undercoverage. Theorem \ref{th 2} provides an example of this possibility in the random-design setting, where stronger dependence and moment assumptions yield substantially sharper control of undercoverage error \(o(n^{-2S/(2S+1)})\). Thus, although modeling dependence is unnecessary for first-order validity recovery, additional dependence information can still be valuable for improving the sharpness of theoretical guarantees for coverage.

%In short, one key point of the contributions of validity recovery is therefore not merely the addition of Gaussian noise, but the replacement of endogenous stochastic calibration by exogenous pivotal calibration. 

\subsection{Comparison with other RDD inference with dependence modeling}
\label{sec 3.5}
Several works model dependence in RDD inference directly. For clustered data, \citeA{bartalotti2017regression} derive the properties of local polynomial RDD estimators when the errors are dependent within clusters. \citeA{noack2026inference} develop a general framework for clustered RDD settings and establish asymptotic normality of the local linear estimator under high-level conditions that cover many empirical designs, including settings with growing cluster sizes. They also show that clustered standard errors currently used in practice can be either inconsistent or overly conservative in finite samples, and they propose a nearest-neighbor-type variance estimator to address these issues. For RDiT models such as Example \ref{ex 3}, \citeA{hausman2017regression} point out that standard errors must account for serial dependence in the residuals and note that the existing RDiT literature has generally done so by using clustered standard errors. HAC studentization \cite{NeweyWest1987} is a standard alternative for serially dependent errors and is also the benchmark in Section \ref{sec 7.2}. The finding of \citeA{noack2026inference} shows that correct studentization is delicate even when the cluster structure is correctly specified, in line with the difficulties discussed in Section \ref{sec 2.2}.

Validity recovery approaches the same difficulty from a different direction. It shifts the burden of stochastic calibration from the studentizer to exogenous Gaussian noise, so the studentizer in \eqref{eq noise DA} need not be consistent. Theorem \ref{prop DA} shows that $\widehat{\mathrm{sd}}_{\mathrm{DA}}$ affects the leading coverage term only through the normalized second moment $\E[(A_T/\widehat{\mathrm{sd}}_{\mathrm{DA}})^2]$, and Corollary \ref{corollary DA} shows that a studentizer that reduces the limit of this moment relative to $\hat\sigma\hat C_{K,1}$ yields a strict gain in coverage. A cluster-robust estimator such as that of \citeA{noack2026inference}, after the same normalization as $\hat\sigma\hat C_{K,1}$ in \eqref{eq variance proxy fixed}, could therefore serve as $\widehat{\mathrm{sd}}_{\mathrm{DA}}$. Any reduction of this moment that it achieves would then translate into a coverage gain through Corollary \ref{corollary DA}, and validity recovery could guard against any remaining inconsistency. The two lines of work are thus complementary rather than competing, and combining them appears promising. Establishing this combination formally for clustered designs, including primitive conditions for the fourth-moment condition of Theorem \ref{prop DA}, is left for future research.

%%%%%%%%%%%%%%%%%%%%%%%%%%%%%%%%%%%%%%%%%%%%%%%%%%%%%%%%%%%%%%%%%%%%%%%%%%%%%%%%%%%%%%%%%%%%%%%%%%%%%%%%%%%%%%%%%%%%%%%%%%%%%%%%%%%%%%%%%%%%%%%%%%%%%%%%%%%%%%%%%%%%%%%%%%%%%%%%%%%%%%%%%%%%%%%%%%%%%%%%%%%%%%%%%%%%%%%%%%%%%%%%%%%%%%%%%%%%%%%%%%%%%%%%%%%%%%%%%%%%%%%%%%%%%%%%%%%%%%%%%%%%%%%%%%%%%%%%%%%%%%%%%%%%%%%%%%%%%%%%%%%%%%%%%%%%%%%%%%%%%%%%%%%%%%%%%%%%%%%%%%%%%%%%%%%%%%%%%%%%%%%%%%%%%%%%%%%%%%%%%%%%%%%%%%%%%%%%%%%%%%%%%%%%%%%%%%%%%%%%%%%%%%%%%%%%%%%%%%%%%%%%%%%%%%%%%%%%%%%%%%%%%%%%%%%%%%%%%%%%%%%%%%%%%%%%%%%%%%%%%%%%%%%%%%%%%%%%%%%%
\section{Dependence-Ignorance Validity Recovery for Random-design RDD}
\label{sec 4}
This section studies validity recovery for RDD models with random-design running variables. Although the results in Section \ref{sec 3} can be extended to random-design RDD inference by conditioning on the realized running variables under suitable conditions, our primary interest here is not merely to establish another limiting-coverage result. Instead, we investigate a more refined question concerning the unconditional coverage accuracy of random-design RDD inference. Specifically, this section focuses on the following question:
\begin{center}
    \textit{If we impose more specific dependence conditions, can we obtain the coverage error rate for each nominal level $1-\alpha$ and show the trade-off between this rate and the undersmoothing factor?} 
\end{center}

Analogous to the fixed-design RDD model \eqref{model fixed-design}, our RDD model with a random-design running variable is defined as follows,
\begin{align}
\label{model random-design}
   Y_{i}=f_-(X_{i})1[X_i\leq 0]+f_{+}(X_{i})1[X_{i}>0]+\varepsilon_{i},\ \E[\varepsilon_{i}|X_{i}]=0,\ \E[\varepsilon_i^2|X_i]=\sigma^2,
\end{align}
where, for some fixed $S\geq 3$, $f(x):=f_{-}(x)\1[x\leq 0]+f_{+}(x)\1[x>0]$ is an element of the function class $\mathcal F_S(1,M_S,\rho_S)$, i.e., the class in Assumption \ref{as smoothness} with $\varepsilon_0=1$. Throughout this section, we set $\varepsilon_0=1$, so that the smoothness in Assumption \ref{as smoothness} holds on the whole support $[-1,1]$ of the running variable (see Assumption \ref{as random design} below). This strengthens Assumption \ref{as smoothness} away from the cutoff, in exchange for a studentizer that uses all observations; see \eqref{eq variance proxy random}.
The parameter of interest is again 
\begin{align}
\label{eq tau 2}
    \theta(f):=\lim_{\varepsilon\to 0+}f_{+}(\varepsilon)-\lim_{\varepsilon\to 0-}f_-(\varepsilon),
\end{align}
Meanwhile, we impose the following assumptions of random-design running variables, moment conditions, and the misspecification of dependence.

\begin{assumption}[Random-design]
\label{as random design}
    $\{X_i\}_{i\geq 1}$ is a sequence of random variables sharing the same distribution as random variable $X_0$ and $X_0$ takes values in $[-1,1]$ with density $f_{X}$ such that $0<c_X\leq f_X(x)\leq C_X<\infty$ holds for all $x\in[-1,1]$ and some fixed $c_X,C_X>0$.
\end{assumption}

\begin{assumption} [Moments]
    \label{as moment}
    We assume the following moment conditions hold:
    \begin{itemize}
        \item [M1] $\sup_{i\geq 1}\sup_{u\in[-1,1]}E\bigl[|\varepsilon_i|^\varsigma\mid X_i=u\bigr]f_X(u)=: B_1<\infty$ holds for some $\varsigma>4+\frac{1}{S}$, where $S$ is given in Assumption \ref{as smoothness}.
        \item [M2] There exist a fixed integer $k_0\geq 1$, a fixed constant $B_2<\infty$ and a non-negative sequence $\{c_k\}_{k\geq k_0}$ with $\sum_{k\geq k_0}c_k<\infty$ such that, for every $i\geq 1$, every $k\geq k_0$ and every interval $A\subset[-1,1]$ of length $|A|$, we have
        \begin{align}
        \label{eq M2}
            E\left[|\varepsilon_i\varepsilon_{i+k}|\,\1\{X_i\in A,\ X_{i+k}\in A\}\right]\leq c_k|A|+B_2|A|^2.
        \end{align}
    \end{itemize}
   \end{assumption}

\begin{assumption} [Misspecified Data Dependence] 
\label{as misspecificaiton 2}
The following two points can summarize the gap between the assumed dependence used in implementation and the true dependence.
\begin{itemize}
    \item [D4] \rm{(Assumed Dependence)} For the implementation of the confidence interval for $\theta(f)$,  we treat $\{(X_i,\varepsilon_i)\}_{i\geq 1}$ as mutually independent.
    \item [D5] \rm{(True Dependence)} For the actual dependence of $\{(X_i,\varepsilon_i)\}_{i\ge 1}$, we assume its $\beta$-mixing coefficient $\beta_k$ satisfies $\beta_k\lesssim k^{-\gamma}$, for some $\gamma>0$ such that
\begin{align*}
    \frac{1}{\varsigma}+\frac{1}{\gamma+1}<\frac{S}{4S+1},
\end{align*}
where $\varsigma$ is given in Assumption \ref{as moment} M1 and $S$ is defined in Assumption \ref{as smoothness}. This joint condition implies $\varsigma>4+\frac{1}{S}$ and $\gamma>3+\frac{1}{S}$, and hence $\gamma>\frac{2\varsigma-2}{\varsigma-2}$. It trades moments against mixing: if M1 holds for every $\varsigma>0$ (e.g., bounded errors), it reduces to $\gamma>3+\frac{1}{S}$; if $\beta_k$ decays geometrically, it reduces to $\varsigma>4+\frac{1}{S}$. The definition of $\beta$-mixing is introduced in Appendix \ref{app F Random}.
\end{itemize}    
\end{assumption}
Assumptions \ref{as random design}-\ref{as misspecificaiton 2} are standard assumptions in local polynomial regression for weakly dependent data. Assumption \ref{as moment} M1 and Assumption \ref{as misspecificaiton 2} follow those in \citeA{hansen2008uniform}. Assumption \ref{as random design} only requires the running variables to share a marginal density, and M2 only bounds how often two observations at least $k_0$ periods apart fall into a common short interval. In particular, M2 does not require $(X_i,X_{i+k})$ to have a joint density. If such a joint density $f_{i,i+k}$ exists and $\sup_{u,v\in[-1,1]}E[|\varepsilon_i\varepsilon_{i+k}|\mid X_i=u,X_{i+k}=v]f_{i,i+k}(u,v)\leq B_2$ for all $i\geq1$ and $k\geq k_0$, which is the joint-density form of this condition used in \citeA{hansen2008uniform}, then \eqref{eq M2} holds with $c_k=0$. The term $c_k|A|$ allows the joint law of $(X_i,X_{i+k})$ to put positive mass on the diagonal. For instance, if the running variables follow \eqref{eq markov} in Example \ref{ex 1} and $\{\varepsilon_i\}_{i\geq1}$ is independent of $\{X_i\}_{i\geq1}$ with $\E[\varepsilon_i^2]=\sigma^2$, then $\P(X_i\in A,X_{i+k}\in A)=\rho^k|A|/2+(1-\rho^k)|A|^2/4$, so that \eqref{eq M2} holds with $c_k=\sigma^2\rho^k/2$ and $B_2=\sigma^2/4$. We emphasize that introducing \(\beta\)-mixing is not necessary for validity recovery itself. As established in Section \ref{sec 3}, we can recover asymptotically exact coverage without specifying a particular dependence framework. Our objective here, however, is more refined: obtaining explicit coverage-error convergence rates requires quantitative concentration inequalities for the underlying stochastic terms. To this end, we adopt the familiar joint \(\beta\)-mixing framework as a convenient and tractable example. This choice is by no means intrinsic to our procedure. With more elaborate concentration arguments and suitable approximation conditions, the results of this section can potentially be extended to broader dependence structures, such as processes that are near-epoch dependent (NED) on an underlying \(\beta\)-mixing sequence (e.g., \citeA{davidson1994stochastic}).

As for the estimation of $\theta(f)$, similar to \eqref{eq local linear}, by denoting $\r^\top(u)=(1,u)\in \R^2$, we consider the following local linear estimator for the random-design RDD,
\begin{align}
    \label{eq local linear random}
    & \tilde \theta_h:=\tilde f_{+,h}(0)-\tilde f_{-,h}(0):=\sum_{i=1}^n(L_{ih,+}(0)-L_{ih,-}(0))Y_i,\notag\\
&L_{ih,-}=\frac{1}{nh}K(X_i/h)\1[X_i\in[-1, 0]] \Big(\e_0^{\top}S_{0,-}^{-1}\r(X_i/h)\Big),\notag\\
&L_{ih,+}=\frac{1}{nh}K(X_i/h)\1[X_i\in(0, 1]] \Big(\e_0^{\top}S_{0,+}^{-1}\r(X_i/h)\Big),\notag\\
&  S_{0,-}=\frac{1}{nh}\sum_{i=1}^nK(X_i/h)\1[X_i\in[-1,0]]\r(X_i/h)\r^\top(X_i/h),\notag\\
   &  S_{0,+}=\frac{1}{nh}\sum_{i=1}^nK(X_i/h)\1[X_i\in(0,1]]\r(X_i/h)\r^\top(X_i/h).
\end{align}

Similar to Section \ref{sec 3}, motivated by the full-bias correction method of improved RBC inference introduced in \citeA{cavaliere2026improvedinferencenonparametricregression}, our bias estimator is formulated as follows,
\begin{align}
\label{eq bias estimator random}
\tilde B_h
:&=\sum_{i=1}^nL_{ih,+}(0)[\tilde f_{+,b}(X_i)-\tilde f_{+,b}(0)]-\sum_{i=1}^nL_{ih,-}(0)[\tilde f_{-,b}(X_i)-\tilde f_{-,b}(0)],\\
&\tilde f_{+,b}(x)=\sum_{j=1}^nW_{jb}(x;0,1)Y_j,\ \ \tilde f_{-,b}(x)=\sum_{j=1}^nW_{jb}(x;-1,0)Y_j;\notag
\end{align}
where the weight function $W_{jb}(x;a,b)$ is the weight function of $S$-order local polynomial with truncated inputs, and its definition is introduced in Appendix \ref{app concentration: random}. Moreover, Assumption \ref{as misspecificaiton 2} indicates that the discussion is built on the situation in which the underlying $\beta$-mixing dependence has been ignored, and we misspecify the data as independent observations. Analogous to \eqref{eq variance proxy fixed}, our independence-assumption-based studentizer of $\text{Var}_{\text{ind}}(\tilde \theta_h)$ is defined as  
\begin{align}
\label{eq variance proxy random}
 \tilde \sigma^2\tilde C_{K,1}^2=&\Big\{\frac{1}{n}\sum_{i=1}^n(Y_i-\tilde f_{b}(X_i))^2\Big\}\Big\{nh\sum_{i=1}^n\Big(L_{ih,+}(0)-L_{ih,-}(0)\Big)^2\Big\},\\
 &\qquad \tilde f_{b}(x)=\tilde f_{+,b}(x)\1[x\in(0,1]]+\tilde f_{-,b}(x)\1[x\in[-1,0]],\notag
\end{align}
\eqref{eq variance proxy random} is a simple plug-in type studentizer that averages the pilot residuals of all $n$ observations. Since $\varepsilon_0=1$ in this section, Assumption \ref{as smoothness} controls the bias of the pilot fit $\tilde f_b$ at every observation, so $\frac{1}{n}\sum_{i=1}^n(Y_i-\tilde f_{b}(X_i))^2$ is a residual-based estimator of the marginal error variance $\sigma^2$.

According to $(\tilde B_h, \tilde \sigma^2\tilde C_{K,1}^2)$, for each given $n\geq 1$ and a user-specified $\tau\in(0,\frac{S}{2S+1})$, we generate auxiliary Gaussian noise $\{G_{in}\}_{i=1}^{m_n}=:\mathcal{G}_{m_n}$ as follows,
\begin{align}
\label{eq noise random}
    G_{in}\overset{\text{i.i.d}}{\sim} G_0,\ &\bar G_n:=\frac{1}{m_n}\sum_{i=1}^{m_n}G_{in},\ \ i=1,\ldots,m_n,\\
\text{where}\ & G_0\sim N(\tilde B_h,\tilde \sigma^2\tilde C_{K,1}^2),\ \    m_n=[n^{\frac{2S}{2S+1}-2\tau}]+1.\notag
\end{align}
Consequently, the independence-assumption-based feasible confidence interval and its recovered version are written as follows,
\begin{align}
    \label{eq CI feasible random}
    &\ \mathit{CI}_{1}^{\mathrm{ind}}(\alpha):=\Big[\tilde \theta_h-\tilde B_h\pm \underbrace{\{\tilde \sigma\tilde C_{K,1}z_{1-\alpha/2}\}}_{\text{leading constant}}(n_{\text{eff}})^{-\frac{1}{2}}\Big],\\
     \label{eq CI recover random}
    \mathit{CI}_{\mathrm{recover}}^{\mathrm{ind}}&(\alpha,\tau):=\mathcal{R}(\mathit{CI}_1^{\mathrm{ind}}(\alpha);\mathcal{G}_{m_n})=\Big[\tilde \theta_h-\bar G_{n}\pm \underbrace{\{\tilde \sigma\tilde C_{K,1}\text{cv}_{1-\alpha}(1/n^{\frac{\tau}{2}})\}}_{\text{leading constant}}m_n^{-\frac{1}{2}}\Big],
\end{align}
where $m_n=[n^{\frac{2S}{2S+1}-2\tau}]+1$ and $\mathcal{R}(\mathit{CI}_1^{\mathrm{ind}}(\alpha); \mathcal{G}_{m_n})$ is the \textit{validity-recovery} mapping defined in \eqref{eq recovery mapping}. $n_{\text{eff}}$ denotes the effective sample size determined by the choice of $h$.

\begin{theorem}
    \label{th 2}
   Suppose $0<h<1$ satisfies $n^{-\frac{1}{2S+1}}/h=o(1)$ and $b=n^{-\frac{1}{2S+1}}$. Under Assumptions \ref{as random design}--\ref{as misspecificaiton 2}, the following three properties hold for every fixed $\alpha\in (0,1)$ and $0<\tau<\frac{S}{2S+1}$,
   \begin{itemize}
  \item \rm{(Undercoverage-error)} For every fixed $S\geq 3$, we have
  \begin{align*}
      \inf_{f\in \mathcal F_S(1,M_S,\rho_S)}\P(\theta(f)\in \mathit{CI}_{\mathrm{recover}}^{\mathrm{ind}}(\alpha,\tau))\geq 1-\alpha-o(n^{-\frac{2S}{2S+1}}).
  \end{align*}
  \item \rm{(Overcoverage-error)} For every fixed $S\geq 3$, we have
  \begin{align*}
      \sup_{f\in\mathcal F_S(1,M_S,\rho_S)}\P(\theta(f)\in \mathit{CI}_{\mathrm{recover}}^{\mathrm{ind}}(\alpha,\tau))\leq 1-\alpha+O(n^{-\tau}).
  \end{align*}
   \item \rm{(Recovery of leading constant)} $\frac{\text{Leading constant of $\mathit{CI}_{\mathrm{recover}}^{\mathrm{ind}}(\alpha,\tau)$}}{\text{Leading constant of $\mathit{CI}_1^{\mathrm{ind}}(\alpha)$}}\xrightarrow[n\to\infty]{\P}1$.
   \end{itemize}
   Moreover, $\mathit{CI}_{\mathrm{recover}}^{\mathrm{ind}}(\alpha,\tau)$ is non-trivial in the sense of Definition \ref{def nontrivial}.
\end{theorem}
The polynomial relaxation entails only a mild efficiency cost. Since our theory accommodates every prespecified \(\tau\in(0,S/(2S+1))\), the recovered interval shrinks at rate \(n^{-S/(2S+1)+\tau}\), where \(\tau\) can be a fixed constant arbitrarily close to zero. For sufficiently small \(\tau>0\), this rate can be more favorable than those associated with conventional MSE-optimal or coverage-error-optimal RBC implementations and standard undersmoothing choices, where MSE-optimal refers to optimality with respect to the debiasing-order of local polynomial regression.

Remarkably, this limited efficiency loss does not come at the expense of coverage reliability. By replacing endogenous sampling-based calibration with exogenous Gaussian auxiliary samples, our procedure changes both the probability structure underlying inference and the tools available for coverage analysis. This yields particularly strong control of undercoverage, of a kind not generally guaranteed by conventional RBC, undersmoothing (e.g., Theorem 2 in \citeA{calonico2018effect}), or bias-aware inference (e.g., Proposition 5 in \citeA{yuan2026safesharphonestinference}), even under independent sampling. Our control of overcoverage is deliberately less stringent. The folded-normal calibration explicitly accommodates the data-dependent displacement, reflecting a clear inferential priority of this paper, which is first to safeguard against undercoverage, and then to ensure asymptotically exact \(1-\alpha\) coverage.

As in Theorem \ref{th 1}, the recovered interval in Theorem \ref{th 2} has a randomized center, a direct consequence of supplying stochastic calibration through auxiliary randomness (Remark \ref{remark 2}). But this also introduces additional stochastic fluctuation into finite-sample performance. To address this, Section \ref{sec 5} shows that averaging over repetitive generations stabilizes this center and protects against undercoverage in finite samples, without changing the radius.

\section{Averaging over Repetitive Generation (AoRG): Finite-Sample Protection Against Non-Negligible Displacement}
\label{sec 5}
For convenience, we inherit the settings and notations used in Section \ref{sec 3.4} for fixed-design RDD inference with the dependence-aware studentizer $\widehat{\mathrm{sd}}_{\mathrm{DA}}$ and the corresponding Gaussian noises $\{G_{iT}^{\mathrm{DA}}\}$. Since AoRG operates only on the auxiliary Gaussian randomization rather than on the original running-variable design, the results of this section translate directly to the independence-assumption-based interval of Section \ref{sec 3.3}, where $\widehat{\mathrm{sd}}_{\mathrm{DA}}=\hat\sigma\hat C_{K,1}$, and to the random-design interval of Section \ref{sec 4}.

Recall that the Gaussian noise $\{G_{iT}^{\mathrm{DA}}\}_{i=1}^{m^{\mathrm{DA}}_T}=:\mathcal{G}^{\mathrm{DA}}_{T}$ can be \textit{repetitively generated}. Thus, together with \eqref{eq CI recover fixed DA}, for any fixed integer $K\geq 1$, we can construct $K$ different validity-recovered confidence intervals, denoted as $\mathbf{CI}_{\mathrm{recover}}^{\mathrm{DA}\ (k)}(\alpha)$, based on K-fold independent auxiliary samples, $\{\{G_{iT,(k)}^{\mathrm{DA}}\}_{i=1}^{m_T^{\mathrm{DA}}}:k\leq K\}$. Then, an AoRG-type confidence interval is defined as follows,
\begin{align}
    \label{eq CI AoRG}
    \mathbf{CI}_{\mathrm{AoRG}}^{\mathrm{DA}}(\alpha,K)=&\Big[\hat \theta_h-\frac{1}{K}\sum_{k=1}^K\bar G^{\mathrm{DA}}_{T,(k)}\pm \Big\{\widehat{\mathrm{sd}}_{\mathrm{DA}}\text{cv}_{1-\alpha}(1/\sqrt{l^{\mathrm{DA}}_T})\Big\}(m_T^{\mathrm{DA}})^{-\frac{1}{2}}\Big],
\end{align}
where $ \bar G^{\mathrm{DA}}_{T,(k)}=\frac{1}{m_T^{\mathrm{DA}}}\sum_{i=1}^{m_{T}^{\mathrm{DA}}}G^{\mathrm{DA}}_{iT,(k)}$ and $ m_T^{\mathrm{DA}}=\left[T^{2S/(2S+1)}/l_T^{\mathrm{DA}}\right]+1$. Meanwhile, for convenience, we introduce the following notations,
\begin{equation}
\label{eq:scales}
\begin{aligned}
&s_T^{\mathrm{DA}}:=\frac{\widehat{\mathrm{sd}}_{\mathrm{DA}}}{\sqrt{m_T^{\mathrm{DA}}}},\quad r_T^{\mathrm{DA}}:=s_T^{\mathrm{DA}}\,\cv_{1-\alpha}\big(1/\sqrt{l^{\mathrm{DA}}_T}\big),\quad c_T:=\frac{r^{\mathrm{DA}}_T}{s^{\mathrm{DA}}_T}=\cv_{1-\alpha}\big(1/\sqrt{l^{\mathrm{DA}}_T}\big),\\
&d_T(f):=\frac{\hat \theta_h-\check B_h-\theta(f)}{s_T^{\mathrm{DA}}},\quad C^{\mathrm{DA}}_{T,K}:=\widehat\theta_h-\frac1K\sum_{k=1}^K\bar G_{T,(k)}^{\mathrm{DA}},\quad \mathbf{CI}_{\mathrm{AoRG}}^{\mathrm{DA}}(\alpha,K)=[\,C^{\mathrm{DA}}_{T,K}\pm r^{\mathrm{DA}}_T\,].
\end{aligned}
\end{equation}

In terms of this notation, the results in Sections \ref{sec 3} and \ref{sec 4} are asymptotic. Their basic premise is that the displacement is asymptotically negligible, $d_T(f)\to 0$ in probability, so that the auxiliary Gaussian component alone governs coverage. In finite samples, this premise need not be adequately satisfied. In particular, strong dependence can inflate the stochastic scale of $\hat\theta_h-\check B_h$ relative to $s^{\mathrm{DA}}_T$, so that $d_T(f)$ converges slowly and may remain \textit{non-negligible} at empirically relevant sample sizes. Throughout this section, we therefore focus solely on the case where $|d_T(f)|$ is \textit{non-negligible}.

In Section \ref{sec 5.1}, we show that, \textit{without changing the interval length}, the AoRG interval achieves the following two properties simultaneously:
\begin{itemize}
    \item[(1)] \textbf{Stability}: it stabilizes the randomized center;
    \item[(2)] \textbf{Finite-sample protection}: it protects against undercoverage in a non-asymptotic sense and enlarges the range of displacements $d_T(f)$ that the interval tolerates.
\end{itemize}
%%%%%%%%%%%%%%%%%%%%%%%%%%%%%%%%%%%%%%%%%%%%%%%%%%%%%%%%%%%%%%%%

\subsection{Stability and finite-sample protection}
\label{sec 5.1}
For $K=1$, $\mathbf{CI}^{\mathrm{DA}}_{\mathrm{AoRG}}(\alpha,1)$ coincides with $\mathbf{CI}^{\mathrm{DA}}_{\mathrm{recover}}(\alpha)$ in \eqref{eq CI recover fixed DA}; for general $K$, AoRG changes only the center. Conditionally on $\mathcal{D}_T$, the injected noise is the only source of randomness, and the data enter the coverage only through the displacement $d_T(f)$ in \eqref{eq:scales}, that is, the estimation error of $\hat{\theta}_h-\check B_h$ measured in units of $s^{\mathrm{DA}}_T$. Any dependence that $\widehat{\mathrm{sd}}_{\mathrm{DA}}$ ignores or misjudges reaches the coverage through the law of $d_T(f)$ and nowhere else. Since the radius equals $c_T$ in the same units, finite-sample undercoverage is a question of how large $d_T(f)$ is relative to $c_T$. Theorem~\ref{thm:aorg-finite} answers this question for every sample size $T$.

\begin{theorem}[Non-asymptotic stability and finite-sample protection]\label{thm:aorg-finite}
Fix $\alpha\in(0,1/2)$, $T\ge1$, $f\in\mathcal{F}_S(\varepsilon_0,M_S,\rho_S)$ and an integer $K\ge1$. Suppose $\widehat{\mathrm{sd}}_{\mathrm{DA}}>0$ $\mathbb{P}$-a.s., and let $\mathcal{D}_T:=\{(x_{tT},Y_{tT})\}_{t=1}^{T}$. Then, based on the $c_T$ defined in \eqref{eq:scales}, the following properties hold.
\begin{itemize}
\rm{\item[(i)](\textbf{Stability enhancement})}
\begin{equation}\label{eq:aorg-stability}
\mathbb{E}\big[C^{\mathrm{DA}}_{T,K}\mid\mathcal{D}_T\big]=\hat{\theta}_h-\check B_h,
\qquad
\mathrm{Var}\big(C^{\mathrm{DA}}_{T,K}\mid\mathcal{D}_T\big)=\frac{(\widehat{\mathrm{sd}}_{\mathrm{DA}})^2}{K m^{\mathrm{DA}}_T}.
\end{equation}

\item[(ii)] (\textbf{Undercoverage protection})
\begin{equation}\label{eq:aorg-protection}
\mathbb{P}\big(\theta(f)\in\mathbf{CI}^{\mathrm{DA}}_{\mathrm{AoRG}}(\alpha,K)\big)
\;\ge\;(1-\alpha)\,
\mathbb{P}\Big(|d_T(f)|\le K^{-1/2}\,\mathrm{cv}^{-1}_{1-\alpha}\big(\sqrt{K}\,c_T\big)\Big),
\end{equation}
\textit{where $\mathrm{cv}^{-1}_{1-\alpha}$ is the inverse of the strictly increasing function $\mathrm{cv}_{1-\alpha}$ on $[0,\infty)$.}

\item[(iii)] (\textbf{Expanding tolerance}) \textit{The tolerance $K^{-1/2}\mathrm{cv}^{-1}_{1-\alpha}(\sqrt{K}\,c_T)$ in \eqref{eq:aorg-protection} equals $1/\sqrt{l^{\mathrm{DA}}_T}$ for $K=1$, is strictly increasing in $K$, and satisfies}
\[
c_T-z_{1-\alpha/2}K^{-1/2}\;\le\;K^{-1/2}\,\mathrm{cv}^{-1}_{1-\alpha}\big(\sqrt{K}\,c_T\big)\;<\;c_T .
\]
\textit{Consequently, the lower bound in \eqref{eq:aorg-protection} is non-decreasing in $K$.}

\item[(iv)] (\textbf{Invariant radius})
\begin{equation}\label{eq:aorg-radius}
\frac{\mathrm{radius}\big(\mathbf{CI}^{\mathrm{DA}}_{\mathrm{recover}}(\alpha)\big)}{\mathrm{radius}\big(\mathbf{CI}^{\mathrm{DA}}_{\mathrm{AoRG}}(\alpha,K)\big)}=1,
\quad a.s.\text{-}\mathbb{P}.
\end{equation}
\end{itemize}
\end{theorem}

\paragraph{Undercoverage occurs only beyond the tolerance}
Theorem~\ref{thm:aorg-finite}(ii) identifies the only source of finite-sample undercoverage: displacements larger than the tolerance $K^{-1/2}\mathrm{cv}^{-1}_{1-\alpha}(\sqrt{K}\,c_T)$. As long as $|d_T(f)|$ stays within the tolerance, the conditional coverage given $\mathcal{D}_T$ is at least $1-\alpha$, whatever the dependence (see the proof of Theorem~\ref{thm:aorg-finite} in Appendix~\ref{app D}). A single generation tolerates displacements only up to $1/\sqrt{l^{\mathrm{DA}}_T}$, which is exactly the standardized bias bound of the folded-normal calibration. AoRG extends the tolerance beyond this bound, to within $z_{1-\alpha/2}K^{-1/2}$ of $c_T$ (Theorem~\ref{thm:aorg-finite}(iii)), while the radius stays exactly that of $\mathbf{CI}^{\mathrm{DA}}_{\mathrm{recover}}(\alpha)$ in \eqref{eq CI recover fixed DA} and of its special case $\mathbf{CI}^{\mathrm{ind}}_{\mathrm{recover}}(\alpha)$ in \eqref{eq CI recover} (Theorem~\ref{thm:aorg-finite}(iv)). For example, with $\alpha=0.05$, $T=2000$ and $l^{\mathrm{DA}}_T=\log T$ as in Remark \ref{remark 1}, $c_T=2.082$, and the tolerance equals $0.363$ for $K=1$, $0.919$ for $K=2$, $1.133$ for $K=3$ and $1.562$ for $K=10$. At $K=10$, the interval keeps conditional coverage of at least $1-\alpha$ for every displacement up to three-quarters of its own radius. This is more than four times the tolerance of a single generation, at no cost in length. In Figure \ref{fig:coverage_stress}, under severe dependence, the empirical coverage at $K=10$ ranges from $90.0\%$ to $95.2\%$ at $T=2000$.

\paragraph{Division of labour between $K$ and $l^{\mathrm{DA}}_T$.}
$K$ sets the tolerance relative to the radius, whereas $l^{\mathrm{DA}}_T$ sets the scale of the displacement relative to the radius. Indeed, by \eqref{eq:scales}, the definition of $A_T$ in Theorem~\ref{prop DA} and the definition of $m^{\mathrm{DA}}_T$,
\[
\mathbb{E}\big[d_T(f)^2\big]=\frac{m^{\mathrm{DA}}_T}{T^{2S/(2S+1)}}\,\mathbb{E}\big[(A_T/\widehat{\mathrm{sd}}_{\mathrm{DA}})^2\big]
\le\Big(\frac{1}{l^{\mathrm{DA}}_T}+\frac{1}{T^{2S/(2S+1)}}\Big)\mathbb{E}\big[(A_T/\widehat{\mathrm{sd}}_{\mathrm{DA}})^2\big].
\]
Displacements beyond the whole radius cannot be absorbed by averaging.  Such displacements reflect severe under-studentization. They are controlled by $l^{\mathrm{DA}}_T$, or by a dependence-aware $\widehat{\mathrm{sd}}_{\mathrm{DA}}$ as in Section \ref{sec 3.4}, rather than by $K$. The two devices therefore act in sequence: $l^{\mathrm{DA}}_T$ (or a dependence-aware studentizer) brings the displacement within the radius, and $K$ then extends the tolerance from $1/\sqrt{l^{\mathrm{DA}}_T}$ towards $c_T$, at no cost in length.

\paragraph{Protection without trivial coverage}
The protection in \eqref{eq:aorg-protection} never comes from trivial coverage. For every finite $K$, the injected component of the center is exactly Gaussian with conditional variance $(s^{\mathrm{DA}}_T)^2/K$ and is independent of $\mathcal D_T$, so the pivot at the radius scale always retains a non-degenerate Gaussian component. Whenever this pivot converges in law, its limit is a Gaussian convolution and therefore has a positive Lebesgue density. The AoRG interval is thus non-trivial in the sense of Definition \ref{def nontrivial}, whether or not $d_T(f)$ is negligible. Only $K=\infty$ removes this component: it yields the deterministic interval $[\hat{\theta}_h-\check B_h\pm r^{\mathrm{DA}}_T]$ with the same radius, which is trivial and therefore excluded; see Section \ref{sec 5.3}.

\paragraph{Choice of $K$.}
$K$ is not a tuning parameter. By Theorem~\ref{thm:aorg-finite}, setting $K$ as a larger but fixed integer never weakens any of the guarantees: the conditional variance of the center decreases in $K$, the tolerance is strictly increasing in $K$, the lower bound in \eqref{eq:aorg-protection} is non-decreasing in $K$, and the radius does not depend on $K$. Every finite $K$ yields a non-trivial interval, so the only restriction is that $K$ be fixed (Section~\ref{sec 5.3}). There is therefore no trade-off to balance, and the only cost of a larger $K$ is generating more independent auxiliary samples. By Theorem~\ref{thm:aorg-finite}(iii), the gap between the tolerance and $c_T$ is at most $z_{1-\alpha/2}K^{-1/2}$, so moderate values already reach a large share of the attainable tolerance: in the example above, $K=10$, the value used in Section~\ref{sec 7}, tolerates displacements up to three-quarters of the radius.

%%%%%%%%%%%%%%%%%%%%%%%%%%%%%%%%%%%%%%%%%%%%%%%%%%%%%%%%%%%%%%

%%%%%%%%%%%%%%%%%%%%%%%%%%%%%%%%%%%%%%%%%%%%%%%%%%%%%%%%%%%%%%%%%%%%%%%%%%%%%%%%%%%%%%%%%%%%%%%%%%%%%%%%%%%%%%%%%%%%%%%%%%%%%%%%%%%%%%%%%%%%%%%%%%%%%%%%%%%%%%%%%%%%%%%%%%%%%%%%%%%%%%%%%%%%%%%%%%%%%%%%%%%%%%%%%%%%%%%%%%%%%%%%%%%%%%%%%%%%%%%%%%%%%%%%%%%%%%%%%%%%%%%%%%%%%%%%%%%%%%%%%%%%%%%%%%%%%%%%%%%%%%%%%%%%%%%%%%%%%%%%%%%%%%%%%%%%%%%%%%%%%%%%%%%%%%%%%%%%%%%%%%%%%%%%%%%%%%%%%%%%%%%%%%%%%%%%%%%%%%%%%%%%%%%%%%%%%%%%%%%%%%%%%%%%%%%%%%%%%%%%%%%%%%%%%%%%%%%%%%%%%%%%%%%%%%%%%%%%%%%%%%%%%%%%%%%%%%%%%%%%%%%%%%%%%%%%%%%%%%%%%%

%%%%%%%%%%%%%%%%%%%%%%%%%%%%%%%%%%%%%%%%%%%%%%%%%%%%%%%%%%%%%%%%%%%%%%%%%%%%%%%%%%%%%%%%%%%%%%%%%%%%%%%%%%%%%%%%%%%%%%%%%%%%%%%%%%%%%%%%%%%%%%%%%%%%%%%%%%%%%%%%%%%%%%%%%%%%%%%%%%%%%%%%%%%%%%%%%%%%%%%%%%%%%%%%%%%%%%%%%%%%%%%%%%%%%%%%%%%%%%%%%%%%%%%%%%%%%%%%%%%%%%%%%%%%%%%%%%%%%%%%%%%%%%%%%%%%%%%%%%%%%%%%%%%%%%%%%%%%%%%%%%%%%%%%%%%%%%%%%%%%%%%%%%%%%%%%%%%%%%%%%%%%%%%%%%%%%%%%%%%%%%%%%%%%%%%%%%%%%%%%%%%%%%%%%%%%%%%%%%%%%%%%%%%%%%%%%%%%%%%%%%%%%%%%%%%%%%%%%%%%%%%%%%%%%%%%%%%%%%%%%%%%%%%%%%%%%%%%%%%%%%%%%%%%%%%%%%%%%%%%%%%%%%%%%%%%%%%%%%%%

%%%%%%%%%%%%%%%%%%%%%%%%%%%%%%%%%%%%%%%%%%%%%%%%%%%%%%%%%%%%%%%%%%%%%%
% Section 5.3. Insert at the end of Section 5, i.e. after the last
% paragraph of Section 5.2 and before \section{Simulation: ...}.
%%%%%%%%%%%%%%%%%%%%%%%%%%%%%%%%%%%%%%%%%%%%%%%%%%%%%%%%%%%%%%%%%%%%%%
\subsection{Why $K$ is kept finite}
\label{sec 5.3}
By Theorem~\ref{thm:aorg-finite}(i), the conditional variance $(\widehat{\mathrm{sd}}_{\mathrm{DA}})^2/(Km_T^{\mathrm{DA}})$ of the center $C^{\mathrm{DA}}_{T,K}$ decreases in $K$, while the radius $r^{\mathrm{DA}}_T$ does not depend on $K$. Removing the injected noise from the center altogether, which corresponds to $K=\infty$, yields the deterministic interval
\begin{align}
\label{eq det interval}
\big[\hat\theta_h-\check B_h\pm r^{\mathrm{DA}}_T\big]
\end{align}
with the same radius. Its center converges faster than its radius, as in Observation 3, and its coverage tends to one (Proposition~\ref{prop K infinity} below). A natural objection is therefore: the radius is the same and the coverage is higher, so why not use \eqref{eq det interval}?

\paragraph{The criterion: coverage may exceed $1-\alpha$, but the interval must be non-trivial.}
Coverage above $1-\alpha$ is not a defect in itself. It comes with the protection against undercoverage of Section \ref{sec 5.1}, at no cost in length. What an interval must not be is trivial in the sense of Definition \ref{def nontrivial}: its pivot at the radius scale must not degenerate, so that $\alpha$ keeps its calibration meaning. Proposition~\ref{prop K infinity} shows that \eqref{eq det interval} fails this requirement, whereas every finite $K$ meets it (Section~\ref{sec 5.1}).

\begin{proposition}[The deterministic limit is trivial]
\label{prop K infinity}
Suppose the conditions of Theorem~\ref{prop DA} hold. Then, for every $\alpha\in(0,1)$ and every $f\in\mathcal{F}_S(\varepsilon_0,M_S,\rho_S)$, the pivot of \eqref{eq det interval} at the radius scale satisfies
\[
\frac{\hat\theta_h-\check B_h-\theta(f)}{r^{\mathrm{DA}}_T}=\frac{d_T(f)}{c_T}\xrightarrow[T\to\infty]{\mathbb P}0 .
\]
Consequently, \eqref{eq det interval} is trivial in the sense of Definition \ref{def nontrivial}, and $\mathbb{P}\big(\theta(f)\in[\hat\theta_h-\check B_h\pm r^{\mathrm{DA}}_T]\big)\to1$.
\end{proposition}
By Definition \ref{def nontrivial} and the discussion following it, the limit law of this pivot is a point mass. The nominal level therefore loses its calibration meaning, the induced test rejects with probability tending to zero for every $\alpha$, and decisions are tilted systematically towards the null. This is why $K$ is kept finite.

\paragraph{Conditional coverage in finite samples}
Within the AoRG family, the contrast already holds for every $T$, conditionally on the data. By \eqref{eq:scales}, $\theta(f)\in[\hat\theta_h-\check B_h\pm r^{\mathrm{DA}}_T]$ if and only if $|d_T(f)|\le c_T$. Conditionally on $\mathcal{D}_T$, the coverage of \eqref{eq det interval} is therefore either zero or one: the level $\alpha$ only moves the threshold $c_T$ and never enters as a probability. For every finite $K$, by contrast, the conditional coverage of $\mathbf{CI}^{\mathrm{DA}}_{\mathrm{AoRG}}(\alpha,K)$ is a non-degenerate probability, and it is at least $1-\alpha$ whenever $|d_T(f)|$ does not exceed the tolerance in \eqref{eq:aorg-protection}.

\paragraph{A negligible displacement does not make the AoRG interval trivial.}
Theorem~\ref{thm:aorg-finite} holds for every $T$, and this section concerns the regime in which $|d_T(f)|$ is not negligible. Passing to the opposite regime, $d_T(f)\to 0$, changes the question, but it does not change the answer. Conditionally on the original data set $\mathcal D_T$, the pivot of $\mathbf{CI}^{\mathrm{DA}}_{\mathrm{AoRG}}(\alpha,K)$ at the radius scale equals $\big(d_T(f)-Z/\sqrt{K}\big)/c_T$, where $Z\sim N(0,1)$ is independent of $\mathcal D_T$. Whenever $d_T(f)\to0$ in probability, as under the conditions of Proposition~\ref{prop K infinity}, this pivot converges in law to $-Z/(\sqrt{K}\,z_{1-\alpha/2})$ for every fixed $K$. Its limit law has a positive Lebesgue density, so the interval is non-trivial in the sense of Definition~\ref{def nontrivial}. Triviality is a property of the limit law of the pivot, not of how close the coverage probability is to one. The conditional coverage makes the same point for every $T$: at $d_T(f)=0$ it equals $2\Phi(\sqrt{K}\,c_T)-1$, a probability fixed by $\alpha$, $K$ and $l^{\mathrm{DA}}_T$, whereas the deterministic interval \eqref{eq det interval} covers with conditional probability one, whatever $\alpha$ is. Nothing is lost in this regime either: by Theorem~\ref{thm:aorg-finite}, the conditional variance of the center still falls by the factor $1/K$, the radius is unchanged, and the conditional coverage is at least $1-\alpha$ once $d_T(f)$ lies within the tolerance in \eqref{eq:aorg-protection}. Only $K=\infty$ is trivial, so we exclude it. However, in practice, this \textit{overcoverage is hard to reach}, especially under strong dependence. At $T=2000$ with $K=10$, the empirical coverage in Figure~\ref{fig:coverage_stress} rises from $90.0\%$ at $\gamma=1.01$ to $95.2\%$ at $\gamma=1.3$, far below the value $2\Phi(\sqrt{10}\,c_T)-1>1-10^{-10}$ that a negligible displacement would produce, and Figure~\ref{fig:length_stress} shows that the radius is the same for every $K$.

\section{Simulation: Stress Test Under Strong Dependence}
\label{sec 7}
 This section conducts a series of stress tests to examine the finite-sample performance of our validity-recovery procedure under severe dependence misspecification. We conduct two complementary simulation exercises. First, Section \ref{sec 7.1} investigates whether validity recovery can restore reliable coverage without modeling the underlying dependence, and evaluates its finite-sample efficiency cost against both independence-based and dependence-aware oracle benchmarks. Second, Section \ref{sec 7.2} investigates the potential benefits of incorporating dependence information into validity recovery, focusing on whether dependence-aware studentization can further improve coverage accuracy and the associated coverage–length trade-off. Together, these exercises illustrate two points: (i) modeling dependence is not necessary for validity recovery (Theorem \ref{th 1}); when it is incorporated through a dependence-aware studentizer (Corollary \ref{th DA}), validity recovery delivers substantial additional coverage gains over conventional HAC-based undersmoothing that models dependence in the same way (Section \ref{sec 7.2}); (ii) The averaging over repetitive generations (AoRG) proposed in Section \ref{sec 5} enhances coverage without increasing interval length. See Theorem \ref{thm:aorg-finite} for details.

For both simulation exercises, we consider the fixed-design RDD model introduced in Example \ref{ex 3},
\begin{align*}
    Y_{tT}&=f_-(x_{tT})1[x_{tT}\leq0]+f_+(x_{tT})1[x_{tT}>0]+\varepsilon_{tT},\\
   f_-(x)&=0.48+1.27x+7.18x^2+20.21x^3+21.54x^4+7.33x^5,\\
f_+(x)&=0.52+0.84x-3.00x^2+7.99x^3-9.01x^4+3.56x^5,
\end{align*}
where $x_{tT}=\frac{t-T/2}{T}$, \(t=1,\ldots,T\) and the value of discontinuity is \(\theta(f)=0.04\). The regression functions \(f_-\) and \(f_+\) follow the widely used \citeA{lee2008randomized}-based simulation design in the RDD literature (see, e.g., \citeA{imbens2012optimal}; \citeA{calonico2014robust}; \citeA{pei2022local}). The disturbances are jointly Gaussian with mean zero and covariance structure

\[
\operatorname{Cov}(\varepsilon_{tT},\varepsilon_{sT})
=
\sigma^2(1+|t-s|)^{-\gamma},
\qquad \sigma^2=1.
\]

We consider \(\gamma\in\{1.01,1.05,1.08,1.1,1.2,1.3\}\), all lying close to the summability boundary \(\gamma=1\). These choices generate particularly challenging dependence environments while remaining within the summable-covariance regime. This design retains a familiar regression specification from the classical RDD simulation literature while deliberately introducing severe dependence misspecification into inference implementation.

\subsection{Validity recovery without modeling dependence}
\label{sec 7.1}
Based on sample sizes \(T\in\{300,500,1000,1500,2000\}\) with 2,000 Monte Carlo replications for each configuration, we do a \textit{double-oracle comparison} which involves the following three confidence intervals,
\begin{align*}
    \mathcal{C}_{\mathrm{oracle}}^{\mathrm{ind}}(\alpha),\ \ \mathcal{C}_{\mathrm{oracle}}^{\mathrm{true}}(\alpha), \ \ \mathbf{CI}_{\mathrm{AoRG}}^{\mathrm{ind}}(\alpha,K).
\end{align*}
The first, similar to \eqref{eq CI oracle}, \(\mathcal{C}_{\mathrm{oracle}}^{\mathrm{ind}}(\alpha)\), is an independence-based oracle interval constructed using the local-linear estimator with MSE-optimal bandwidth \(h_0=T^{-1/5}\), the exact smoothing bias, and the true marginal error variance \(\sigma^2\), but ignoring all cross-observation covariances. This setting isolates the consequences of dependence misspecification from bias estimation and marginal variance estimation errors. The second, \(\mathcal{C}_{\mathrm{oracle}}^{\mathrm{true}}(\alpha)\), uses the same estimator, bandwidth, and exact bias correction, but replaces the independence-assumption-based variance with the exact variance of the same estimator under the true covariance matrix, \(\sigma^2\mathbf{w}^\top\boldsymbol{\Sigma}_\gamma\mathbf{w}\), where \(\mathbf{w}=(l^+_{th_0}(0)-l^-_{th_0}(0))_{t=1}^T\) and \((\boldsymbol{\Sigma}_\gamma)_{ts}=(1+|t-s|)^{-\gamma}\). It therefore serves as an infeasible, correctly calibrated oracle benchmark. The third, \(\mathbf{CI}_{\mathrm{AoRG}}^{\mathrm{ind}}(\alpha,K)\), is our feasible validity-recovered interval defined in \eqref{eq CI AoRG}, based on the conventional independence-assumption-based bias corrector and studentizer. We set \(S=3\), \(h=T^{-0.001}\), \(b=T^{-1/7}\), \(l_T=\log T\), and consider \(K\in\{1,3,5,10\}\). Importantly, implementing AoRG uses no information about \(\gamma\) or the underlying covariance matrix.

Our simulations pursue two \textit{objectives} as follows,
\begin{itemize}
    \item  We investigate whether \(\mathbf{CI}_{\mathrm{AoRG}}^{\mathrm{ind}}(\alpha,K)\) can deliver reliable finite-sample validity recovery relative to \(\mathcal{C}_{\mathrm{oracle}}^{\mathrm{ind}}(\alpha)\), even when ignoring dependence produces severe variance underestimation. At the same time, we examine the central promise of Section \ref{sec 5.1}: increasing \(K\) improves coverage without increasing the interval radius.
    \item We assess the finite-sample efficiency cost of dependence-ignorant validity recovery by comparing the average length of \(\mathbf{CI}_{\mathrm{AoRG}}^{\mathrm{ind}}(\alpha,K)\) with that of \(\mathcal{C}_{\mathrm{oracle}}^{\mathrm{true}}(\alpha)\). Since the latter exploits the complete and correctly specified dependence structure, this comparison provides a demanding benchmark for evaluating how much additional interval length is required to recover validity without modeling data dependence. 
\end{itemize}
Together, these two exercises examine whether dependence-ignorant RDD inference induced by the validity recovery procedure can simultaneously achieve substantial robustness to severe dependence misspecification and competitive finite-sample interval length.

\begin{figure}[H]
    \centering
    \includegraphics[width=\textwidth]{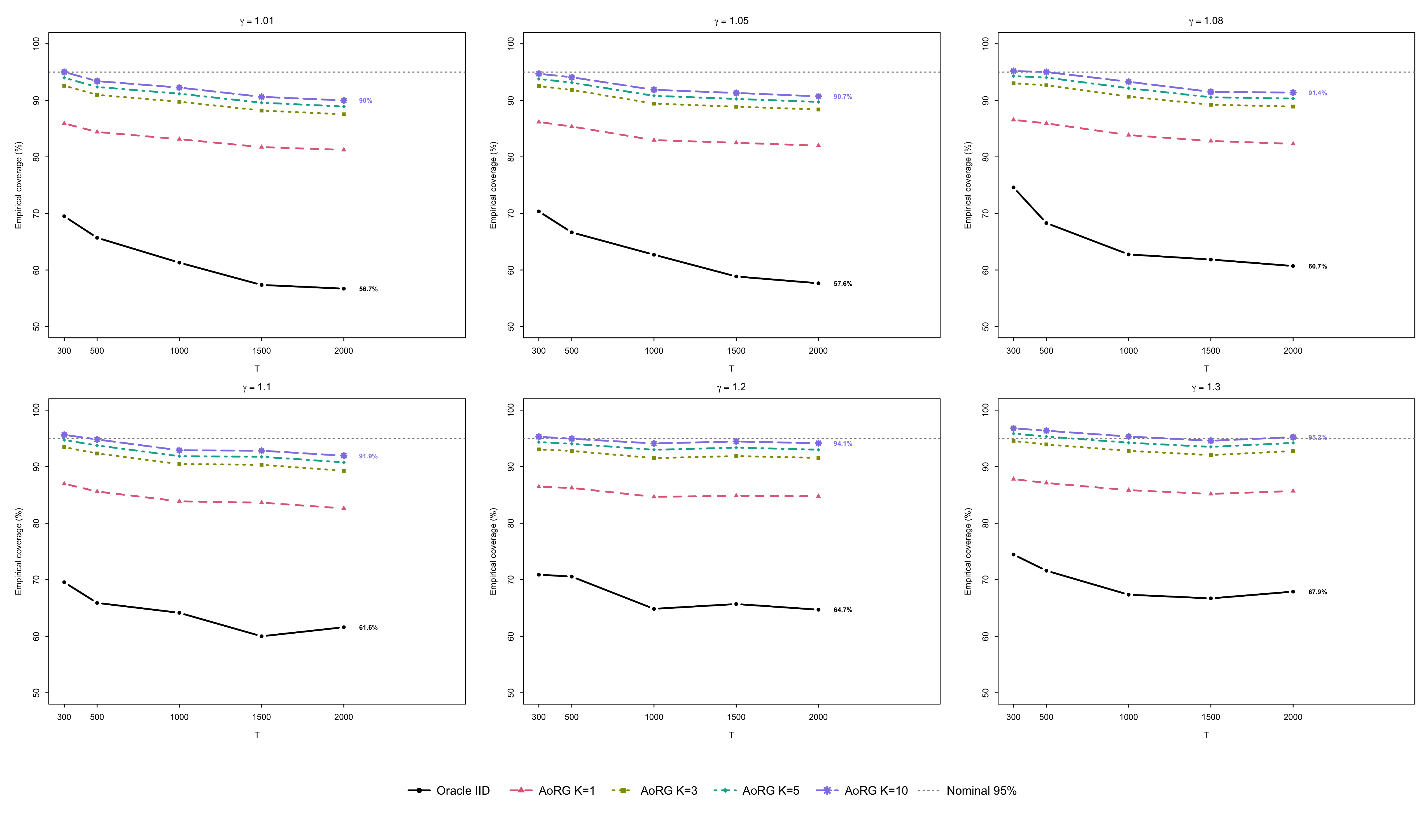}
    \caption{The figure compares the coverage of $\mathbf{CI}_{\mathrm{AoRG}}^{\mathrm{ind}}(\alpha,K)$,
    $K\in\{1,3,5,10\}$, with infeasible but independence-assumption-based interval, $\mathcal{C}_{\mathrm{oracle}}^{\mathrm{ind}}(\alpha)$. The horizontal dotted line indicates the nominal coverage level of $95\%$.}
    \label{fig:coverage_stress}
\end{figure}
Figure \ref{fig:coverage_stress} demonstrates the validity-recovery capability of $\mathbf{CI}_{\mathrm{AoRG}}^{\mathrm{ind}}(\alpha,K)$ under severe dependence misspecification. Even with exact bias correction and knowledge of the true marginal variance $\sigma^2$, the independence-assumption-based interval $\mathcal{C}_{\mathrm{oracle}}^{\mathrm{ind}}(\alpha)$ still suffers from substantial undercoverage, with empirical coverage falling to $56.7\%$–$67.9\%$ at large sample size \(T=2000\). In contrast, $\mathbf{CI}_{\mathrm{AoRG}}^{\mathrm{ind}}(\alpha,K)$ uses only the conventional independence-assumption-based bias corrector and studentizer as inputs but delivers much more reliable coverage without modeling or estimating the underlying dependence structure. More importantly, increasing $K$ consistently improves coverage without changing the interval radius: at $T=2000$, the coverage at $K=10$ ranges from $90.0\%$ to $95.2\%$ across the six configurations. This directly supports the central promise of Section \ref{sec 5.1}: repetitive Gaussian generation provides a free improvement in coverage, even in highly unfavorable dependence environments.

\begin{figure}[H]
    \centering
    \includegraphics[width=\textwidth]{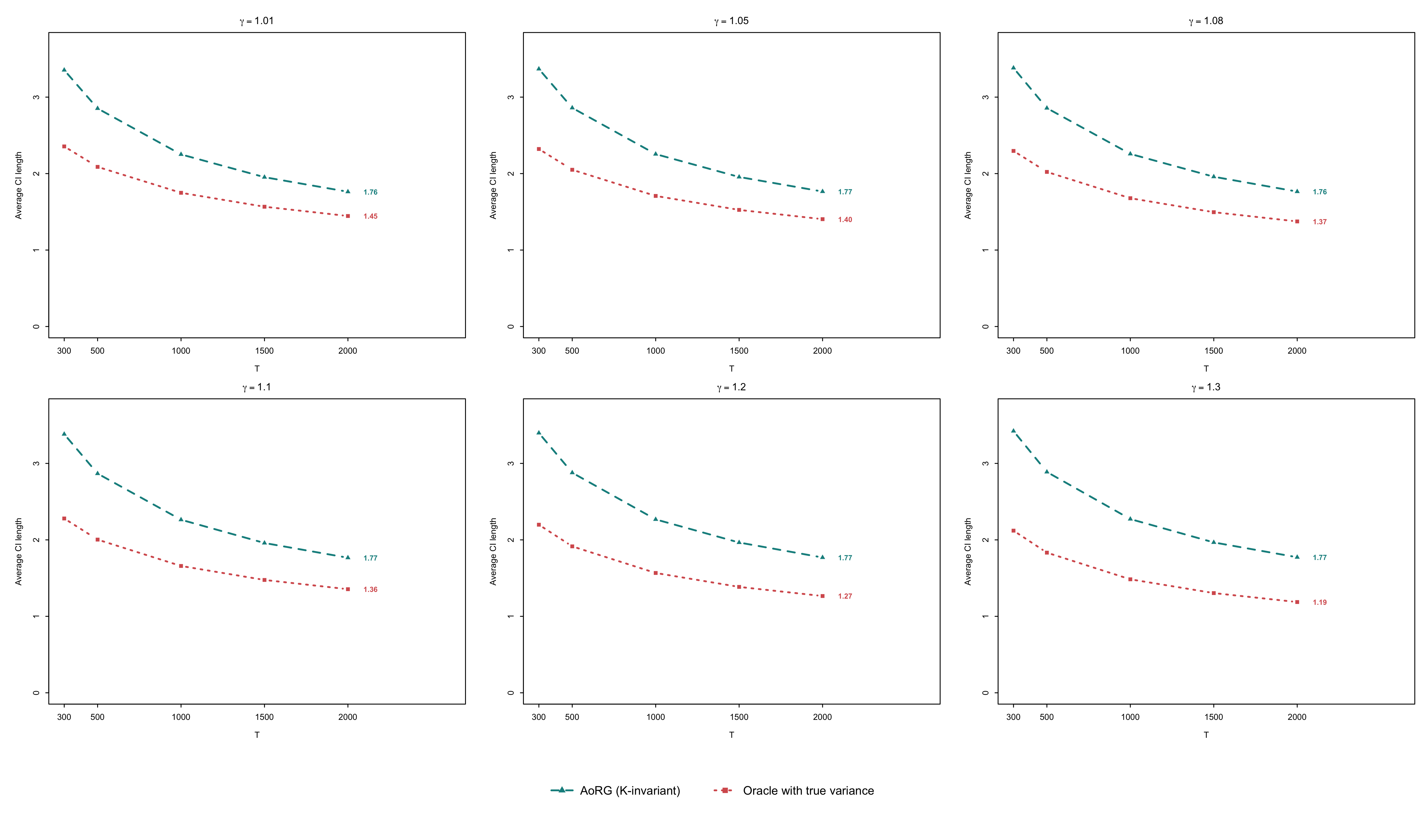}
    \caption{ The figure compares the average interval length of $\mathbf{CI}_{\mathrm{AoRG}}^{\mathrm{ind}}(\alpha,K)$ and the infeasible but correct interval, $\mathcal{C}_{\mathrm{oracle}}^{\mathrm{true}}(\alpha)$.
    The length of $\mathbf{CI}_{\mathrm{AoRG}}^{\mathrm{ind}}(\alpha,K)$ is invariant with respect
    to $K$. The numerical labels at $T=2000$
    report the average lengths of $\mathbf{CI}_{\mathrm{AoRG}}^{\mathrm{ind}}(\alpha,K)$ and $\mathcal{C}_{\mathrm{oracle}}^{\mathrm{true}}(\alpha)$.
    }
    \label{fig:length_stress}
\end{figure}

In addition to validity recovery, another important criterion is interval length. A natural benchmark for comparison is $\mathcal{C}_{\mathrm{oracle}}^{\mathrm{true}}(\alpha)$, which is equipped with oracle bias-correction, MSE-optimal bandwidth and perfect studentization. Figure \ref{fig:length_stress} shows that the interval-length premium remains modest even under severe ignorance of dependence.  

Figure \ref{fig:length_stress} examines the efficiency cost of this validity recovery. At \(T=2000\), the average interval length of $\mathbf{CI}_{\mathrm{AoRG}}^{\mathrm{ind}}(\alpha,K)$ is approximately $1.76–1.77$, compared with $1.19$–$1.45$ for the $\mathcal{C}_{\mathrm{oracle}}^{\mathrm{true}}(\alpha)$. Thus, $\mathbf{CI}_{\mathrm{AoRG}}^{\mathrm{ind}}(\alpha,K)$ incurs only about a $21\%$–$49\%$ length premium relative to the benchmark interval $\mathcal{C}_{\mathrm{oracle}}^{\mathrm{true}}(\alpha)$ which is equipped with complete knowledge of the dependence structure. Moreover, as dependence becomes stronger, $\mathcal{C}_{\mathrm{oracle}}^{\mathrm{true}}(\alpha)$ requires increasingly longer intervals, whereas the length of $\mathbf{CI}_{\mathrm{AoRG}}^{\mathrm{ind}}(\alpha,K)$  remains nearly unchanged. Consequently, the relative length penalty is smallest precisely when dependence misspecification is most severe. These findings show that dependence-ignorant validity recovery can be achieved at a relatively modest finite-sample efficiency cost.

Taken together, these simulations provide a compelling dual-oracle assessment of our validity-recovery procedure.  $\mathcal{C}_{\mathrm{oracle}}^{\mathrm{ind}}(\alpha)$, equipped with exact bias correction and the true marginal error variance, demonstrates that even nearly ideal knowledge of conventional nuisance quantities cannot prevent severe undercoverage when data dependence is misspecified.  $\mathcal{C}_{\mathrm{oracle}}^{\mathrm{true}}(\alpha)$, which additionally exploits the true dependence structure, provides a benchmark for evaluating the efficiency cost of our procedure. Against these two benchmarks, our $\mathbf{CI}_{\mathrm{AoRG}}^{\mathrm{ind}}(\alpha,K)$ intervals achieve substantial coverage improvements without modeling the underlying dependence, while incurring only a modest interval-length premium relative to the fully informed oracle. This dual-oracle comparison therefore isolates the fundamental limitations of conventional dependence-sensitive calibration and shows that we can recover reliable inference without learning the true dependence structure, at a relatively mild efficiency cost.

\subsection{Validity recovery assisted by modeling dependence}
\label{sec 7.2}
Section \ref{sec 7.2} focuses on the finite-sample performance of dependence-aware validity recovery relative to conventional HAC-based undersmoothing inference. Corollary \ref{th DA} establishes that the dependence-aware validity recovery permits the use of the “\(S=2\)”-optimal local-linear bandwidth without explicit bias correction, transferring the role of conventional undersmoothing to the Gaussian calibration factor. To examine this theoretical prediction, we compare the dependence-aware AoRG interval with conventional undersmoothing intervals based on HAC studentization. Using the same data-generating process and dependence configurations specified above, we set \(\check B_h=0\), \(h_{\mathrm{DA}}=T^{-1/5}\), and \(l_T^{\mathrm{DA}}=T^{0.05}\), yielding \(m_T^{\mathrm{DA}}=\lfloor T^{3/4}\rfloor+1\). As benchmarks, we construct conventional local-linear undersmoothing intervals with bandwidths \(h=T^{-a}\), \(a\in\{0.25,0.35,0.5\}\), using the same Newey–West HAC lag rule and standard-normal critical values. Notably, the weakest undersmoothing choice has effective sample size \(Th=T^{3/4}\), matching \(m_T^{\mathrm{DA}}\) up to the integer adjustment. We evaluate empirical coverage and average interval length across \(T\in\{1000,2000,3000,4000,5000\}\), with 2,000 Monte Carlo replications per configuration, and examine whether increasing \(K\) further improves coverage without increasing interval length, in line with Theorem \ref{thm:aorg-finite}. For HAC variance estimation, we employ the Newey--West estimator with Bartlett weights and truncation lag \(L_T=\max\{1,\lfloor 4(T/100)^{2/9}\rfloor\}\). See \citeA{NeweyWest1987,newey1994automatic} for examples. This lag-selection rule is applied uniformly to dependence-aware validity recovery and all conventional HAC-based undersmoothing benchmarks, regardless of the bandwidth choice or the underlying dependence parameter \(\gamma\). In particular, no knowledge of the true dependence structure is used in selecting \(L_T\).

The \textit{objective} of this simulation is as follows:
\begin{itemize}
    \item Assess whether dependence-aware validity recovery can achieve substantially better finite-sample coverage than conventional HAC-based undersmoothing at a comparable effective sample size and a modest interval-length cost, while further exploiting AoRG to improve coverage without increasing interval length.
\end{itemize}

\begin{figure}[H]
    \centering
    \includegraphics[width=\textwidth]{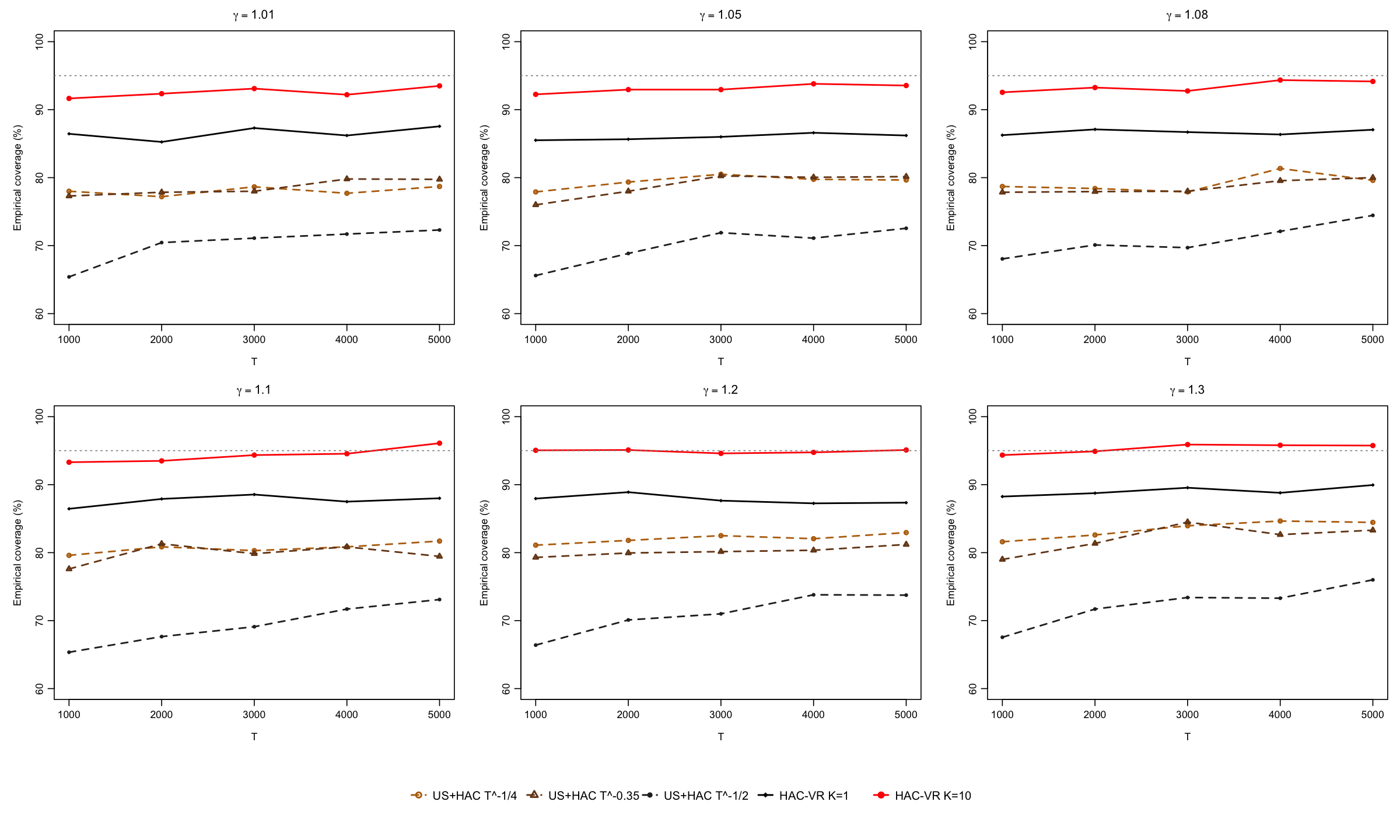}
    \caption{Empirical coverage probabilities across six dependence configurations indexed by $\gamma$. In the legend, “US+HAC” denotes conventional undersmoothing inference with Newey--West HAC studentization, where $T^{-1/4}$, $T^{-0.35}$, and $T^{-1/2}$ indicate the respective bandwidth choices. “HAC-VR” denotes HAC-based dependence-aware validity recovery, $\mathrm{CI}_{\mathrm{AoRG}}^{\mathrm{DA}}(\alpha,K)$,
where $K=1$ and $K=10$ indicate the numbers of independent
Gaussian generations used in averaging over repetitive generation (AoRG). The dashed curves represent US+HAC, while the black and red solid curves represent HAC-VR with $K=1$ and $K=10$, respectively. The horizontal dotted line indicates the nominal coverage level of $95\%$.}
    \label{fig:coverage_stress_2}
\end{figure}

\begin{figure}[H]
    \centering
    \includegraphics[width=\textwidth]{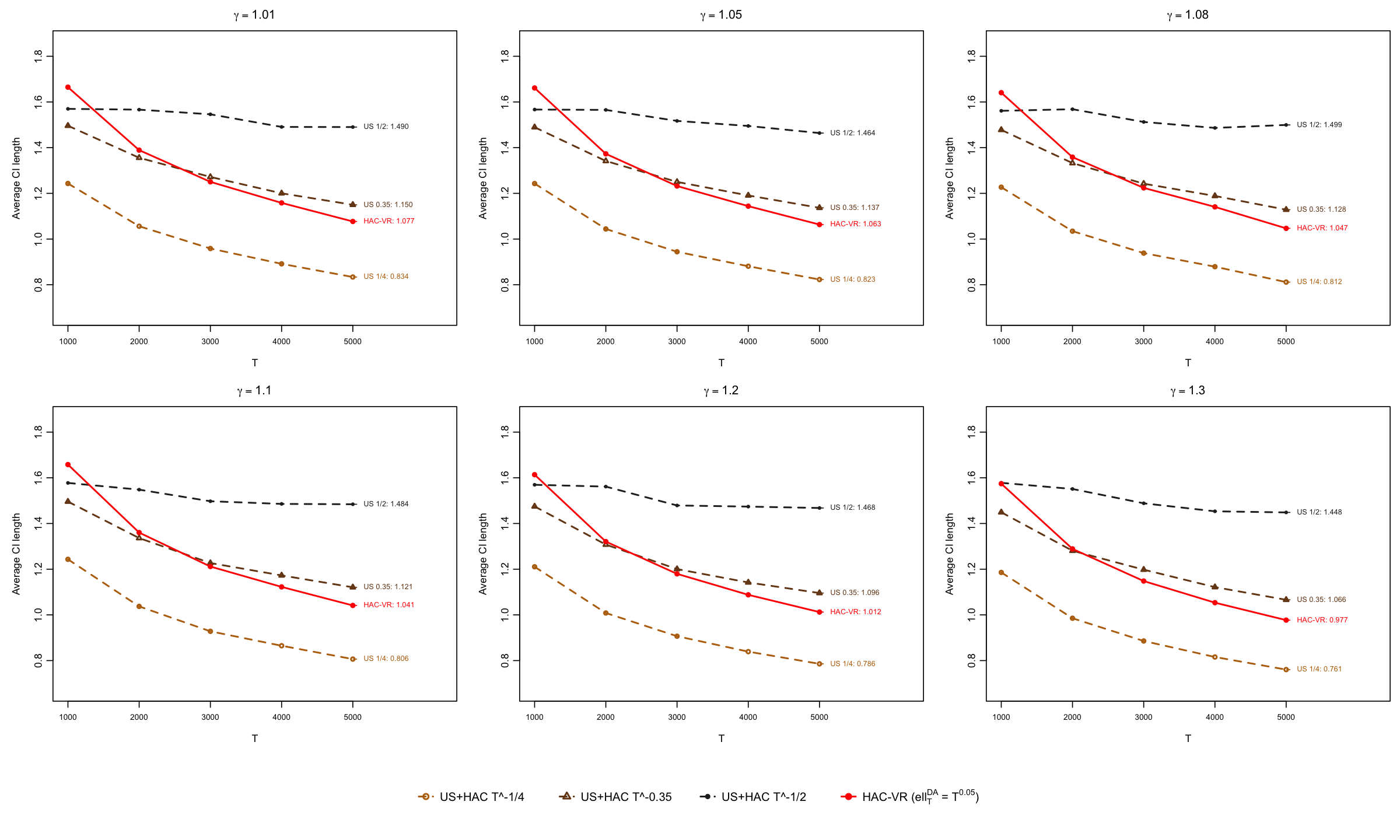}
    \caption{The figure compares the average confidence interval lengths of
$\mathrm{CI}_{\mathrm{AoRG}}^{\mathrm{DA}}(\alpha,K)$ and three conventional HAC-based undersmoothing intervals. For the weakest undersmoothing benchmark $h=T^{-1/4}$, the effective sample size $Th=T^{3/4}$ matches $m_T^{\mathrm{DA}}=\lfloor T^{3/4}\rfloor+1$ up to integer rounding. The leading difference in their radius calibration is therefore the folded-normal critical value
$\operatorname{cv}_{1-\alpha}(T^{-0.025})$ versus the standard-normal critical value $z_{1-\alpha/2}$, yielding a nearly constant-level gap in finite-sample interval lengths. The length of $\mathrm{CI}_{\mathrm{AoRG}}^{\mathrm{DA}}(\alpha,K)$ is invariant in $K$. Numerical labels at $T=5000$ report the average interval lengths.}
    \label{fig:length_stress_2}
\end{figure}

We now turn to the simulation results: Figure \ref{fig:coverage_stress_2} reports empirical coverage, and Figure \ref{fig:length_stress_2} examines the corresponding average confidence interval lengths across the six dependence configurations.

Figure \ref{fig:coverage_stress_2} shows the substantial finite-sample coverage improvement from dependence-aware validity recovery under strong serial dependence. Across all six dependence configurations, conventional HAC-based undersmoothing suffers from persistent undercoverage, with empirical coverage generally remaining below $85\%$ even at \(T=5000\). In contrast, \(\mathrm{CI}_{\mathrm{AoRG}}^{\mathrm{DA}}(\alpha,K)\) delivers considerably higher coverage even with \(K=1\). More importantly, increasing \(K\) from 1 to 10 consistently raises empirical coverage by several percentage points, bringing it close to the nominal $95\%$ level across all configurations. This improvement is particularly noteworthy because \(l_T^{\mathrm{DA}}=T^{0.05}\) diverges very slowly, so finite-sample displacement control need not yet be sufficiently accurate. These findings are consistent with Theorem \ref{thm:aorg-finite}: AoRG can substantially alleviate finite-sample undercoverage by reducing auxiliary Gaussian uncertainty, without enlarging interval length or requiring faster convergence of the random bias term defined in \eqref{eq new decomposition}.

\paragraph{Where the extra length comes from.}
Figure \ref{fig:length_stress_2} shows that the substantial coverage improvement in Figure \ref{fig:coverage_stress_2} is achieved at a modest interval-length cost, and this cost can be traced exactly. The weakest conventional undersmoothing benchmark, \(h=T^{-1/4}\), has effective sample size \(Th=T^{3/4}\), which matches \(m_T^{\mathrm{DA}}=\lfloor T^{3/4}\rfloor+1\) up to integer rounding. At the leading-order radius level, the two intervals therefore differ only in their critical values: the folded-normal critical value \(\operatorname{cv}_{1-\alpha}(T^{-0.025})\) for validity recovery versus the standard-normal critical value \(z_{1-\alpha/2}\) for the conventional interval. Table \ref{tab cv ratio} reports their ratio at \(\alpha=0.05\). This ratio alone implies a length premium of \(25\%\)--\(27\%\) at every sample size considered, which accounts for most of the \(28\%\)--\(30\%\) premium observed at \(T=5000\) in Figure \ref{fig:length_stress_2}. The extra length is thus the price of the folded-normal calibration, which absorbs the random bias term in \eqref{eq new decomposition}, rather than the result of any additional interval enlargement.

\begin{table}[H]
    \centering
    \begin{tabular}{cccccc}
        \hline
        $T$ & $1000$ & $2000$ & $3000$ & $4000$ & $5000$ \\
        \hline
        $T^{-0.025}$ & $0.841$ & $0.827$ & $0.819$ & $0.813$ & $0.808$ \\
        $\operatorname{cv}_{0.95}(T^{-0.025})$ & $2.490$ & $2.476$ & $2.468$ & $2.463$ & $2.458$ \\
        $\operatorname{cv}_{0.95}(T^{-0.025})/z_{0.975}$ & $1.271$ & $1.264$ & $1.259$ & $1.257$ & $1.254$ \\
        \hline
    \end{tabular}
    \caption{Leading-order radius ratio between dependence-aware validity recovery and the matched undersmoothing benchmark \(h=T^{-1/4}\) at \(\alpha=0.05\). The ratio equals the folded-normal critical value \(\operatorname{cv}_{0.95}(T^{-0.025})\) divided by the standard-normal critical value \(z_{0.975}=1.960\).}
    \label{tab cv ratio}
\end{table}
Moreover, for \(T\ge 3000\), the dependence-aware validity-recovered interval is shorter than both stronger undersmoothing benchmarks, \(h=T^{-0.35}\) and \(h=T^{-1/2}\), while achieving markedly higher coverage. Finally, increasing \(K\) from 1 to 10 leaves the confidence radius exactly unchanged, confirming that the coverage enhancement from AoRG entails no additional interval-length cost.

Taken together, Figures \ref{fig:coverage_stress_2} and \ref{fig:length_stress_2} provide compelling finite-sample evidence consistent with the theoretical discussion in Sections \ref{sec 3.4} and \ref{sec 5}. First, as Corollary \ref{th DA} establishes, dependence-aware validity recovery lets us use the \(S=2\)-optimal local-linear bandwidth with no explicit bias correction, while achieving substantially better coverage than conventional HAC-based undersmoothing. Importantly, this improvement comes at only a modest, approximately constant-factor interval-length cost relative to the benchmark with a matched effective sample size, rather than relying on aggressive interval enlargement. Second, consistent with Theorem \ref{thm:aorg-finite}, increasing \(K\) further improves empirical coverage without changing interval length, even when \(l_T^{\mathrm{DA}}\) diverges very slowly and finite-sample displacement control may remain imperfect. Overall, these findings reinforce the central message of our theoretical analysis: validity recovery fundamentally changes the role of stochastic calibration, allowing dependence-aware implementation and repetitive Gaussian generation to deliver substantial finite-sample inferential gains without sacrificing interval-length efficiency.

\section{Conclusions}

This paper investigates whether data dependence can be ignored in the implementation of regression discontinuity design (RDD) inference without compromising inferential validity. We first demonstrate a fundamental dilemma of conventional asymptotic-normality-based inference: ignoring dependence may produce persistent undercoverage even with perfect bias control, whereas explicitly modeling dependence can introduce substantial difficulties in stochastic calibration. To overcome this dilemma, we develop a validity-recovery procedure that decouples the dependence structure of the original observations from stochastic calibration by introducing auxiliary Gaussian randomness. For fixed-design RDD, the resulting confidence intervals achieve asymptotically exact nominal coverage without modeling the underlying dependence, while preserving the leading constant of the independence-assumption-based benchmark and attaining near-minimax interval-length rates. For random-design RDD, we further establish explicit coverage-error bounds and characterize their trade-off with interval length under more specific dependence conditions.

Beyond restoring validity, our framework offers several additional inferential benefits. Although dependence modeling is not necessary for validity recovery, dependence-aware studentization can further enhance coverage. The framework also accommodates the optimal local-linear bandwidth without explicit bias correction, transferring the role of conventional undersmoothing to the auxiliary Gaussian calibration. Moreover, averaging over repetitive generation (AoRG) simultaneously stabilizes the randomized interval center and improves coverage without increasing the interval radius. We establish, for every sample size, a finite-sample coverage lower bound that is non-decreasing in the number of repetitions.

Our Monte Carlo experiments provide finite-sample evidence consistent with these theoretical findings. Even under strong serial dependence, validity recovery substantially improves coverage relative to conventional procedures, with comparatively modest interval-length costs. Dependence-aware calibration and AoRG provide further improvements, illustrating that enhanced coverage need not rely on enlarging confidence intervals. The central implication is that data independence need not be a property of the true data-generating process to serve as a working assumption in inference implementation. Under appropriate conditions, validity recovery allows researchers to retain the simplicity of independence-based stochastic calibration while accommodating dependence in the original observations.

\bibliographystyle{apacite}
\bibliography{ref.bib}

\newpage

\appendix

\section{Proof of Results in Section \ref{sec 2}}
\label{app A}
\setcounter{equation}{0}
\renewcommand\theequation{A.\arabic{equation}}
\renewcommand\thetheorem{A.\arabic{theorem}}
\renewcommand\thecorollary{A.\arabic{corollary}}
\renewcommand\thelemma{A.\arabic{lemma}}
\renewcommand\theremark{A.\arabic{remark}}
\setcounter{proposition}{0}
\renewcommand{\theproposition}{B.\arabic{proposition}}

\noindent\textbf{Proof of Proposition \ref{prop underestimation}} 

\noindent\textbf{Step 1} (Asymptotic equivalence) Based on $L_{ih,\pm}(0)$ and $L_{ih,\Psi,\pm}(0)$ introduced in \eqref{eq local linear random} and Appendix \ref{app concentration: random} with $q=1$, we first define
\begin{align*}
    w_{ih}&=L_{ih,+}(0)-L_{ih,-}(0),
v_{ih}=L_{ih,\Psi,+}(0)-L_{ih,\Psi,-}(0),\\
&\Psi(x)=\frac{x+1}{2}\1[-1\leq x\leq 1]+\1[x>1].
\end{align*}
Using basic properties of Markov chains, we know $\Psi$ is the distribution of each $X_i$ in Example \eqref{ex 1}. Since the running-variable process $\{X_i\}_{i\geq 1}$ has a refresh probability $1-\rho>0$, it is stationary and geometrically $\beta$-mixing, with $\beta(k)\le\rho^k$. The VC-dimension argument combined with Lemma \ref{lemma 3}, as in Step 1 of the proof of Theorem \ref{th F3}, gives
$P(D_n>d_n)=o(1)$, for some $d_n=o(h)$. On $\{D_n\le d_n\}$, Theorem \ref{th concentration random} yields
\begin{align}
\label{eq proof of prop 1 eq1}
    \max_{1\le i\le n}|w_{ih}-v_{ih}|=o(\frac{1}{nh}).
\end{align}
Note that both weights vanish whenever $|X_i|>h$, and the number of observations satisfying $|X_i|\le h$ is $O_\P(nh)$, since $D_n=o_\P(h)$. Consequently, $nh\sum_{i=1}^n(w_{ih}-v_{ih})^2
=o_\P(1).$ We thus only need to consider the ratio 
\begin{align*}
    \frac{\sigma^2\sum_{i=1}^nv_{ih}^2(1+o(1))}{\mathrm{Var}(\tilde\theta_h|X_1,...,X_n)}.
\end{align*}

\noindent\textbf{Step 2} (Moments of $v_{ih}$) Recall that $\r(u)^\top \e_0=1$, where $\r(u)=(1,u)^\top$ and $\e_0=(1,0)^\top$. By the definition of $L_{ih,\Psi,\pm}$ and the polynomial reproduction property, we have
\begin{align*}
    \E[L_{ih,\Psi,+}(0)]=\E[L_{ih,\Psi,-}(0)]=\frac1n,
\qquad \E[v_{ih}]=0.
\end{align*}
For every fixed $k\ge1$, the Markov mechanism \eqref{eq markov} implies that $X_{i+k}=X_i$ with probability $\rho^k$. Otherwise, at least one refresh occurs, making $X_{i+k}$ independent of $X_i$ with the same marginal distribution. Therefore,
\begin{align}
\label{eq proof of prop 1 eq2}
    \E[v_{ih}v_{i+k,h}]=\rho^k\E[v_{ih}^2]+(1-\rho^k)(\E[v_{ih}])^2=\rho^k\E[v_{ih}^2].
\end{align}
Note that $|v_{ih}|\le \frac{C}{nh}\mathbf{1}\{|X_i|\le h\}$ and, since $\Psi$ is the distribution function of $U[-1,1]$, $P(|X_i|\le h)=h$. Consequently, $E[v_{ih}^4]\le \frac{C}{n^4h^3}$, $
E[v_{ih}^2]\asymp\frac{1}{n^2h}$. Indeed, by the definition of $L_{ih,\Psi,\pm}(0)$,
\begin{align*}
n^2hE[v_{ih}^2]=\frac12\int_0^1\left[e_0^\top
M_{1h,\Psi}(0;0,1)^{-1}K(u)r(u)\right]^2du+\frac12\int_{-1}^0\left[e_0^\top M_{1h,\Psi}(0;-1,0)^{-1}K(u)r(u)\right]^2du
\asymp 1,
\end{align*}
where both population moment matrices are independent of $h$
under the uniform marginal distribution and are positive definite
by Lemma \ref{lemma 2}.

For every integer $k\ge0$, the transition mechanism \eqref{eq markov} and $E[v_{ih}]=0$ imply
\begin{align*}
    \E[v_{i+k,h}\mid X_i]&=\rho^k v_{ih},\ \E[v_{i+k,h}^2\mid X_i]
=\rho^k v_{ih}^2+(1-\rho^k)E[v_{ih}^2],\ \E[v_{ih}v_{i+k,h}^3]=\rho^kE[v_{ih}^4].
\end{align*}
Meanwhile, for $r>k$, iterated conditioning on $X_{i+r}$ and $X_{i+k}$ gives
\begin{align*}
&\E[v_{ih}v_{i+k,h}v_{i+r,h}v_{i+r+k,h}]=\rho^kE\left[v_{ih}v_{i+k,h}
E[v_{i+r,h}^2\mid X_{i+k}]\right]\\
&\quad=\rho^{r}E[v_{ih}v_{i+k,h}^3]+\rho^k(1-\rho^{r-k})E[v_{ih}v_{i+k,h}]E[v_{ih}^2]\\
&\quad=\rho^{r+k}E[v_{ih}^4]+\rho^{2k}(1-\rho^{r-k})(E[v_{ih}^2])^2.
\end{align*}

\noindent\textbf{Step 3} (variance of partial sum) Together with \eqref{eq proof of prop 1 eq2} (which also holds for $k=0$), this yields the exact covariance identity
\begin{align*}
    \mathrm{Cov}(v_{ih}v_{i+k,h}, v_{i+r,h}v_{i+r+k,h})=\rho^{r+k}(E[v_{ih}^4]-(E[v_{ih}^2])^2),\qquad r>k.
\end{align*}
For $0\le r\le k$, the bound on $v_{ih}$ and the marginal
probability $P(|X_i|\le h)=h$ give
\begin{align*}
    \left|\mathrm{Cov}(v_{ih}v_{i+k,h},v_{i+r,h}v_{i+r+k,h})\right|\le\frac{C}{n^4h^3}.
\end{align*}
For $r>k$, the exact covariance identity gives the sharper bound
\begin{align*}
    \left|\mathrm{Cov}(v_{ih}v_{i+k,h},v_{i+r,h}v_{i+r+k,h})\right|\le\frac{C\rho^{r+k}}{n^4h^3}.
\end{align*}
Summing over all pairs of indices, for each fixed $k\ge0$,
\begin{align*}
\mathrm{Var}\left(nh\sum_{i=1}^{n-k}v_{ih}v_{i+k,h}\right)\le C\frac{(nh)^2n}{n^4h^3}\left(k+1+\sum_{r=k+1}^{\infty}\rho^{r+k}\right)=O\left(\frac{k+1}{nh}\right)=o(1).
\end{align*}

\noindent \textbf{Step 4} By \eqref{eq proof of prop 1 eq2} and stationarity,
\begin{align}
\label{eq proof of prop 1 eq3}
    nh\sum_{i=1}^n v_{ih}^2&=n^2hE[v_{ih}^2]+o_\P(1)=O_\P(1),\notag\\
nh\sum_{i=1}^{n-k}v_{ih}v_{i+k,h}&=(n-k)nh\rho^kE[v_{ih}^2]+o_\P(1)=\rho^k nh\sum_{i=1}^n v_{ih}^2+o_\P(1),
\end{align}
for each fixed $k\ge1$. By Step 1, $nh\sum_{i=1}^n(w_{ih}-v_{ih})^2=o_\P(1)$. The Cauchy--Schwarz inequality and \eqref{eq proof of prop 1 eq3} then imply $ nh\sum_{i=1}^n w_{ih}^2=nh\sum_{i=1}^n v_{ih}^2+o_\P(1)=O_\P(1)$ and, for every fixed $k\ge1$,
\begin{align*}
    nh\left|\sum_{i=1}^{n-k}w_{ih}w_{i+k,h}
-\sum_{i=1}^{n-k}v_{ih}v_{i+k,h}\right|=o_\P(1).
\end{align*}
Consequently,
\begin{align}
\label{eq proof of prop 1 eq4}
    \sum_{i=1}^n w_{ih}^2=O_\P(\frac{1}{nh}),\qquad
\frac{\sum_{i=1}^{n-k}w_{ih}w_{i+k,h}}
{\sum_{i=1}^n w_{ih}^2}\xrightarrow{\P}\rho^k.
\end{align}
Since $\{\varepsilon_i\}$ and $\{X_i\}$ are independent
processes, some simple algebra gives
\begin{align*}
\mathrm{Var}(\tilde\theta_h\mid X_1,\ldots,X_n)=\sigma^2\sum_{i=1}^n w_{ih}^2+2\sigma^2\sum_{k=1}^{n-1}(1+k)^{-\gamma}\sum_{i=1}^{n-k}w_{ih}w_{i+k,h}.
\end{align*}
Recall that, for every $k\ge1$, Cauchy--Schwarz asserts $|\sum_{i=1}^{n-k}w_{ih}w_{i+k,h}|\le\sum_{i=1}^n w_{ih}^2.$ Since $\gamma>1$, the series $\sum_{k\ge1}(1+k)^{-\gamma}$ is summable. Thus \eqref{eq proof of prop 1 eq4}, followed by truncation of that series, yields
\begin{align*}
    \frac{\mathrm{Var}(\tilde\theta_h\mid X_1,\ldots,X_n)}{\sigma^2\sum_{i=1}^n w_{ih}^2}&=1+2\sum_{k=1}^{n-1}(1+k)^{-\gamma}\frac{\sum_{i=1}^{n-k}w_{ih}w_{i+k,h}}{\sum_{i=1}^n w_{ih}^2}\\
&\xrightarrow{\P}1+2\sum_{k=1}^{\infty}
\frac{\rho^k}{(1+k)^\gamma}>1.
\end{align*}
Thus, we finish the proof of Proposition \ref{prop underestimation}.\\

\noindent\textbf{Proof of Proposition \ref{prop bias aware}}
Since
\[
\inf_{f\in\mathcal F_n}
P\left(
\theta(f)\in
\left[
\hat\theta_n(f)\pm
\hat s_n cv_{1-\alpha}(B_n/\hat s_n)
\right]
\right)
\]
is bounded above by the coverage probability evaluated at $f_n^+$, it is sufficient to consider $f_n^+$ defined in Assumption 1 A1. By construction, $E_{f_n^+}[\hat\theta_n(f_n^+)]-\theta(f_n^+)=B_n.$ Under Assumption \ref{as 1} A2,
\[
Z:=
\frac{
\hat\theta_n(f_n^+)-E_{f_n^+}[\hat\theta_n(f_n^+)]
}{
\sigma_n
}
\sim N(0,1),
\]
and hence $\hat\theta_n(f_n^+)-\theta(f_n^+)=\sigma_n Z+B_n.$ Therefore,
\begin{align*}
    P\left(
\theta(f_n^+)\in
\left[
\hat\theta_n(f_n^+)
\pm
\hat s_n \text{cv}_{1-\alpha}(B_n/\hat s_n)
\right]
\right) =
P\left(
|\sigma_n Z+B_n|
\le
\hat s_n \text{cv}_{1-\alpha}(B_n/\hat s_n)
\right).
\end{align*}

\noindent \textbf{Step 1} We first establish a deterministic monotonicity property of the mapping $s\to s\ cv_{1-\alpha}(B_n/s)$, where $s>0$. For every $t\ge0$, the definition of $cv_{1-\alpha}(t)$ and the symmetry of the standard normal distribution give
\[
\Phi\left(\text{cv}_{1-\alpha}(t)-t\right)
+
\Phi\left(\text{cv}_{1-\alpha}(t)+t\right)
=
2-\alpha.
\]
Since $\alpha\in(0,1/2)$ implies $1-\alpha>1/2$, we have
\[
P(|N(t,1)|\le t)
=
P(-2t\le Z\le0)
<
\frac12
<
1-\alpha,
\]
which asserts $\text{cv}_{1-\alpha}(t)>t$, $t\ge0.$ Since $\Phi\left(\text{cv}_{1-\alpha}(t)-t\right)
+
\Phi\left(\text{cv}_{1-\alpha}(t)+t\right)
=
2-\alpha$, implicit differentiation of this equation with respect to $t$ yields
\[
\text{cv}'_{1-\alpha}(t)
=
\frac{
\phi\left(\text{cv}_{1-\alpha}(t)-t\right)
-
\phi\left(\text{cv}_{1-\alpha}(t)+t\right)
}{
\phi\left(\text{cv}_{1-\alpha}(t)-t\right)
+
\phi\left(\text{cv}_{1-\alpha}(t)+t\right)
}.
\]
Because $\text{cv}_{1-\alpha}(t)>t\ge0$, both $\text{cv}_{1-\alpha}(t)-t$ and $\text{cv}_{1-\alpha}(t)+t$ are nonnegative, with $\text{cv}_{1-\alpha}(t)+t\ge\text{cv}_{1-\alpha}(t)-t$. Since the standard-normal density is decreasing on $[0,\infty)$, $0\le \text{cv}'_{1-\alpha}(t)<1$. Consequently, for every $s>0$,
\begin{align*}
    \frac{d}{ds}\left\{s\,\text{cv}_{1-\alpha}(B_n/s)\right\}&=
\text{cv}_{1-\alpha}(B_n/s)-\frac{B_n}{s}\text{cv}'_{1-\alpha}(B_n/s) \\
&\ge\text{cv}_{1-\alpha}(B_n/s)-\frac{B_n}{s}>0.
\end{align*}
Thus, $s\longmapsto s\,\text{cv}_{1-\alpha}(B_n/s)$ is strictly increasing on $(0,\infty)$.

\noindent\textbf{Step 2} Now consider the event $\{\frac{\sigma_n}{\hat s_n}\ge\beta\}$. Since $\hat s_n>0$, this event implies $\hat s_n\le\frac{\sigma_n}{\beta}$. By the monotonicity established above, we obtain $ \hat s_n\text{cv}_{1-\alpha}(B_n/\hat s_n)\le\frac{\sigma_n}{\beta}\text{cv}_{1-\alpha}\left(\frac{\beta B_n}{\sigma_n}\right).$
Hence,
\begin{align*}
    &\P\left(|\sigma_n Z+B_n|\le\hat s_n\text{cv}_{1-\alpha}(B_n/\hat s_n)\right) \\
&\le \P\left(|\sigma_n Z+B_n|\le\hat s_n\text{cv}_{1-\alpha}(B_n/\hat s_n),\frac{\sigma_n}{\hat s_n}\ge\beta\right)+\P\left(\frac{\sigma_n}{\hat s_n}<\beta\right) \\
&\le\P\left(|\sigma_n Z+B_n|\le\frac{\sigma_n}{\beta}\text{cv}_{1-\alpha}\left(\frac{\beta B_n}{\sigma_n}\right)\right)+
\P\left(\frac{\sigma_n}{\hat s_n}<\beta\right).
\end{align*}
Multiplying the inequality inside the first probability by $\beta/\sigma_n$ gives
\begin{align*}
    \P\left(\left|\beta Z+\frac{\beta B_n}{\sigma_n}\right|\le\text{cv}_{1-\alpha}\left(\frac{\beta B_n}{\sigma_n}\right)\right)=G_\alpha\left(\frac{\beta B_n}{\sigma_n},\beta\right).
\end{align*}
Therefore,
\begin{align*}
    &\P\left(\theta(f_n^+)\in\left[
\hat\theta_n(f_n^+)\pm\hat s_n\text{cv}_{1-\alpha}(B_n/\hat s_n)\right]\right) \le
G_\alpha\left(\frac{\beta B_n}{\sigma_n},\beta\right)+\P\left(\frac{\sigma_n}{\hat s_n}<\beta
\right) \\
&\le\sup_{t\ge0}G_\alpha(t,\beta)+\P\left(\frac{\sigma_n}{\hat s_n}<\beta\right)<1-\alpha-\eta_{\alpha,\beta}+\P\left(\frac{\sigma_n}{\hat s_n}<\beta\right),
\end{align*}
where the last inequality is due to Lemma \ref{lemma 1}. 
Together with Assumption \ref{as 1} A3, we finish the proof.\\

%%%%%%%%%%%%%%%%%%%%%%%%%%%%%%%%%%%%%%%%%%%%%%%%%%%%%%%%
\noindent\textbf{Proof of Proposition \ref{prop impossible}} 
Fix any $c\in(0,0.1)$. Consider the two admissible sub-triangular arrays in Example \ref{ex 2} corresponding to $a_T=0$ and $a_T=1$, and denote their respective $T$-dimensional distributions by $P_{0T}$ and $P_{1T}$. By Proposition \ref{prop 1} P3, for every $T\ge2$,
\begin{align*}
\Omega(P_{1T})-\Omega(P_{0T})=\sigma^2\left[\frac{c(T-1)}{T}+\frac{c^2}{T}+\frac{c^2(T-1)(2T-1)}{3T^2}\right]>\frac{\sigma^2c}{2}.
\end{align*}
Set $\eta=\sigma^2c/4$. Then $\Omega_T(P_{1T})-\Omega_T(P_{0T})>2\eta$ for every $T\ge2$.

We next show that $P_{0T}$ and $P_{1T}$ cannot be asymptotically perfectly distinguished. 

Under $\P_{0T}$, the observations are i.i.d.\ $N(0,\sigma^2)$, whereas under $\P_{1T}$, Proposition \ref{prop 1} P2 gives $\sum_{t=1}^T\operatorname{Var}_{P_{1T}}(\varepsilon_{tT}(c))
=\sigma^2(T+c^2).$ Under $\P_{1T}$, we have $\varepsilon_{tT}(c)=g_t+\frac{c}{T}\sum_{j=1}^{T}g_{t-j}.$ Since $g_t$ is independent of all preceding Gaussian innovations, and $\varepsilon_{1T}(c),\ldots,\varepsilon_{t-1,T}(c)$ depend only on these preceding innovations,
\begin{align*}
&\operatorname{Var}_{P_{1T}}\left(
\varepsilon_{tT}(c)\mid
\varepsilon_{1T}(c),\ldots,\varepsilon_{t-1,T}(c)
\right)\\
&=\sigma^2+\frac{c^2}{T^2}
\operatorname{Var}_{P_{1T}}\left(
\sum_{j=1}^{T}g_{t-j}\mid
\varepsilon_{1T}(c),\ldots,\varepsilon_{t-1,T}(c)
\right)
\ge\sigma^2.
\end{align*}
According to the sequential conditional-variance decomposition
of the determinant of a Gaussian covariance matrix,
\begin{align*}
\det\operatorname{Cov}_{P_{1T}}
\bigl((\varepsilon_{tT}(c))_{t=1}^{T}\bigr)=\prod_{t=1}^{T}
\operatorname{Var}_{P_{1T}}\left(
\varepsilon_{tT}(c)\mid
\varepsilon_{1T}(c),\ldots,\varepsilon_{t-1,T}(c)
\right)
\ge\sigma^{2T},
\end{align*}
where the conditioning set is empty when $t=1$.
Consequently, the Gaussian KL-divergence yields
\begin{align*}
D_{\mathrm{KL}}(P_{1T}\Vert P_{0T})=\frac12\left[
c^2-\log\frac{\det\operatorname{Cov}_{P_{1T}}\bigl((\varepsilon_{tT}(c))_{t=1}^T\bigr)}{\sigma^{2T}}\right]\le\frac{c^2}{2}.
\end{align*}
By Pinsker's inequality,
\begin{align*}
    \|P_{1T}-P_{0T}\|_{\mathrm{TV}}\le\sqrt{\frac12D_{\mathrm{KL}}(P_{1T}\Vert P_{0T})}\le\frac c2\le\frac c{\sqrt2}<1.
\end{align*}

Now let $\widehat\Omega_T:\mathbb R^T\to\mathbb R$ be any measurable estimator. Since the two target values differ by more than $2\eta$, Le Cam's two-point inequality gives
\begin{align*}
&\max\Big\{
P_{0T}\bigl(|\widehat\Omega_T-\Omega(P_{0T})|>\eta\bigr),
P_{1T}\bigl(|\widehat\Omega_T-\Omega(P_{1T})|>\eta\bigr)
\Big\}\\
&\quad\ge\frac12\left(1-\|P_{1T}-P_{0T}\|_{\mathrm{TV}}\right)
\ge\frac12\left(1-\frac c{\sqrt2}\right)>0.
\end{align*}
This lower bound holds for every $T\ge2$ and is independent of $T$. If $\widehat\Omega_T$ satisfied UPC as defined in Definition \ref{def universal pointwise consistent}, the same estimator would have to be consistent along both admissible sub-triangular arrays. Consequently, both error probabilities in the preceding maximum would converge to zero, contradicting the strictly positive lower bound. Therefore, no estimator $\widehat\Omega_T$ can possess universal pointwise consistency on $\{\mathcal E_T(c):T\ge1,c>0\}$.

%%%%%%%%%%%%%%%%%%%%%%%%%%%%%%%%%%%%%%%%%%%%%%%%%%%

\newpage
\section{Proof of Results in Section \ref{sec 3}}
\label{app B}
\setcounter{equation}{0}
\renewcommand\theequation{B.\arabic{equation}}
\renewcommand\thetheorem{B.\arabic{theorem}}
\renewcommand\thecorollary{B.\arabic{corollary}}
\renewcommand\thelemma{B.\arabic{lemma}}
\renewcommand\theremark{B.\arabic{remark}}
\setcounter{proposition}{0}
\renewcommand{\theproposition}{C.\arabic{proposition}}

\noindent\textbf{Proof of Theorem \ref{th 1}} The proof is divided into five steps.

\noindent \textbf{Step 1} (MSE of $S_{1T}+S_{2T}$).
Recall from \eqref{eq new decomposition} that $S_{1T}=\widehat\theta_h-\E[\widehat\theta_h]$ and $ S_{2T}=B_h-\widehat B_h$.
Under Assumption \ref{as misspecification 1} D2,
\begin{align*}
    \E[S_{1T}^2] =\mathrm{Var}(\widehat\theta_h)=O\left(\frac{1}{Th}\right).
\end{align*}
Since $b=T^{-1/(2S+1)}$ and
$T^{-1/(2S+1)}/h=o(1)$, using Corollary \ref{corollary F2} gives
\[
\mathrm{Var}(\widehat B_h)=O\left(\frac{1}{Tb}\right),
\qquad
\sup_{f\in\mathcal F_S(\varepsilon_0,M_S,\rho_S)}
\left|\E[\widehat B_h]-B_h\right|
\lesssim b^S\rho_S(b).
\]
Thus, uniformly over $\mathcal F_S(\varepsilon_0,M_S,\rho_S)$,
\begin{align*}
&\E[(S_{1T}+S_{2T})^2]
\leq
2\E[S_{1T}^2]+2\E[S_{2T}^2]\\
&\lesssim
\frac{1}{Th}+\frac{1}{Tb}
+b^{2S}\rho_S^2(b)\lesssim T^{-2S/(2S+1)},
\end{align*}
where the last inequality follows from
$b=T^{-1/(2S+1)}$, $b/h=o(1)$, and $\rho_S(b)\to0$.

\noindent \textbf{Step 2}(Negligibility of the normalized displacement) Let $N_{T+}$ and $N_{T-}$ denote the cardinalities of the sets $\{x_{tT}\in (0, h]\}$ and $\{x_{tT}\in [-h, 0]\}$ respectively. Assumption \ref{as fixed-design} implies that $\textbf{Card}(N_{T\pm})\lesssim Th$. Together with the fact that $\sum_{t=1}^Tl_{th}^{\pm}(0)=1$, using the Cauchy--Schwarz inequality yields
\begin{align*}
    1=\Big(\sum_{t=1}^Tl_{th}^{\pm}(0)\Big)^2\leq N_{T\pm}\sum_{t=1}^T(l_{th}^{\pm}(0))^2\leq T(\bar c_{\psi}h+D_T)\sum_{t=1}^T(l_{th}^{\pm}(0))^2,
\end{align*}
where $D_T$ is defined in Assumption \ref{as fixed-design}. Recall that $\widehat C_{K,1}^2=Th\sum_{t=1}^T(l_{th}^+(0)-l_{th}^-(0))^2$. Then,  for sufficiently large $T$, Assumption \ref{as misspecification 1} D3 asserts
\begin{align*}
  \lim_{T\to\infty}\inf_{f\in\F} &\P\Big(  \hat C^2_{K,1}\geq \frac{2Th}{T(\bar c_{\psi}h+2D_T)}\Big)= 1.
\end{align*}
Since $ \frac{Th}{T(\bar c_{\psi}h+D_T)}=\bar c_{\psi}^{-1}+o(1)>0$, we know $\hat C_{K,1}$ is asymptotically bounded away from zero.
For every fixed $\eta>0$, Assumption \ref{as misspecification 1} D3
and Markov's inequality yield
\begin{align*}
&\sup_{f\in\mathcal F_S(\varepsilon_0,M_S,\rho_S)}
\P\left(
\frac{|S_{1T}+S_{2T}|}{s_T}>\eta
\right)\\
\leq &\sup_{f\in\F}\P\left(\Big\{
\frac{|S_{1T}+S_{2T}|}{s_T}>\eta\Big\}\bigcap\{s_T\neq 0\}
\right)+\sup_{f\in \F}\P(s_T=0)\\
\lesssim &
\sup_{f\in\mathcal F_S(\varepsilon_0,M_S,\rho_S)}
\P(\widehat\sigma\leq\sigma_0)
+\frac{m_T}{\eta^2\sigma_0^2}
\sup_{f\in\mathcal F_S(\varepsilon_0,M_S,\rho_S)}
\E[(S_{1T}+S_{2T})^2]+\sup_{f\in\F}\P(s_T=0).
\end{align*}
Assumption \ref{as misspecification 1} D3 also ensures that
$\sup_{f\in\F}\P(s_T=0)\to0$. Therefore, for every fixed $\eta>0$,
\begin{align*}
    \sup_{f\in\mathcal F_S(\varepsilon_0,M_S,\rho_S)}
\P\left(\frac{|S_{1T}+S_{2T}|}{s_T}>\eta\right)\xrightarrow[T\to\infty]{}0.
\end{align*}

\noindent\textbf{Step 3 }(Uniform Coverage)
By \eqref{eq noise},
conditional on the original data $\mathcal D_T$, $\bar G_T\mid\mathcal D_T\sim N(\widehat B_h,s_T^2).$ By \eqref{eq CI recover} and \eqref{eq:scales}, when event $\{s_T>0\}$ happens,
\begin{align*}
    c_T=\frac{r_T}{s_T}=\mathrm{cv}_{1-\alpha}\left(\frac{1}{\sqrt{l_T}}\right)\xrightarrow[T\to\infty]{} z_{1-\frac{\alpha}{2}}
\end{align*}
By \eqref{eq noise}, when $\{s_T>0\}$ happens, conditional Gaussian integration
gives
\begin{align*}
&\P\left(\theta(f)\in CI_{\mathrm{recover}}^{\mathrm{ind}}(\alpha)\mid\mathcal D_T\right)\\
&=\Phi\left(c_T-\frac{S_{1T}+S_{2T}}{s_T}\right)+\Phi\left(c_T+\frac{S_{1T}+S_{2T}}{s_T}\right)-1.
\end{align*}
Using $|\Phi(x)-\Phi(y)|\leq\varphi(0)|x-y|$,
for every fixed $\eta>0$ we obtain
\begin{align*}
&\sup_{f\in\mathcal F_S(\varepsilon_0,M_S,\rho_S)}
\left|
\P\left(
\theta(f)\in CI_{\mathrm{recover}}^{\mathrm{ind}}(\alpha)
\right)-(1-\alpha)
\right|\\
&\leq
2\varphi(0)
\left(\eta+
|c_T-z_{1-\alpha/2}|\right)+
\sup_{f\in\F}
\P(s_T=0)\\
&\quad+\sup_{f\in\mathcal F_S(\varepsilon_0,M_S,\rho_S)}\P\left(\frac{|S_{1T}+S_{2T}|}{s_T}>\eta\right)
\end{align*}
By Step 2, Assumption \ref{as misspecification 1} D3, and
$c_T\to z_{1-\alpha/2}$, first letting
$T\to\infty$ and then $\eta\downarrow0$ establishes uniformly exact coverage
\begin{align*}
    \sup_{f\in\mathcal F_S(\varepsilon_0,M_S,\rho_S)}\left|\P\left(\theta(f)\in \mathbf{CI}_{\mathrm{recover}}^{\mathrm{ind}}(\alpha)\right)-(1-\alpha)\right|\longrightarrow0.
\end{align*}

For non-triviality, on the event $\{s_T>0\}$, write
\begin{align*}
    \frac{\hat\theta_h-\bar G_T-\theta(f)}{r_T}=\frac{1}{c_T}\left(\frac{S_{1T}+S_{2T}}{s_T}-\frac{\bar G_T-\hat B_h}{s_T}\right).
\end{align*}
By Step 2, the first term in the parentheses converges to zero in probability. Conditionally on $\mathcal D_T$, the second term is exactly $N(0,1)$, and Assumption \ref{as misspecification 1} D3 gives $\P(s_T>0)\to1$. Together with $c_T\to z_{1-\alpha/2}$, the pivot at the radius scale converges weakly to $N(0,z_{1-\alpha/2}^{-2})$, which has a positive Lebesgue density. Hence $\mathbf{CI}_{\mathrm{recover}}^{\mathrm{ind}}(\alpha)$ is non-trivial in the sense of Definition \ref{def nontrivial}.

\noindent\textbf{Recovery of leading constant.} By the expansion in Remark \ref{remark 1}, $\mathrm{cv}_{1-\alpha}(1/\sqrt{l_T})\to z_{1-\alpha/2}$. Hence it suffices to show $\hat C_{K,1}/\hat C_{K,1,h_0}\to1$ and $\hat\sigma/\hat\sigma_{h_0}\xrightarrow{\P}1$. Throughout this part, $g$ denotes either $h$ or $h_0$, so that $g\to0$ and $D_T/g\to0$.

\noindent (i) \textit{The design factor.} Since the two one-sided weights have disjoint supports,
\begin{align*}
    \hat C_{K,1,g}^2=Tg\sum_{t=1}^T(l^+_{tg}(0))^2+Tg\sum_{t=1}^T(l^-_{tg}(0))^2.
\end{align*}
Consider the right side. By Theorem \ref{th concentration fixed} and Corollary \ref{corollary concentration fixed} with $[a,b]=[0,1]$, the weight $l^+_{tg}(0)$ differs from its population counterpart $L_{tg,\Psi}(0;0,1)$ by at most $CD_T/(Tg^2)$, uniformly in $t$. At most $O(Tg)$ of these weights are nonzero and each is $O(1/(Tg))$, so replacing $l^+_{tg}(0)$ by $L_{tg,\Psi}(0;0,1)$ changes $Tg\sum_t(l^+_{tg}(0))^2$ by $O(D_T/g)=o(1)$. Moreover, $Tg\sum_t L_{tg,\Psi}^2(0;0,1)=\e_0^\top M_g^{-1}\widetilde N_g M_g^{-1}\e_0$, where $\widetilde N_g$ is a Stieltjes integral against $\Psi_T$ of an integrand of bounded variation, so replacing $\Psi_T$ by $\Psi$ costs another $O(D_T/g)$. After the change of variables $z=gu$,
\begin{align*}
&Tg\sum_{t=1}^T(l^+_{tg}(0))^2=\e_0^\top M_g^{-1}N_gM_g^{-1}\e_0+o(1),\\
&M_g:=\int_0^1K(u)\r(u)\r^\top(u)\,d\Psi_g(u),\qquad N_g:=\int_0^1K^2(u)\r(u)\r^\top(u)\,d\Psi_g(u),
\end{align*}
where $\Psi_g(u):=(\Psi(gu)-\Psi(0))/g$. Since $\Psi$ is right-differentiable at $0$ with derivative $\psi_+\in[\underline c_\psi,\bar c_\psi]$, $\Psi_g(u)\to\psi_+u$ uniformly on $[0,1]$. Integrating by parts, since $K$ and $\r$ are of bounded variation on $[0,1]$, gives $M_g\to\psi_+\Gamma_1^+$ and $N_g\to\psi_+\Gamma_2^+$, where $\Gamma_k^+:=\int_0^1K^k(u)\r(u)\r^\top(u)du$ is positive definite by Lemma \ref{lemma 2}. Hence
\begin{align*}
    Tg\sum_{t=1}^T(l^+_{tg}(0))^2\longrightarrow\psi_+^{-1}\e_0^\top(\Gamma_1^+)^{-1}\Gamma_2^+(\Gamma_1^+)^{-1}\e_0,
\end{align*}
a limit that does not depend on $g$. The left side is identical, with the left derivative of $\Psi$ at $0$. Therefore, $\hat C_{K,1}$ and $\hat C_{K,1,h_0}$ have the same positive limit.

\noindent (ii) \textit{The scale estimator.} For $g\in\{h,h_0\}$, let $l_{sg}(x)$ denote $l^+_{sg}(x)$ for $x>0$ and $l^-_{sg}(x)$ for $x\leq0$, and write $Y_{tT}-\hat f_g(x_{tT})=\varepsilon_{tT}-\eta_{t,g}+r_{t,g}$, where $\eta_{t,g}:=\sum_{s}l_{sg}(x_{tT})\varepsilon_{sT}$ and $r_{t,g}:=f(x_{tT})-\sum_{s}l_{sg}(x_{tT})f(x_{sT})$. With $A_g:=\frac1T\sum_t(\eta_{t,g}-r_{t,g})^2$, the Cauchy--Schwarz inequality gives
\begin{align*}
    \Big|\hat\sigma_g^2-\frac1T\sum_{t=1}^T\varepsilon_{tT}^2\Big|\leq A_g+2\Big(\frac1T\sum_{t=1}^T\varepsilon_{tT}^2\Big)^{1/2}A_g^{1/2}.
\end{align*}
By the argument leading to \eqref{3}--\eqref{4}, applied at every evaluation point, $\max_{s,x}|l_{sg}(x)|\lesssim1/(Tg)$ and $\sup_x\sum_s|l_{sg}(x)|\leq C$. Hence Assumption \ref{as misspecification 1} gives
\begin{align*}
    \E\Big[\frac1T\sum_{t=1}^T\eta_{t,g}^2\Big]\leq\max_{s,x}|l_{sg}(x)|\,\sup_x\sum_s|l_{sg}(x)|\,\Big(E_2^2+\sup_{s}\sum_{k\neq s}|\mathrm{Cov}(\varepsilon_{sT},\varepsilon_{kT})|\Big)\lesssim\frac{1}{Tg}\to0.
\end{align*} The weights reproduce constants and are supported on same-side design points within distance $g$ of $x_{tT}$. Hence $|r_{t,g}|\leq C\omega_t(g)$, where $\omega_t(g)$ is the oscillation of $f$ over such points, and $|r_{t,g}|\leq(1+C)\cdot2\sup|f|$. Since $f_\pm$ are bounded and piecewise continuous, $\omega_t(g)$ is bounded by the moduli of continuity of $f_\pm$ on their finitely many pieces, except for design points within distance $g$ of a discontinuity, whose fraction is $O(g+D_T)$. Thus $\frac1T\sum_tr_{t,g}^2\to0$, so $A_g\xrightarrow{\P}0$. Since $\E[\frac1T\sum_t\varepsilon_{tT}^2]\leq E_2^2$, we obtain $\hat\sigma_g^2-\frac1T\sum_t\varepsilon_{tT}^2\xrightarrow{\P}0$ for both $g=h$ and $g=h_0$, and hence $\hat\sigma^2-\hat\sigma_{h_0}^2\xrightarrow{\P}0$. By Assumption \ref{as misspecification 1} D3, $\hat\sigma>\sigma_0$ with probability tending to one. Therefore, $\hat\sigma/\hat\sigma_{h_0}\xrightarrow{\P}1$, which completes the proof.\\

%%%%%%%%%%%%%%%%%%%%%%%%%%%%%%%%%%%%%%%%%%%%%%%%%%%%%%%%%

\noindent\textbf{Proof of Theorem \ref{prop DA}} By the dependence-aware Gaussian injection in \eqref{eq noise DA}, conditionally on $\mathcal D_T$,
\begin{align*}
    \overline G_T^{\mathrm{DA}}-\check B_h
\sim\
\frac{\widehat{\mathrm{sd}}_{\mathrm{DA}}}{\sqrt{m_T^{\mathrm{DA}}}}Z,
\qquad Z\sim N(0,1).
\end{align*}
Since $A_T=T^{S/(2S+1)}(\widehat\theta_h-\check B_h-\theta(f))$,
the conditional coverage probability equals
\begin{align*}
&\P\!\left(\theta(f)\in\mathrm{CI}_{\mathrm{recover}}^{\mathrm{DA}}(\alpha)
\mid\mathcal D_T\right)\\
=&\Phi\!\left(
\operatorname{cv}_{1-\alpha}\!\left((l_T^{\mathrm{DA}})^{-1/2}\right)+\frac{\sqrt{m_T^{\mathrm{DA}}}\,A_T}
{T^{S/(2S+1)}\widehat{\mathrm{sd}}_{\mathrm{DA}}}\right)
\\
&+\Phi\!\left(
\operatorname{cv}_{1-\alpha}\!\left((l_T^{\mathrm{DA}})^{-1/2}\right)
-\frac{\sqrt{m_T^{\mathrm{DA}}}\,A_T}
{T^{S/(2S+1)}\widehat{\mathrm{sd}}_{\mathrm{DA}}}\right)-1.
\end{align*}
The folded-normal quantile expansion, as $t\to0$, is
\[
\operatorname{cv}_{1-\alpha}(t)
=z_{1-\alpha/2}+\frac{z_{1-\alpha/2}}{2}t^2+O(t^4).
\]
Consequently, as $l_T^{\mathrm{DA}}\to\infty$,
\begin{align}
&2\Phi\!\left(\operatorname{cv}_{1-\alpha}
\left((l_T^{\mathrm{DA}})^{-1/2}\right)\right)-1
=1-\alpha+\frac{z_{1-\alpha/2}\varphi(z_{1-\alpha/2})}
{l_T^{\mathrm{DA}}}
+O\!\left((l_T^{\mathrm{DA}})^{-2}\right),\label{eq:prop6-a}\\
&\operatorname{cv}_{1-\alpha}\!\left((l_T^{\mathrm{DA}})^{-1/2}\right)
\varphi\!\left(\operatorname{cv}_{1-\alpha}
\left((l_T^{\mathrm{DA}})^{-1/2}\right)\right)
=z_{1-\alpha/2}\varphi(z_{1-\alpha/2})
+O\!\left((l_T^{\mathrm{DA}})^{-1}\right).
\label{eq:prop6-b}
\end{align}
For all $x,y\in\mathbb R$, Taylor's theorem and
$\sup_{x\in\mathbb R}|\Phi^{(4)}(x)|<\infty$ give the
uniform expansion
\[
\Phi(x+y)+\Phi(x-y)-1
=2\Phi(x)-1-x\varphi(x)y^2+O(y^4).
\]
Applying this expansion to the conditional coverage probability and
using \eqref{eq:prop6-a}--\eqref{eq:prop6-b} yields
\begin{align*}
&P\!\left(\theta(f)\in\mathrm{CI}_{\mathrm{recover}}^{\mathrm{DA}}(\alpha)
\mid\mathcal D_T\right)-(1-\alpha)\\
&=z_{1-\alpha/2}\varphi(z_{1-\alpha/2})
\left[\frac{1}{l_T^{\mathrm{DA}}}
-\frac{m_T^{\mathrm{DA}}}{T^{2S/(2S+1)}}
\frac{A_T^2}{\widehat{\mathrm{sd}}_{\mathrm{DA}}^2}\right]\\
&\quad+O\!\left(
\frac{1}{(l_T^{\mathrm{DA}})^2}
+\frac{m_T^{\mathrm{DA}}A_T^2}
{T^{2S/(2S+1)}l_T^{\mathrm{DA}}
 \widehat{\mathrm{sd}}_{\mathrm{DA}}^2}
+\frac{(m_T^{\mathrm{DA}})^2A_T^4}
{T^{4S/(2S+1)}\widehat{\mathrm{sd}}_{\mathrm{DA}}^4}
\right).
\end{align*}
The remainder here is bounded in absolute value by a deterministic
constant times the displayed sum. By Cauchy--Schwarz and the
fourth-moment assumption in Theorem \ref{prop DA},
\[
\sup_{T\ge1}E\!\left(
\frac{A_T^2}{\widehat{\mathrm{sd}}_{\mathrm{DA}}^2}\right)
\le\left[\sup_{T\ge1}E\!\left(
\frac{A_T^4}{\widehat{\mathrm{sd}}_{\mathrm{DA}}^4}\right)\right]^{1/2}<\infty.
\]
Taking expectations of the conditional expansion and bounding the expected absolute remainder therefore gives
\begin{align*}
&\P\!\left(\theta(f)\in
\mathrm{CI}_{\mathrm{recover}}^{\mathrm{DA}}(\alpha)\right)\\
&=(1-\alpha)+\frac{z_{1-\alpha/2}\varphi(z_{1-\alpha/2})}
{l_T^{\mathrm{DA}}}
\left[1-
\frac{m_T^{\mathrm{DA}}l_T^{\mathrm{DA}}}{T^{2S/(2S+1)}}
E\!\left(\frac{A_T^2}{\widehat{\mathrm{sd}}_{\mathrm{DA}}^2}\right)
\right]\\
&\quad+O\!\left(
\frac{1}{(l_T^{\mathrm{DA}})^2}
+\frac{m_T^{\mathrm{DA}}}
{T^{2S/(2S+1)}l_T^{\mathrm{DA}}}
+\frac{(m_T^{\mathrm{DA}})^2}{T^{4S/(2S+1)}}
\right),
\end{align*}
which is precisely the claimed decomposition into
$\mathrm{CE}_{\mathrm{lead}}$ and $\mathrm{CE}_{\mathrm{remainder}}$.

\noindent\textbf{Proof of Corollaries \ref{corollary DA} and \ref{th DA}} These two corollaries are immediate consequences of Theorem \ref{prop DA}. We thus omit their proof here.

\newpage

\section{Proof of Results in Section \ref{sec 4}}
\label{app C}
\setcounter{equation}{0}
\renewcommand\theequation{C.\arabic{equation}}
\renewcommand\thetheorem{C.\arabic{theorem}}
\renewcommand\thecorollary{C.\arabic{corollary}}
\renewcommand\thelemma{C.\arabic{lemma}}
\renewcommand\theremark{C.\arabic{remark}}
\setcounter{proposition}{0}
\renewcommand{\theproposition}{C.\arabic{proposition}}
Appendix \ref{app C} demonstrates the proof of Theorem \ref{th 2}. We further decompose the proof into two parts, analysis of overcoverage and undercoverage error. During the proof, $c_X,C_X>0$ are used to denote some fixed constants independent of $n,b,h$.

\subsection{Proof of Theorem \ref{th 2} (Overcoverage)}
\label{app C.1}
\textbf{Step 1} (Non-degeneracy of $\tilde\sigma\tilde C_{K,1}$)
By model \eqref{model random-design}, $\E[\varepsilon_i^2\mid X_i=u]=\sigma^2>0.$ For every $u\in(0,\varepsilon_0]$, Assumptions \ref{as random design} and \ref{as moment} M1 imply
\[
\E[|\varepsilon_i|^4\mid X_i=u]
\le
\left(\frac{B_1}{c_X}\right)^{4/\varsigma}.
\]
Moreover, using the Cauchy--Schwarz inequality yields
\begin{align*}
\frac{3\sigma^2}{4}
\le\E\left[\varepsilon_i^2\mathbf 1\{|\varepsilon_i|>\sigma/2\}| X_i=u
\right]\le
\left(\E[|\varepsilon_i|^4\mid X_i=u]\right)^{1/2}\mathbb P\left(|\varepsilon_i|>\sigma/2\mid X_i=u\right)^{1/2}.
\end{align*}
Consequently,
\[
\E[\1[|\varepsilon_i|>\sigma/2]|X_i=u]=\mathbb P\left(|\varepsilon_i|>\sigma/2\mid X_i=u\right)\ge\frac{9\sigma^4}{16(B_1/c_X)^{4/\varsigma}}.
\]
Integrating with respect to $X_i$ gives
\[
\E\left[\mathbf 1\{0<X_i\le\varepsilon_0,\,|\varepsilon_i|>\sigma/2\}\right]\ge\frac{9\sigma^4c_X\varepsilon_0}{16(B_1/c_X)^{4/\varsigma}}>0.
\]

Consider the bounded indicators appearing above. By Assumption \ref{as misspecificaiton 2} D5, their $\beta$-mixing
coefficients are bounded by those of $\{(X_i,\varepsilon_i)\}_{i\ge1}$. Since $\sum_{k\ge1}\beta_k<\infty$,
the covariance inequality for bounded $\beta$-mixing variables gives
\begin{align*}
\mathrm{Var}\left(
\frac1n\sum_{i=1}^n\1[0<X_i\le\varepsilon_0,\,|\varepsilon_i|>\sigma/2]\right)\lesssim
\frac1{n^2}\left(n+\sum_{k=1}^{n-1}(n-k)\beta_k\right)=O(n^{-1}).
\end{align*}
Therefore, Chebyshev's inequality yields
\begin{align}
\label{eq proof overcoverage eq1}
\mathbb P\Bigg(\frac1n\sum_{i=1}^n\1\{0<X_i\le\varepsilon_0,\,|\varepsilon_i|>\sigma/2\}<\frac{9\sigma^4c_X\varepsilon_0}{32(B_1/c_X)^{4/\varsigma}}\Bigg)=O(n^{-1}).
\end{align}

Next, by defining $\triangle_+=\sup_{x\in (0,\varepsilon_0]}|\tilde f_{+,b}(x)-f(x)|$, on event $\{\triangle_+\le\sigma/4\}$, whenever $0<X_i\le\varepsilon_0$ and $|\varepsilon_i|>\sigma/2$, we have $|Y_i-\tilde f_b(X_i)|=|\varepsilon_i+f_+(X_i)
-\tilde f_{+,b}(X_i)|\ge
|\varepsilon_i|-\triangle_+>\frac{\sigma}{4},$
which, together with the definition of
$\tilde\sigma^2$ in \eqref{eq variance proxy random}, asserts that, when $\{\triangle_+\le\sigma/4\}$ happens,
\[
\tilde\sigma^2
\ge
\frac{\sigma^2}{16n}
\sum_{i=1}^n
\mathbf 1\{0<X_i\le\varepsilon_0,\,
|\varepsilon_i|>\sigma/2\}.
\]
Meanwhile, since Theorem \ref{th F3} implies $\P(\triangle_+>\sigma/4)=o\left(n^{-2S/(2S+1)}\right)$,
together with \eqref{eq proof overcoverage eq1}, uniformly
over $\mathcal F(\varepsilon_0,M_S,\rho_S)$, we have
\begin{align*}
\mathbb P(\tilde\sigma<c_{\sigma})\le\P(\triangle_+>\sigma/4)+O(n^{-1})=
o\left(n^{-2S/(2S+1)}\right),\qquad c_{\sigma}^2:=\frac{\sigma^2}{16}\cdot\frac{9\sigma^4c_X\varepsilon_0}{32(B_1/c_X)^{4/\varsigma}}>0.
\end{align*}
In particular, $\mathbb P(\tilde\sigma=0)=o\left(n^{-2S/(2S+1)}\right)$.

\noindent\textbf{Step 2}(Non-degeneracy of
$\tilde C_{K,1}$).
Step 1 of the proof of Theorem \ref{th F3} gives $\mathbb P(D_n>b^2)
=o\left(n^{-2S/(2S+1)}\right).$
On event $\{D_n\le b^2\}$, since $\frac{D_n}{h}\le\frac{b^2}{h}=o(1)$,
 Theorem \ref{th concentration random} and Corollary \ref{corollary concentration random} ensure that
the local-linear design matrices in \eqref{eq local linear}
are invertible for all sufficiently large
$n$. Consequently, the local-linear
weights are well-defined and reproduce
constants, i.e.,
\begin{align*}
    \sum_{i=1}^n L_{ih,+}(0)=\sum_{i=1}^n L_{ih,-}(0)=1.
\end{align*}
Since the positive- and negative-side weights have disjoint supports, 
\begin{align*}
\tilde C_{K,1}^{\,2}=nh\sum_{i=1}^n
\bigl(L_{ih,+}(0)-L_{ih,-}(0)\bigr)^2=nh\left[\sum_{i=1}^nL_{ih,+}^2(0)+\sum_{i=1}^nL_{ih,-}^2(0)\right].
\end{align*}
Moreover, $L_{ih,+}(0)\neq0$ only if $0<X_i\le h$, and on $\{D_n\le b^2\}$ the number of such observations satisfies $N_{n,+}\le n(\Psi(h)-\Psi(0)+2D_n)\le n(C_Xh+2b^2)$. By $\sum_{i=1}^nL_{ih,+}(0)=1$ and the Cauchy--Schwarz inequality, $1\le N_{n,+}\sum_{i=1}^nL_{ih,+}^2(0)$. Hence, on $\{D_n\le b^2\}$,
\begin{align*}
    \tilde C_{K,1}^{\,2}\geq \frac{nh}{N_{n,+}}\geq \frac{1}{C_X+2b^2/h}\geq\frac{1}{2C_X}
\end{align*}
for all sufficiently large $n$, because $b^2/h=o(1)$. Therefore,
\begin{align*}
    \mathbb P\big(\tilde C_{K,1}^{\,2}<(2C_X)^{-1}\big)\le\mathbb \P(D_n>b^2)=o\left(n^{-2S/(2S+1)}\right).
\end{align*}

\noindent\textbf{Step 3}(Conditional Gaussian coverage bound)
By the Gaussian generation scheme \eqref{eq noise random},
conditional on the original observations
$\{(X_i,Y_i)\}_{i=1}^n=:\mathcal{D}_n$, when event $\{\tilde\sigma\tilde C_{K,1}>0\}$ happens, we have $\frac{\sqrt{m_n}
(\overline G_n-\tilde B_h)}
{\tilde\sigma\tilde C_{K,1}}\Big|\mathcal{D}_n
\sim N(0,1)$. Consequently, by \eqref{eq CI recover random},
\begin{align*}
&\mathbb P\left(
\theta(f)\in CI_{\mathrm{recover}}^{\mathrm{ind}}(\alpha,\tau)
\mid\{(X_i,Y_i)\}_{i=1}^n
\right)\\
&=\Phi\left(\frac{\sqrt{m_n}(\tilde\theta_h-\tilde B_h-\theta(f))}{\tilde\sigma\tilde C_{K,1}}+\operatorname{cv}_{1-\alpha}(n^{-\tau/2})\right)\\
&\quad-
\Phi\left(\frac{\sqrt{m_n}(\tilde\theta_h-\tilde B_h-\theta(f))}{\tilde\sigma\tilde C_{K,1}}-\operatorname{cv}_{1-\alpha}(n^{-\tau/2})
\right).
\end{align*}
By the symmetry and unimodality of the
standard Gaussian distribution, the
probability of an interval with a fixed
radius is maximized when its center is
zero. Therefore,
\begin{align*}
&\mathbb P\left(\theta(f)\in CI_{\mathrm{recover}}^{\mathrm{ind}}(\alpha,\tau)|
\mathcal{D}_n\right)\leq2\Phi\left(\operatorname{cv}_{1-\alpha}(n^{-\tau/2})\right)-1.
\end{align*}
Thus, taking expectations yields
\begin{align*}
&\mathbb P\left(\theta(f)\in CI_{\mathrm{recover}}^{\mathrm{ind}}(\alpha,\tau)\right)\le2\Phi\left(\operatorname{cv}_{1-\alpha}(n^{-\tau/2})\right)-1+\mathbb P(\tilde\sigma\tilde C_{K,1}=0).
\end{align*}
Finally, we finish the proof of overcoverage by using the Lipschitz property of $\Phi$  and Steps 1 and 2.

\subsection{Proof of Theorem \ref{th 2} (Undercoverage)}
\label{app C.2}
\textbf{Step 1}(Concentration of the original stochastic
component) By \eqref{eq local linear} and \eqref{eq bias correction}, and the definition of $B_h$, we have the exact decomposition
\begin{align*}
\tilde\theta_h-\tilde B_h-\theta(f)=
\sum_{i=1}^n
\bigl(L_{ih,+}(0)-L_{ih,-}(0)\bigr)\varepsilon_i+(B_h-\tilde B_h).
\end{align*}
Recall from Step 1 of the proof of Theorem \ref{th F3} that $\P(D_n>b^2)=o\left(n^{-2S/(2S+1)}\right).$ On $\{D_n\le b^2\}$, since $D_n/h=o(1)$, Theorem \ref{th concentration random} and Corollary \ref{corollary concentration random} ensure that
the local-linear design matrices in \eqref{eq local linear random}
are uniformly invertible for all sufficiently large $n$. Consequently,
\begin{align*}
&\left|\sum_{i=1}^n\bigl(L_{ih,+}(0)-L_{ih,-}(0)\bigr)\varepsilon_i\right|\\
&\quad\lesssim\max_{0\le j\le1}\left|\frac1{nh}\sum_{i=1}^nK(X_i/h)\mathbf1\{X_i>0\}(X_i/h)^j\varepsilon_i\right|\\
&\qquad+
\max_{0\le j\le1}\left|\frac1{nh}\sum_{i=1}^nK(X_i/h)\mathbf1\{X_i\le0\}(X_i/h)^j\varepsilon_i\right|.
\end{align*}

The concentration argument of Lemma \ref{lemma 5} applies to these local-linear scores by restricting the polynomial indices to
$j=0,1$ and using bandwidth $h$ in place
of the pilot bandwidth $b$. Indeed, $b=o(h)$, and the corresponding
local-score variance, truncation, and mixing bounds used in that argument remain valid at bandwidth $h$. In particular, with the same $L_n$ and $r$, the range condition in Step 4 of that proof holds a fortiori: for $t_h:=(\log n)^{\eta}/\sqrt{nh}$, $rL_n\cdot nh\,t_h\lesssim n^{\lambda+\rho}(\log n)^{\eta}\sqrt{nh}=o(nh)$, because $n^{\lambda+\rho}(\log n)^{\eta}=o(\sqrt{nb})$ and $b=o(h)$.
Thus, for any fixed $\eta\in(0.5,1)$,
\begin{align*}
&\mathbb P\Bigg(\max_{0\le j\le1}
\left|\frac1{nh}\sum_{i=1}^nK(X_i/h)\mathbf1\{X_i>0\}(X_i/h)^j\varepsilon_i\right|>\frac{(\log n)^\eta}{\sqrt{nh}}\Bigg)=o\left(n^{-2S/(2S+1)}\right).
\end{align*}
The same bound holds for the negative-side scores. Since $(\log n)^\eta\le\log n$ for sufficiently large $n$, it follows
that, with probability $1-o(n^{-2S/(2S+1)})$,
\begin{align*}
    \left|\sum_{i=1}^n\bigl(L_{ih,+}(0)-L_{ih,-}(0)\bigr)\varepsilon_i\right|\lesssim\frac{\log n}{\sqrt{nh}}.
\end{align*}
Moreover, Corollary \ref{corollary F4} gives $\mathbb P\left(|\tilde B_h-B_h|>\frac{\log n}{n^{S/(2S+1)}}\right)=o\left(n^{-2S/(2S+1)}\right).$ Combining these results yields
\begin{align}
\label{eq proof undercoverage eq1}
\inf_{f\in\F}\P\Bigg(|\tilde\theta_h-\tilde B_h-\theta(f)|\lesssim\frac{\log n}{n^{S/(2S+1)}}\Bigg)\geq 1-o(n^{-2S/(2S+1)}).
\end{align}

\noindent\textbf{Step 2}(Control of the normalized
displacement)
We directly use the non-degeneracy established in Steps 1 and 2 of Appendix \ref{app C.1}. In particular, with $c_0:=c_{\sigma}/\sqrt{2C_X}>0$,
\begin{align}
\label{eq proof undercoverage eq2}
    \sup_{f\in\F}\P(\tilde \sigma\tilde C_{K,1}< c_0)=o(n^{-\frac{2S}{2S+1}})
\end{align}
Recall that $m_n=[n^{\frac{2S}{2S+1}-2\tau}]+1$. Combining \eqref{eq proof undercoverage eq1} and \eqref{eq proof undercoverage eq2} yields
\begin{align}
\label{eq proof undercoverage eq3}
   &\sup_{f\in\F}\P(\mathcal{E}_n(f)):= \sup_{f\in\F}\P\Bigg(\frac{
\sqrt{m_n}|\tilde\theta_h-\tilde B_h-\theta(f)|}{\tilde\sigma\tilde C_{K,1}}> n^{-\tau/2}\Bigg) \notag\\
\leq &\sup_{f\in\F}\P\Bigg(\frac{
\sqrt{m_n}|\tilde\theta_h-\tilde B_h-\theta(f)|}{c_0}> n^{-\tau/2}\Bigg)+ \sup_{f\in\F}\P(\tilde \sigma\tilde C_{K,1}< c_0)\notag \\
&\qquad = o(n^{-\frac{2S}{2S+1}})
\end{align}

\noindent\textbf{Step 3}(Conditional Gaussian
coverage lower bound)
By (31), conditional on $\{(X_i,Y_i)\}_{i=1}^n=\mathcal{D}_n$, whenever $\{\tilde\sigma\tilde C_{K,1}>c_0\}$, $\frac{\sqrt{m_n}(\overline G_n-\tilde B_h)}{\tilde\sigma\tilde C_{K,1}}\sim N(0,1).$
Meanwhile, \eqref{eq CI recover random} yields that, for all $f\in\F$,
\begin{align*}
&\P\left(\theta(f)\in CI_{\mathrm{recover}}^{\mathrm{ind}}(\alpha,\tau)\right)\\
\geq &\P\left(\theta(f)\in CI_{\mathrm{recover}}^{\mathrm{ind}}(\alpha,\tau)\bigcap\{\tilde \sigma\tilde C_{K,1}\geq c_0\}\right)\\
=&\E\left[\1[\tilde \sigma\tilde C_{K,1}\geq c_0]\P(\theta(f)\in CI_{\mathrm{recover}}^{\mathrm{ind}}(\alpha,\tau)|\mathcal{D}_n)\right]\\
\geq &\E\left[\1[\tilde \sigma\tilde C_{K,1}\geq c_0]\E[\1[\theta(f)\in CI_{\mathrm{recover}}^{\mathrm{ind}}(\alpha,\tau)]\1[\mathcal{E}^c_n(f)]|\mathcal{D}_n]\right]\\
=&\E\left[\1[\tilde \sigma\tilde C_{K,1}\geq c_0]\1[\mathcal{E}^c_n(f)]\E[\1[\theta(f)\in CI_{\mathrm{recover}}^{\mathrm{ind}}(\alpha,\tau)]|\mathcal{D}_n]\right],
\end{align*}
where the last equality is due to the fact that $\1[\mathcal{E}^c_n(f)]$ is measurable with respect to the sigma algebra generated by $\mathcal{D}_n$ and $\mathcal{E}^c_n(f)$ is the complement of $\mathcal{E}_n(f)$.

Note that, for every $f\in \F$, 
\begin{align*}
   &\1[\mathcal{E}^c_n(f)] \E[\1[\theta(f)\in CI_{\mathrm{recover}}^{\mathrm{ind}}(\alpha,\tau)]|\mathcal{D}_n]\\
=&\1[\mathcal{E}^c_n(f)]\P\left(\left|Z+\frac{\sqrt{m_n}|\tilde\theta_h-\tilde B_h-\theta(f)|}{\tilde\sigma\tilde C_{K,1}}\right|\leq \mathrm{cv}_{1-\alpha}(n^{-\tau/2})\right)\geq (1-\alpha)\1[\mathcal{E}^c_n(f)],\ \ \text{a.s.}
\end{align*}
where $Z\sim N(0,1)$. Therefore, we have 
\begin{align*}
    &\inf_{f\in\F}\P\left(\theta(f)\in CI_{\mathrm{recover}}^{\mathrm{ind}}(\alpha,\tau)\right)\\
    \geq &(1-\alpha)\Big(1-\sup_{f\in \F}\P(\mathcal{E}_n(f))-\sup_{f\in\F}\P(\tilde\sigma\tilde C_{K,1}<c_0)\Big)\\
    \geq & 1-\alpha-o(n^{-\frac{2S}{2S+1}}).
\end{align*}
This completes the proof of the undercoverage statement in Theorem \ref{th 2}.

\noindent\textbf{Non-triviality} Let $s_n=\tilde\sigma\tilde C_{K,1}/\sqrt{m_n}$ and $c_n=\mathrm{cv}_{1-\alpha}(n^{-\tau/2})$, so that the radius of $\mathit{CI}_{\mathrm{recover}}^{\mathrm{ind}}(\alpha,\tau)$ equals $c_ns_n$. On the event $\{\tilde\sigma\tilde C_{K,1}\geq c_0\}$,
\begin{align*}
    \frac{\tilde\theta_h-\bar G_n-\theta(f)}{c_ns_n}=\frac{1}{c_n}\left(\frac{\tilde\theta_h-\tilde B_h-\theta(f)}{s_n}-\frac{\bar G_n-\tilde B_h}{s_n}\right).
\end{align*}
By \eqref{eq proof undercoverage eq3}, the first term in the parentheses converges to zero in probability. By Step 3, the second term is exactly $N(0,1)$ conditionally on $\mathcal D_n$, and \eqref{eq proof undercoverage eq2} shows that the event above has probability tending to one. Since $c_n\to z_{1-\alpha/2}$, the pivot at the radius scale converges weakly to $N(0,z_{1-\alpha/2}^{-2})$, which has a positive Lebesgue density. Hence $\mathit{CI}_{\mathrm{recover}}^{\mathrm{ind}}(\alpha,\tau)$ is non-trivial in the sense of Definition \ref{def nontrivial}.

%%%%%%%%%%%%%%%%%%%%%%%%%%%%%%%%%%%%%%%%%%%%%%%%%%%%
\newpage
\section{Proof of Results in Section \ref{sec 5}}

\label{app D}
\setcounter{equation}{0}
\renewcommand\theequation{D.\arabic{equation}}
\renewcommand\thetheorem{D.\arabic{theorem}}
\renewcommand\thecorollary{D.\arabic{corollary}}
\renewcommand\thelemma{D.\arabic{lemma}}
\renewcommand\theremark{D.\arabic{remark}}
\setcounter{proposition}{0}
\renewcommand{\theproposition}{D.\arabic{proposition}}
%%%%%%%%%%%%%%%%%%%%%%%%%%%%%%%%%%%%%%%%%%%%%%%%%%%%%%%%%%%%%%%%%%%%%%
\noindent\textbf{Proof of Theorem \ref{thm:aorg-finite}}
Fix $\alpha\in(0,1/2)$, $T\ge1$, $f\in\mathcal{F}_S(\varepsilon_0,M_S,\rho_S)$ and an integer $K\ge1$. Throughout, $\hat\theta_h$, $\check B_h$ and $\widehat{\mathrm{sd}}_{\mathrm{DA}}$ are functions of $\mathcal D_T$, whereas $\theta(f)$, $l_T^{\mathrm{DA}}$, $m_T^{\mathrm{DA}}$ and $c_T$ are deterministic. By \eqref{eq:scales},
\begin{align}
\label{eq aorg proof scales}
s_T^{\mathrm{DA}}=\frac{\widehat{\mathrm{sd}}_{\mathrm{DA}}}{\sqrt{m_T^{\mathrm{DA}}}},\qquad
d_T(f)=\frac{\hat\theta_h-\check B_h-\theta(f)}{s_T^{\mathrm{DA}}},\qquad
r_T^{\mathrm{DA}}=s_T^{\mathrm{DA}}\,c_T .
\end{align}
Since $\widehat{\mathrm{sd}}_{\mathrm{DA}}>0$ $\mathbb{P}$-a.s., we work on this event of probability one, on which $s_T^{\mathrm{DA}}>0$ and $d_T(f)$ is well defined; conditional probabilities given $\mathcal D_T$ are understood $\mathbb{P}$-a.s. Steps 1--3 establish three auxiliary facts. Properties (i)--(iv) are then proved one by one, in the order in which they are stated.

\noindent\textbf{Step 1} (Properties of $\mathrm{cv}_{1-\alpha}$)
Fix $t\ge0$. On $[0,\infty)$, the distribution function of $|N(t,1)|$ is $\Phi(\cdot-t)-\Phi(-\cdot-t)=\Phi(\cdot-t)+\Phi(\cdot+t)-1$. It is continuous, has the positive derivative $\phi(\cdot-t)+\phi(\cdot+t)$, equals $0$ at $0$ and tends to $1$ at infinity. Hence $\mathrm{cv}_{1-\alpha}(t)$, the $(1-\alpha)$-quantile of $|N(t,1)|$, is the unique solution of
\begin{align}
\label{eq aorg proof cv identity}
\Phi\big(\mathrm{cv}_{1-\alpha}(t)-t\big)+\Phi\big(\mathrm{cv}_{1-\alpha}(t)+t\big)=2-\alpha ,
\end{align}
and a point of $[0,\infty)$ lies strictly below $\mathrm{cv}_{1-\alpha}(t)$ if and only if $\Phi(\cdot-t)+\Phi(\cdot+t)$ is strictly smaller than $2-\alpha$ at that point. This gives the following five properties.

\textit{(a)} For $t=0$, \eqref{eq aorg proof cv identity} reads $2\Phi(\mathrm{cv}_{1-\alpha}(0))=2-\alpha$. Hence $\mathrm{cv}_{1-\alpha}(0)=z_{1-\alpha/2}$.

\textit{(b)} At the point $t$, $\Phi(0)+\Phi(2t)<1/2+1<2-\alpha$, because $\alpha<1/2$; hence $t<\mathrm{cv}_{1-\alpha}(t)$. At the point $t+z_{1-\alpha/2}$, $\Phi(z_{1-\alpha/2})+\Phi(2t+z_{1-\alpha/2})\ge2\Phi(z_{1-\alpha/2})=2-\alpha$; hence $\mathrm{cv}_{1-\alpha}(t)\le t+z_{1-\alpha/2}$. Therefore,
\begin{align}
\label{eq aorg proof cv bounds}
t<\mathrm{cv}_{1-\alpha}(t)\le t+z_{1-\alpha/2},\qquad t\ge0 .
\end{align}

\textit{(c)} The function $\Phi(\cdot-t)+\Phi(\cdot+t)$ is continuously differentiable jointly in its argument and in $t$, and its derivative in its argument is positive. By the implicit function theorem, $\mathrm{cv}_{1-\alpha}$ is continuously differentiable on $[0,\infty)$, and differentiating \eqref{eq aorg proof cv identity} in $t$ gives
\[
\mathrm{cv}'_{1-\alpha}(t)=\frac{\phi\big(\mathrm{cv}_{1-\alpha}(t)-t\big)-\phi\big(\mathrm{cv}_{1-\alpha}(t)+t\big)}{\phi\big(\mathrm{cv}_{1-\alpha}(t)-t\big)+\phi\big(\mathrm{cv}_{1-\alpha}(t)+t\big)} .
\]
For $t>0$, \eqref{eq aorg proof cv bounds} gives $0<\mathrm{cv}_{1-\alpha}(t)-t<\mathrm{cv}_{1-\alpha}(t)+t$. Since $\phi$ is strictly decreasing on $[0,\infty)$, the numerator is positive. Hence $\mathrm{cv}'_{1-\alpha}(t)>0$ for $t>0$, and $\mathrm{cv}_{1-\alpha}$ is continuous and strictly increasing on $[0,\infty)$.

\textit{(d)} By (a), (c) and the lower bound in \eqref{eq aorg proof cv bounds}, which makes $\mathrm{cv}_{1-\alpha}$ unbounded, the intermediate value theorem shows that $\mathrm{cv}_{1-\alpha}$ maps $[0,\infty)$ onto $[z_{1-\alpha/2},\infty)$. Hence the inverse $\mathrm{cv}^{-1}_{1-\alpha}$ in Theorem \ref{thm:aorg-finite}(ii) is well defined on $[z_{1-\alpha/2},\infty)$, takes values in $[0,\infty)$, is strictly increasing, and satisfies $\mathrm{cv}^{-1}_{1-\alpha}(z_{1-\alpha/2})=0$.

\textit{(e)} Since $1/\sqrt{l_T^{\mathrm{DA}}}>0$, \eqref{eq:scales}, (a) and (c) give
\begin{align}
\label{eq aorg proof cT}
c_T=\mathrm{cv}_{1-\alpha}\big(1/\sqrt{l_T^{\mathrm{DA}}}\big)>\mathrm{cv}_{1-\alpha}(0)=z_{1-\alpha/2}>0 .
\end{align}

\noindent\textbf{Step 2} (An auxiliary function)
For $x\in\mathbb{R}$, and with $K$ temporarily regarded as a real number in $[1,\infty)$, consider
\begin{align}
\label{eq aorg proof g}
\Phi\big(\sqrt{K}\,[c_T-x]\big)+\Phi\big(\sqrt{K}\,[c_T+x]\big)-1 .
\end{align}

\textit{(a)} (Shape in $x$) The function \eqref{eq aorg proof g} is even in $x$. For $x>0$, its derivative with respect to $x$ is
\begin{align}
\label{eq aorg proof dx}
\sqrt{K}\,\Big\{\phi\big(\sqrt{K}\,[c_T+x]\big)-\phi\big(\sqrt{K}\,[c_T-x]\big)\Big\}.
\end{align}
By \eqref{eq aorg proof cT}, $c_T>0$, so $|c_T-x|<c_T+x$. Since $\phi$ is even and strictly decreasing on $[0,\infty)$, \eqref{eq aorg proof dx} is negative. Hence \eqref{eq aorg proof g} depends on $x$ only through $|x|$, is strictly decreasing in $|x|$, and attains its maximum $2\Phi(\sqrt{K}\,c_T)-1$ at $x=0$.

\textit{(b)} (Monotonicity in $K$) For fixed $x$, the derivative of \eqref{eq aorg proof g} with respect to $K\in[1,\infty)$ is
\begin{align}
\label{eq aorg proof dK}
\frac{1}{2\sqrt{K}}\Big\{[c_T-x]\,\phi\big(\sqrt{K}\,[c_T-x]\big)+[c_T+x]\,\phi\big(\sqrt{K}\,[c_T+x]\big)\Big\}.
\end{align}
If $|x|<c_T$, then $c_T-x>0$ and $c_T+x>0$, so \eqref{eq aorg proof dK} is positive. Hence, for every fixed $x$ with $|x|<c_T$, \eqref{eq aorg proof g} is strictly increasing in $K\in[1,\infty)$.

\textit{(c)} (The level-$(1-\alpha)$ point) By \eqref{eq aorg proof cT}, $\sqrt{K}\,c_T\ge c_T>z_{1-\alpha/2}$. By Step 1(d), $\mathrm{cv}^{-1}_{1-\alpha}(\sqrt{K}\,c_T)$ is therefore well defined and positive. Evaluating \eqref{eq aorg proof cv identity} at $t=\mathrm{cv}^{-1}_{1-\alpha}(\sqrt{K}\,c_T)$, for which $\mathrm{cv}_{1-\alpha}(t)=\sqrt{K}\,c_T$, gives
\begin{align}
\label{eq aorg proof level}
\Phi\Big(\sqrt{K}\,\big[c_T-K^{-1/2}\mathrm{cv}^{-1}_{1-\alpha}(\sqrt{K}\,c_T)\big]\Big)
+\Phi\Big(\sqrt{K}\,\big[c_T+K^{-1/2}\mathrm{cv}^{-1}_{1-\alpha}(\sqrt{K}\,c_T)\big]\Big)-1=1-\alpha .
\end{align}
Thus \eqref{eq aorg proof g} equals $1-\alpha$ when $|x|=K^{-1/2}\mathrm{cv}^{-1}_{1-\alpha}(\sqrt{K}\,c_T)$. Combined with (a), \eqref{eq aorg proof g} is at least $1-\alpha$ if and only if $|x|\le K^{-1/2}\mathrm{cv}^{-1}_{1-\alpha}(\sqrt{K}\,c_T)$, and it is strictly larger than $1-\alpha$ if and only if $|x|<K^{-1/2}\mathrm{cv}^{-1}_{1-\alpha}(\sqrt{K}\,c_T)$.

\noindent\textbf{Step 3} (Conditional law of the center and conditional coverage)
By \eqref{eq noise DA} and the construction of the $K$ independent auxiliary samples preceding \eqref{eq CI AoRG}, conditionally on $\mathcal D_T$, the variables $G^{\mathrm{DA}}_{iT,(k)}$, $i=1,\ldots,m_T^{\mathrm{DA}}$, $k=1,\ldots,K$, are independent with common law $N\big(\check B_h,(\widehat{\mathrm{sd}}_{\mathrm{DA}})^2\big)$. Hence, by the definition of $\bar G^{\mathrm{DA}}_{T,(k)}$ below \eqref{eq CI AoRG}, $\bar G^{\mathrm{DA}}_{T,(1)},\ldots,\bar G^{\mathrm{DA}}_{T,(K)}$ are conditionally i.i.d.\ $N\big(\check B_h,(\widehat{\mathrm{sd}}_{\mathrm{DA}})^2/m_T^{\mathrm{DA}}\big)$, and their average is conditionally $N\big(\check B_h,(\widehat{\mathrm{sd}}_{\mathrm{DA}})^2/(Km_T^{\mathrm{DA}})\big)$. Since $\hat\theta_h$ is a function of $\mathcal D_T$, \eqref{eq:scales} and \eqref{eq aorg proof scales} give
\begin{align}
\label{eq aorg proof cond}
C^{\mathrm{DA}}_{T,K}\mid\mathcal D_T\;\sim\;N\Big(\hat\theta_h-\check B_h,\;\frac{(\widehat{\mathrm{sd}}_{\mathrm{DA}})^2}{Km_T^{\mathrm{DA}}}\Big)
=N\Big(\hat\theta_h-\check B_h,\;\frac{(s_T^{\mathrm{DA}})^2}{K}\Big).
\end{align}
Since $\theta(f)$ is deterministic, \eqref{eq aorg proof cond} and \eqref{eq aorg proof scales} give, conditionally on $\mathcal D_T$,
\[
C^{\mathrm{DA}}_{T,K}-\theta(f)\;\sim\;N\Big(s_T^{\mathrm{DA}}d_T(f),\;\frac{(s_T^{\mathrm{DA}})^2}{K}\Big).
\]
By \eqref{eq:scales}, $\theta(f)\in\mathbf{CI}^{\mathrm{DA}}_{\mathrm{AoRG}}(\alpha,K)$ if and only if $|C^{\mathrm{DA}}_{T,K}-\theta(f)|\le r_T^{\mathrm{DA}}$, and $r_T^{\mathrm{DA}}=s_T^{\mathrm{DA}}c_T$ by \eqref{eq aorg proof scales}. Standardizing by $s_T^{\mathrm{DA}}/\sqrt{K}>0$ yields
\begin{align*}
\mathbb{P}\big(\theta(f)\in\mathbf{CI}^{\mathrm{DA}}_{\mathrm{AoRG}}(\alpha,K)\mid\mathcal D_T\big)
&=\mathbb{P}\big(-s_T^{\mathrm{DA}}c_T\le C^{\mathrm{DA}}_{T,K}-\theta(f)\le s_T^{\mathrm{DA}}c_T\mid\mathcal D_T\big)\\
&=\Phi\big(\sqrt{K}\,[c_T-d_T(f)]\big)-\Phi\big(-\sqrt{K}\,[c_T+d_T(f)]\big),
\end{align*}
and the symmetry of $\Phi$ gives
\begin{align}
\label{eq aorg proof coverage}
\mathbb{P}\big(\theta(f)\in\mathbf{CI}^{\mathrm{DA}}_{\mathrm{AoRG}}(\alpha,K)\mid\mathcal D_T\big)
=\Phi\big(\sqrt{K}\,[c_T-d_T(f)]\big)+\Phi\big(\sqrt{K}\,[c_T+d_T(f)]\big)-1 .
\end{align}
That is, the conditional coverage is \eqref{eq aorg proof g} evaluated at $x=d_T(f)$.

\noindent\textbf{Proof of (i)}
The conditional mean and variance in \eqref{eq aorg proof cond} are exactly those in \eqref{eq:aorg-stability}. This proves (i).

\noindent\textbf{Proof of (ii)}
By the tower property and \eqref{eq aorg proof coverage},
\[
\mathbb{P}\big(\theta(f)\in\mathbf{CI}^{\mathrm{DA}}_{\mathrm{AoRG}}(\alpha,K)\big)
=\mathbb{E}\Big[\Phi\big(\sqrt{K}\,[c_T-d_T(f)]\big)+\Phi\big(\sqrt{K}\,[c_T+d_T(f)]\big)-1\Big].
\]
The integrand is \eqref{eq aorg proof g} at $x=d_T(f)$. By Step 2(c), it is at least $1-\alpha$ on the event $\{|d_T(f)|\le K^{-1/2}\mathrm{cv}^{-1}_{1-\alpha}(\sqrt{K}\,c_T)\}$. Off this event, it is a conditional probability and hence non-negative. Therefore,
\[
\mathbb{P}\big(\theta(f)\in\mathbf{CI}^{\mathrm{DA}}_{\mathrm{AoRG}}(\alpha,K)\big)
\ge(1-\alpha)\,\mathbb{P}\Big(|d_T(f)|\le K^{-1/2}\mathrm{cv}^{-1}_{1-\alpha}\big(\sqrt{K}\,c_T\big)\Big),
\]
which is \eqref{eq:aorg-protection}, with $\mathrm{cv}^{-1}_{1-\alpha}$ given by Step 1(d). This proves (ii).

\noindent\textbf{Proof of (iii)}
We prove the four assertions of (iii) in turn.

\textit{The case $K=1$.} By \eqref{eq:scales} and Step 1(d), $\mathrm{cv}^{-1}_{1-\alpha}(c_T)=\mathrm{cv}^{-1}_{1-\alpha}\big(\mathrm{cv}_{1-\alpha}(1/\sqrt{l_T^{\mathrm{DA}}})\big)=1/\sqrt{l_T^{\mathrm{DA}}}$.

\textit{The two bounds.} Apply \eqref{eq aorg proof cv bounds} at $t=\mathrm{cv}^{-1}_{1-\alpha}(\sqrt{K}\,c_T)\ge0$, for which $\mathrm{cv}_{1-\alpha}(t)=\sqrt{K}\,c_T$:
\[
\mathrm{cv}^{-1}_{1-\alpha}(\sqrt{K}\,c_T)\;<\;\sqrt{K}\,c_T\;\le\;\mathrm{cv}^{-1}_{1-\alpha}(\sqrt{K}\,c_T)+z_{1-\alpha/2}.
\]
Dividing by $\sqrt{K}$ gives $c_T-z_{1-\alpha/2}K^{-1/2}\le K^{-1/2}\mathrm{cv}^{-1}_{1-\alpha}(\sqrt{K}\,c_T)<c_T$.

\textit{Strict increase in $K$.} Set $x=K^{-1/2}\mathrm{cv}^{-1}_{1-\alpha}(\sqrt{K}\,c_T)$. By Step 2(c) and the two bounds, $0<x<c_T$. By \eqref{eq aorg proof level}, \eqref{eq aorg proof g} equals $1-\alpha$ at this $x$. Since $|x|<c_T$, Step 2(b) shows that \eqref{eq aorg proof g} with $K+1$ in place of $K$ is strictly larger than $1-\alpha$ at the same $x$. By Step 2(c) applied with $K+1$ in place of $K$, this happens if and only if $x<(K+1)^{-1/2}\mathrm{cv}^{-1}_{1-\alpha}(\sqrt{K+1}\,c_T)$. Hence
\[
(K+1)^{-1/2}\,\mathrm{cv}^{-1}_{1-\alpha}\big(\sqrt{K+1}\,c_T\big)\;>\;K^{-1/2}\,\mathrm{cv}^{-1}_{1-\alpha}\big(\sqrt{K}\,c_T\big).
\]

\textit{Monotonicity of the lower bound.} By the strict increase just shown, the event $\{|d_T(f)|\le K^{-1/2}\mathrm{cv}^{-1}_{1-\alpha}(\sqrt{K}\,c_T)\}$ is contained in the corresponding event with $K+1$ in place of $K$. Hence the lower bound in \eqref{eq:aorg-protection} is non-decreasing in $K$. This proves (iii).

\noindent\textbf{Proof of (iv)}
By \eqref{eq CI recover fixed DA} and \eqref{eq:scales}, the radius of $\mathbf{CI}^{\mathrm{DA}}_{\mathrm{recover}}(\alpha)$ is $\widehat{\mathrm{sd}}_{\mathrm{DA}}\,\mathrm{cv}_{1-\alpha}(1/\sqrt{l_T^{\mathrm{DA}}})(m_T^{\mathrm{DA}})^{-1/2}=r_T^{\mathrm{DA}}$. By \eqref{eq:scales}, the radius of $\mathbf{CI}^{\mathrm{DA}}_{\mathrm{AoRG}}(\alpha,K)$ is also $r_T^{\mathrm{DA}}$, which does not involve $K$. On the event $\{\widehat{\mathrm{sd}}_{\mathrm{DA}}>0\}$, \eqref{eq aorg proof scales} and \eqref{eq aorg proof cT} give $r_T^{\mathrm{DA}}=s_T^{\mathrm{DA}}c_T>0$, so the ratio in \eqref{eq:aorg-radius} is well defined and equals $1$. Since this event has probability one, \eqref{eq:aorg-radius} holds $\mathbb{P}$-a.s. This proves (iv), and we thus finish the proof.\\

%%%%%%%%%%%%%%%%%%%%%%%%%%%%%%%%%%%%%%%%%%%%%%%%%%%%%%%%%%%%%%%%%%%%%%%%%%%%%%%%%%%%%%%%%%%%%%%%%%%%%%%%%%%%%%%%%%%%%%%%%
%%%%%%%%%%%%%%%%%%%%%%%%%%%%%%%%%%%%%%%%%%%%%%%%%%%%%%%%%%%%%%%%%%%%%%
% Proof of the proposition in Section 5.3 (label prop K infinity).

%%%%%%%%%%%%%%%%%%%%%%%%%%%%%%%%%%%%%%%%%%%%%%%%%%%%%%%%%%%%%%%%%%%%%%

\noindent\textbf{Proof of Proposition \ref{prop K infinity}}
Fix $\alpha\in(0,1)$ and $f\in\mathcal{F}_S(\varepsilon_0,M_S,\rho_S)$. Under the conditions of Theorem \ref{prop DA}, $\widehat{\mathrm{sd}}_{\mathrm{DA}}>0$ $\mathbb{P}$-a.s., so $s_T^{\mathrm{DA}}>0$ $\mathbb{P}$-a.s.

\noindent\textbf{Step 1} ($c_T\ge z_{1-\alpha/2}$ for every $T$)
For $t\ge0$, we have $|z_{1-\alpha/2}-t|\le z_{1-\alpha/2}+t$, so $\phi(z_{1-\alpha/2}+t)\le\phi(z_{1-\alpha/2}-t)$. Hence the function $\Phi(z_{1-\alpha/2}-\cdot)+\Phi(z_{1-\alpha/2}+\cdot)$ has the non-positive derivative $\phi(z_{1-\alpha/2}+\cdot)-\phi(z_{1-\alpha/2}-\cdot)$ on $[0,\infty)$, and
\[
\Phi(z_{1-\alpha/2}-t)+\Phi(z_{1-\alpha/2}+t)\le2\Phi(z_{1-\alpha/2})=2-\alpha,\qquad t\ge0 .
\]
As in Step 1 of the proof of Theorem \ref{thm:aorg-finite}, the distribution function $\Phi(\cdot-t)+\Phi(\cdot+t)-1$ of $|N(t,1)|$ is strictly increasing on $[0,\infty)$; this part of that step does not use $\alpha<1/2$. By the display, its value at $z_{1-\alpha/2}$ is at most $1-\alpha$, so $z_{1-\alpha/2}$ does not exceed its $(1-\alpha)$-quantile, that is, $\mathrm{cv}_{1-\alpha}(t)\ge z_{1-\alpha/2}$ for every $t\ge0$. By \eqref{eq:scales}, $c_T=\mathrm{cv}_{1-\alpha}(1/\sqrt{l_T^{\mathrm{DA}}})\ge z_{1-\alpha/2}>0$.

\noindent\textbf{Step 2} (A moment bound)
By \eqref{eq:scales} and the definition of $A_T$ in Theorem \ref{prop DA},
\[
d_T(f)=\frac{\sqrt{m_T^{\mathrm{DA}}}\,\big(\hat\theta_h-\check B_h-\theta(f)\big)}{\widehat{\mathrm{sd}}_{\mathrm{DA}}}
=\Big(\frac{m_T^{\mathrm{DA}}}{T^{2S/(2S+1)}}\Big)^{1/2}\frac{A_T}{\widehat{\mathrm{sd}}_{\mathrm{DA}}} .
\]
Hence, by the conditions $m_T^{\mathrm{DA}}=o(T^{2S/(2S+1)})$ and $\sup_{T\ge1}\sup_{f}\mathbb{E}(A_T/\widehat{\mathrm{sd}}_{\mathrm{DA}})^4<\infty$ in Theorem \ref{prop DA},
\begin{align}
\label{eq det moment}
\mathbb{E}\big[d_T(f)^4\big]
&=\Big(\frac{m_T^{\mathrm{DA}}}{T^{2S/(2S+1)}}\Big)^2\mathbb{E}\big[(A_T/\widehat{\mathrm{sd}}_{\mathrm{DA}})^4\big]\notag\\
&\le\Big(\frac{m_T^{\mathrm{DA}}}{T^{2S/(2S+1)}}\Big)^2\sup_{T\ge1}\sup_{f\in\mathcal{F}_S(\varepsilon_0,M_S,\rho_S)}\mathbb{E}\big(A_T/\widehat{\mathrm{sd}}_{\mathrm{DA}}\big)^4
\xrightarrow[T\to\infty]{}0 .
\end{align}

\noindent\textbf{Step 3} (Degeneracy of the pivot)
By \eqref{eq:scales}, the pivot of \eqref{eq det interval} at the radius scale is $(\hat\theta_h-\check B_h-\theta(f))/r^{\mathrm{DA}}_T=d_T(f)/c_T$. For every $\eta>0$, Step 1 and Markov's inequality give
\[
\mathbb{P}\big(|d_T(f)/c_T|>\eta\big)
\le\mathbb{P}\big(|d_T(f)|>\eta\,z_{1-\alpha/2}\big)
\le\frac{\mathbb{E}\big[d_T(f)^4\big]}{\eta^4z_{1-\alpha/2}^4},
\]
which tends to zero by \eqref{eq det moment}. Hence $d_T(f)/c_T\to0$ in probability, and \eqref{eq det interval} is trivial in the sense of Definition \ref{def nontrivial}. Finally, $\theta(f)\in[\hat\theta_h-\check B_h\pm r^{\mathrm{DA}}_T]$ if and only if $|d_T(f)/c_T|\le1$, so taking $\eta=1$ gives $\mathbb{P}\big(\theta(f)\notin[\hat\theta_h-\check B_h\pm r^{\mathrm{DA}}_T]\big)\to0$. This finishes the proof.\\

\newpage

\section{Uniform Coupling of Weight Function with Truncated Inputs}
\label{app E}

\def\theequation{E.\arabic{equation}}
\setcounter{equation}{0}
\def\thelemma{E.\arabic{lemma}}
\setcounter{lemma}{0}
\def\thetheorem{E.\arabic{theorem}}
\setcounter{theorem}{0}

%%%%%%%%%%%%%%%%%%%%%%%%%%%%%%%%%%%%%%%%%%%%%%%%%%%%%%%%%%%%%%%%%%%%%%%%%%%%%%%%%%%%%%%%%%%%%%%%%%%%%%%%%%%%
\subsection{Fixed Design}
\label{app concentration: fixed}
 Let $K(u)$ be a Lipschitz continuous density with compact support $[-1,1]$ and Lipschitz constant $L_K>0$. We assume $\sup_{u\in [-1,1]}K(u)=: \bar K$ and $\textbf{Leb}(\{K(u)>0\})>0$, where $\textbf{Leb}(A)$ is the Lebesgue measure of set $A\subset \mathbb{R}$. For any given integer $q\geq 1$, interval $[a,b]\subset [-C_{\psi},C_{\psi}]$ and point $x\in [a,b]$, define
\begin{align}
    u_{th,x}=\frac{x_{tT}-x}{h},\ &\r_q(u)=(1,u,...,u^q)\in \R^{q+1},\ \e_0=(1,0,...,0)\in \R^{q+1},\\
    M_{1h}(x;a,b)&=\frac{1}{Th} \sum_{t=1}^nK(u_{th,x})\1[x_{tT}\in [a,b]] \r_q(u_{th,x})\r_q(u_{th,x})^\top,\notag\\
     M_{1h,\Psi}(x;a,b)&=\frac{1}{h} \int_{-1}^1K(\frac{z-x}{h})\1[z\in [a,b]] \r_q(\frac{z-x}{h})\r_q(\frac{z-x}{h})^\top d\Psi(z),\notag\\
     w_{th}(x;a,b)&=\frac{1}{nh}\e_0^{\top} M_{1h}^{-1}(x;a,b)K(u_{th,x})\1[x_{tT}\in [a,b]]\r_q(u_{th,x}),\notag\\
     L_{th,\Psi}(x;a,b)&=\frac{1}{nh}\e_0^{\top} M_{1h,\Psi}^{-1}(x;a,b)K(u_{th,x})\1[x_{tT}\in [a,b]]\r_q(u_{th,x}),\notag
\end{align}
where $\Psi$ is introduced in Assumption \ref{as fixed-design}. 

\begin{theorem}
    \label{th concentration fixed}
    Suppose $\{x_{tT}\}_{t=1}^T$ satisfies Assumption \ref{as fixed-design} and $h>0$ satisfies $T^{-\frac{1}{2S+1}}/h=o(1)$, for some integer $S\geq 1$. There exists a fixed constant $C>0$ independent of $T,h$ such that 
    \begin{align*}
        \sup_{x\in [a,b]}\max_{1\leq t\leq T}|w_{th}(x;a,b)-L_{th,\Psi}(x;a,b)|\leq \frac{CD_T}{Th^2}
    \end{align*}
    holds for all $T\geq 1, h>0$ such that $ C_{q,K}D_T/h\leq c_{\psi}c_{1,q,\mu}/2$, where $c_{\psi}$ and $c_{1,q,\mu}$ are introduced in Assumption \ref{as fixed-design} and Lemma \ref{lemma 2}, respectively.
\end{theorem}

\begin{corollary}
    \label{corollary concentration fixed}
    Theorem \ref{th concentration fixed} still holds if we replace the indicator function $\1[x_{tT}\in [a,b]]$ in $w_{th}(x;a,b)$ and $L_{th,\Psi}(x;a,b)$ with $\1[x_{tT}\in (a,b]]$.
\end{corollary}

\noindent\textbf{Proof of Theorem \ref{th concentration fixed}}

\noindent\textbf{Step 1} For any given $[a,b]\subset [-C_{\psi},C_{\psi}]$ and $x\in [a,b]$, for the $(k,l)$-th entry of matrix $M_{1h}(x;a,b)-M_{1h,\Psi}(x;a,b)$, denoted as $M(k,l)$, by letting $k+l=m$, we have
\begin{align*}
    M(k,l)=\int_{-C_{\psi}}^{C_{\psi}}\frac{1}{h}K(\frac{z-x}{h})\1[z\in [a,b]](\frac{z-x}{h})^md(\Psi_T(z)-\Psi(z)).
\end{align*}
Recall that, for an arbitrary given bounded variation function $g$, together with Assumption \ref{as fixed-design}, doing integration by parts with respect to the Stieltjes integral yields
\begin{align}
    \label{eq proof of th D2 eq1}
    \Big|\int_{-1}^1g(z)d(\Psi_T(z)-\Psi(z))\Big|\leq D_T(\text{TV}_{[-1,1]}(g)+2||g||_{\infty}),
\end{align}
where $\text{TV}_{[-1,1]}(g)$ denotes the total variation of $g$ on interval $[-1,1]$. Meanwhile, recall that the mapping $u\to u^mK(u)$ has finite total variation on $[-1,1]$ and rescaling $u=\frac{z-x}{h}$ only adds one additional coefficient $1/h$. Together with the fact that $1[z\in [a,b]]$ only brings finitely many jumps with jump values no larger than $\bar K/h$, we obtain 
\begin{align*}
    \text{TV}_{[-1,1]}&\Big\{\frac{1}{h}K(\frac{z-x}{h})\1[z\in[a,b]](\frac{z-x}{h})^m\Big\}\leq \frac{1}{h}\Big(\frac{2L_K}{m+1}+4\bar K\Big),\\
     &\Big\|\frac{1}{h}K(\frac{z-x}{h})\1[z\in[a,b]](\frac{z-x}{h})^m\Big\|_{\infty}\leq \frac{1}{h}\bar K,
\end{align*}
which asserts $|M(k,l)|\leq (\frac{2L_K}{m+1}+4\bar K)D_T/h$ holds uniformly over $x\in [a,b]$. Then, since the dimension of the matrix $M_{1h}(x;a,b)-M_{1h,\Psi}(x;a,b)$ is $q+1$, we have
\begin{align}
    \label{eq proof of th D2 eq2}
    \sup_{x\in [a,b]}||M_{1h}(x;a,b)-M_{1h,\Psi}(x;a,b)||_{\text{op}}\leq C_{q,K}D_T/h=o(1),
\end{align}
where $C_{q,K}>0$ is a constant depending only on the choice of $q, K$ and the second equality is based on Assumption \ref{as fixed-design} and $T^{-\frac{1}{2S+1}}/h=o(1)$.

\noindent\textbf{Step 2} Note that Assumption \ref{as fixed-design} implies the following measure domination,
\begin{align*}
    d\Psi(u)\geq c_{\psi}d\mu,
\end{align*}
where $\mu$ is the Lebesgue measure. This further asserts that, for arbitrary non-zero $\nu\in \R^{q+1}$,
\begin{align*}
    &\nu^\top M_{1h,\Psi}(x;a,b)\nu=\frac{1}{h}\int_{a}^b K(\frac{z-x}{h})\Big[\nu^\top r\Big(\frac{z-x}{h}\Big)\Big]d\Psi(z)\\
    &\geq  \frac{c_{\psi}}{h}\int_{a}^b K(\frac{z-x}{h})\Big[\nu^\top r\Big(\frac{z-x}{h}\Big)\Big]dz=c_{\psi}\int_{I_{x,h}}K(u)(\nu^\top \r_{q}(u))^2du.
\end{align*}
Together with Lemma \ref{lemma 2}, we thus obtain 
\begin{align}
     \label{eq proof of th D2 eq3}
     \inf_{x\in [a,b]}\lambda_{\min}(M_{1h,\Psi}(x;a,b))\geq c_{\psi}\lambda_{\min}(\Gamma_{1,q,\mu}(x;a,b))\geq c_{\psi}c_{1,q,\mu}>0.
\end{align}
Thus, for every  $T\geq 1,h>0$ such that $ C_{q,K}D_T/h<\frac{c_{\psi}c_{1,q,\mu}}{2}$, using Weyl's inequality yields 
\begin{align*}
    \inf_{x\in [a,b]}\lambda_{\min}(M_{1h}(x;a,b))&\geq\inf_{x\in [a,b]} \lambda_{\min}(M_{1h,\Psi}(x;a,b))- \sup_{x\in [a,b]}||M_{1h}(x;a,b)-M_{1h,\Psi}(x;a,b)||_{\text{op}}\geq \frac{c_{\psi}c_{1,q,\mu}}{2},\\
    &\sup_{x\in [a,b]}||M_{1h}^{-1}(x;a,b)||_{\text{op}}\leq \frac{2}{c_{\psi}c_{1,q,\mu}},\qquad \sup_{x\in [a,b]}||M_{1h,\Psi}^{-1}(x;a,b)||_{\text{op}}\leq \frac{1}{c_{\psi}c_{1,q,\mu}}.
\end{align*}
Recall that the inverse matrix perturbation identity indicates
\begin{align*}
  M_{1h}^{-1}(x;a,b) - M_{1h,\Psi}^{-1}(x;a,b)=M_{1h}^{-1}(x;a,b)(M_{1h}(x;a,b)-M_{1h,\Psi}(x;a,b))M_{1h,\Psi}^{-1}(x;a,b).
\end{align*}
Then, together with the definitions of $w_{th}(x;a,b)$ and $L_{th,\Psi}(x;a,b)$, we immediately obtain
\begin{align*}
   \sup_{x\in [a,b]}\max_{1\leq t\leq T}|w_{th}(x;a,b)-L_{th,\Psi}(x;a,b)|\leq \frac{\bar K\sqrt{q+1}C_{q,K}}{(c_{\psi}c_{1,q,\mu})^2}\frac{D_T}{Th^2}=:\frac{CD_T}{Th^2}    
\end{align*}

\noindent\textbf{Proof of Corollary \ref{corollary concentration fixed}} The proof is nearly the same as that of Theorem \ref{th concentration fixed}. We thus omit it.

\subsection{Random Design}
\label{app concentration: random}
We now extend Theorem \ref{th concentration fixed} to random-design inputs. We retain the notation introduced in Theorem \ref{th concentration fixed}, except that $\{X_i\}_{i=1}^n$ is now a random sample taking values in $[-1,1]$. Define
\[
\Psi_n(x)
=
\frac{1}{n}\sum_{i=1}^n\mathbf{1}\{X_i\le x\},
\qquad
D_n
=
\sup_{x\in[-1,1]}
|\Psi_n(x)-\Psi(x)|.
\]
Suppose $\Psi$ is a continuous distribution function on $[-1,1]$ that
admits a Lebesgue density $\psi$ satisfying $\inf_{x\in[-1,1]}\psi(x)=c_{\psi}>0$. For any given integer $q\ge 1$, interval $[a,b]\subset[-1,1]$, and point $x\in[a,b]$, we introduce the following notation
\begin{align*}
    U_{ih,x}&=\frac{X_i-x}{h},\ \r_q(u)=(1,u,\ldots,u^q)^{\top}\in \R^{q+1},\ \e_0=(1,0,\ldots,0)^{\top}\in \R^{q+1},\\
    &\M_{1h}(x;a,b)=\frac{1}{nh}\sum_{i=1}^nK(U_{ih,x})\mathbf{1}\{X_i\in[a,b]\}r_q(U_{ih,x})r_q(U_{ih,x})^{\top},\\
    \M_{1h,\Psi}&(x;a,b)=\frac{1}{h}\int_{-1}^{1}K\left(\frac{z-x}{h}\right)\mathbf{1}\{z\in[a,b]\}r_q\left(\frac{z-x}{h}\right)r_q\left(\frac{z-x}{h}\right)^{\top}d\Psi(z),\\
    &W_{ih}(x;a,b)=\frac{1}{nh}\e_0^{\top}\M_{1h}^{-1}(x;a,b)K(U_{ih,x})\mathbf{1}\{X_i\in[a,b]\}r_q(U_{ih,x}),\\
    &L_{ih,\Psi}(x;a,b)=\frac{1}{nh}\e_0^{\top}\M_{1h,\Psi}^{-1}(x;a,b)K(U_{ih,x})\mathbf{1}\{X_i\in[a,b]\}r_q(U_{ih,x}).
\end{align*}
\begin{theorem}
\label{th concentration random}
Suppose $K$ satisfies the conditions imposed in Section \ref{app concentration: fixed}, and suppose
$0<h\le (b-a)/2$ satisfies $\frac{n^{-1/(2S+1)}}{h}=o(1)$
for some integer $S\ge 1$. Suppose further that $$P\left(D_n\le d_n\right)\geq 1-O(\xi_n)$$ holds for some $d_n,\xi_n>0$ such that $d_n\lor \xi_n\to 0$.
Then there exists a fixed constant $C>0$, independent of $n$ and $h$,
such that
\begin{align*}
    \P\left(\sup_{x\in[a,b]}\max_{1\le i\le n}\left|W_{ih}(x;a,b)-L_{ih,\Psi}(x;a,b)\right|\leq C\frac{d_n}{nh^2}\right)
\geq 1-O(\xi_n).
\end{align*}
\end{theorem}
\begin{corollary}
    \label{corollary concentration random}
    Theorem \ref{th concentration random} still holds if we replace the indicator function $\1[X_i\in [a,b]]$ in $W_{ih}(x;a,b)$ and $L_{ih,\Psi}(x;a,b)$ with $\1[X_i\in (a,b]]$.
\end{corollary}

\noindent \textbf{Proof of Theorem \ref{th concentration random}} Conditional on any realization $\{X_i\}_{i=1}^n$, the running variables form a fixed design. Moreover, since $\psi(x)\ge c_\psi>0$,
\[\frac{\Psi(y)-\Psi(x)}{y-x}=
\frac{1}{y-x}\int_x^y\psi(u)\,du\ge c_\psi,\qquad -1\le x<y\le1.\]
Hence the population placement condition in Assumption \ref{as fixed-design} is satisfied.

On the event $\left\{D_n\le d_n\right\},$
Theorem \ref{th concentration fixed} applied conditionally on the realized design gives
\[
\sup_{x\in[a,b]} \max_{1\le i\le n}|W_{ih}(x;a,b)-L_{ih,\Psi}(x;a,b)|\le
C\frac{D_n}{nh^2}\le C\frac{d_n}{nh^2},
\]
for all sufficiently large $n$. Therefore,
\begin{align*}
    \P\left(\sup_{x\in[a,b]}\max_{1\le i\le n}|W_{ih}(x;a,b)-L_{ih,\Psi}(x;a,b)|\le C\frac{d_n}{nh^2}\right)\ge \P\left(D_n\le d_n\right)\geq 1-O(\xi_n).
\end{align*}

\noindent\textbf{Proof of Corollary \ref{corollary concentration random}} The proof is nearly the same as that of Theorem \ref{th concentration random}. We thus omit it.

%%%%%%%%%%%%%%%%%%%%%%%%%%%%%%%%%%%%%%%%%%%%%%%%%%%%%%%%%%%%%%%%%%%%%%%%%%%%%%%%%%%%%%%%%%%%%%%%%%%%%%%%%%%%%%%%%%%%%%%%%%%%%%%%%%%%%%%%%%%%%%%%%%%%%%%%%%%%%%%%%%%%%%%%%%%%%%%%%%%%%%%%%%%%%%%%%%%%%%%%%%%%%%%%%%%%%%%%%%%%%%%%%%%%%%%%%%%%%%%%%%%%%%%%%%%%%%%%%%%%%%%%%%%%%%%%%%%%%%%%%%%%%%%%%%%%%%%%%%%%%%%%%%%%%%%%%%%%%%%%%%%%%%%%%%%%%%%%%%%%%%%%%%%%%%%%%%%%%%%%%%%%%%%%%%%%%%%%%%%%%%%%%%%%%%%%%%%%%%%%%%%%%%%%%%%%%%%%%%%%%%%%%%%%%%%%%%%%%%%%%%%%%%%%%%%%%%%%%%%
\section{Asymptotic Properties of Bias Correction}
\label{app F}
\def\theequation{F.\arabic{equation}}
\setcounter{equation}{0}
\def\thelemma{F.\arabic{lemma}}
\setcounter{lemma}{0}
\def\thetheorem{F.\arabic{theorem}}
\setcounter{theorem}{0}
\renewcommand\thecorollary{F.\arabic{corollary}}
\setcounter{corollary}{0}
\subsection{Fixed-Design}
\label{app F Fixed}
This section is dedicated to showing some asymptotic properties for the bias estimator $\hat B_h$ used in the validity recovery procedure for fixed-design RDD inference. For convenience and self-completeness, we still review the notation system first and investigate the asymptotic properties step by step. 

By letting $u_{th,x}=\frac{x_{tT}-x}{h}$, $\r_S=(1,u,...,u^S)\in \R^{S+1}$ and $l_{th}^{\pm}(0)$ defined in \eqref{eq local linear} of Section \ref{sec 3.1}, define 
\begin{align*}
    &\qquad \qquad \qquad \hat B_h=\hat B_{h,+}-\hat B_{h,-},\\
\hat B_{h,+}=&\sum_{t=1}^Tl_{th}^+(0)(\hat f_{+,b}(x_{tT})-\hat f_{-,b}(0)),\ \ \hat f_{+,b}(x)=\sum_{j=1}^Tw_{jb}(x;0,1)Y_{jT},\\
\hat B_{h,-}=&\sum_{t=1}^Tl_{th}^-(0)(\hat f_{-,b}(x_{tT})-\hat f_{-,b}(0)),\ \ \hat f_{-,b}(x)=\sum_{j=1}^Tw_{jb}(x;-1,0)Y_{jT},\\
w_{jb}(x)&= w_{th}(x;-1,0)=\frac{1}{nh}\e_0^{\top} M_{1h}^{-1}(x;a,b)K(u_{th,x})\1[x_{tT}\in (0,1]]\r_S(u_{th,x}),\\
    \Gamma_{1,S,+}:=\int_{0}^1K(u)&\r_S(u)\r^\top_S(u)du,\ \Gamma_{2,S,+}=\int_{0}^1K^2(u)\r_S(u)\r^\top_S(u)du,\ v_{S,+}:=\e_0^\top \Gamma_{1,S,+}^{-1}\Gamma_{2,S,+}\Gamma_{1,S,+}^{-1}\e_0,\\
     \Gamma_{1,S,-}:=\int_{-1}^0K(u)&\r_S(u)\r^\top_S(u)du,\ \Gamma_{2,S,-}=\int_{0}^1K^2(u)\r_S(u)\r^\top_S(u)du,\ v_{S,-}:=\e_0^\top \Gamma_{1,S,-}^{-1}\Gamma_{2,S,-}\Gamma_{1,S,-}^{-1}\e_0,\\
     q_{jT}:=q_{jT}^+-q_{jT}^-:=&\sum_{j=1}^T l_{th}^+(0)(w_{jb}(x_{tT};0,1)-w_{jb}(0;0,1))-\sum_{j=1}^T l_{th}^-(0)(w_{jb}(x_{tT};-1,0)-w_{jb}(0;-1,0))
\end{align*}
We first introduce the following fundamental assumption for bandwidths $h,b$.
\begin{assumption}
    \label{as bandwidth}
    For the bandwidths of baseline and pilot estimators,  $h$ and $b$, we assume $0<h<\varepsilon_0$, $b\to 0$, $Tb\to\infty$ and $b/h\to 0$.
\end{assumption}
\begin{lemma}
    \label{lemma F1}
    Under Assumption \ref{as bandwidth}, for all sufficiently large $T$, there exists a constant $C>0$ independent of $T,h,b$ such that 
    \begin{align*}
        \max_{t}|l^+_{th}(0)|\leq \frac{C}{Th},\ \  \sup_{x\in [0,h]}\max_{j\leq T}|w_{jb}(x;0,1)|\leq \frac{C}{Tb}.
    \end{align*}
    Moreover, $l^+_{th}(0)\neq 0$ implies $0<x_{tT}\leq h$ and $w_{jb}(x_{tT};0,1)\neq 0$ implies $|x_{jT}-x_{tT}|\leq b$.
\end{lemma}
The proof of Lemma \ref{lemma F1} is a direct corollary of Theorem \ref{th concentration fixed} and the definition of $l_{th}^{\pm}$; we thus omit its proof here.

According to the notations above and model \eqref{model fixed-design}, exchanging two finite sums gives
\begin{align*}
    &\hat B_{h,+}-\E[\hat B_{h,+}]=\sum_{j=1}^Tq_{jT}^+\varepsilon_{jT},\ \text{where}\ \ q_{jT}^+:= A_{jT}^+-V_{jT}^+:=\sum_{t=1}^T l_{th}^+(0)w_{jb}(x_{tT};0,1)-w_{jb}(0;0,1).
\end{align*}
\begin{theorem}
    \label{th F1}
    Based on the $q_{jT}^+$ introduced above, under Assumption \ref{as bandwidth}, 
    \begin{align*}
        \sum_{j=1}^T(q_{jT}^+)^2=\frac{v_{S,+}}{Tb}(1+o(1)).
    \end{align*}
    More precisely, 
    \begin{align*}
        \sum_{j=1}^T(V_{jT}^+)^2=\frac{v_{S,+}}{Tb}(1+o(1)),\ \sum_{j=1}^T(A_{jT}^+)^2=O(\frac{1}{Th}),\ \sum_{j=1}^TA^+_{jT}V^+_{jT}=O(\frac{1}{T\sqrt{hb}}).
    \end{align*}
\end{theorem}
 Theorem \ref{th F1} immediately implies the following corollary.
\begin{corollary}
    \label{corollary F1}
    Based on the assumptions and notations used in Theorem \ref{th F1}, we have 
    \begin{align*}
        \sum_{j=1}^Tq_{jT}^2=\frac{v_{S,+}+v_{S,-}}{Tb}(1+o(1)).
    \end{align*}
\end{corollary}

\begin{corollary}
    \label{corollary F2}
    Suppose that $\{\varepsilon_{tT}\}$ satisfies $\E[\varepsilon_{tT}]=0$, $\E[\varepsilon_{tT}^2]=\sigma^2$ and D2 of Assumption \ref{as misspecification 1}. Based on the notations and assumptions used in Theorem \ref{th F1}, we have 
    \begin{align*}
        \mathrm{Var}(\hat B_h)=O(\frac{1}{Tb}),\  \sup_{f\in \mathcal{F}(\varepsilon_0,M_S,\rho_S)}|\E[\hat B_h]-B_h|\leq CM_Sb^S\rho_S(b).
    \end{align*}
\end{corollary}

\noindent\textbf{Proof of Theorem \ref{th F1}} A basic decomposition is 
\begin{align*}
   \sum_{j=1}^T (q^+_{jT})^2=\sum_{j=1}^T(V_{jT}^+)^2+\sum_{j=1}^T(A_{jT}^+)^2-2\sum_{j=1}^TA_{jT}^+V_{jT}^+.
\end{align*}
\noindent\textbf{Step 1}($\sum_{j=1}^T(V_{jT}^+)^2$) At point $x=0$, let 
\begin{align*}
    M_{1,b,+}:&=\frac{1}{Tb}\sum_{j=1}^TK(\frac{x_{jT}}{b})\1[x_{jT}\in [0,1]]\r_S(\frac{x_{jT}}{b})\r_S^\top(\frac{x_{jT}}{b}),\\
        M_{2,b,+}:&=\frac{1}{Tb}\sum_{j=1}^TK^2(\frac{x_{jT}}{b})\1[x_{jT}\in [0,1]]\r_S(\frac{x_{jT}}{b})\r_S^\top(\frac{x_{jT}}{b}).
\end{align*}
Together with the fact that $M_{k,b,+}$ is symmetric, $k=1,2$, some simple algebra yields
\begin{align}
    \label{eq proof of th F1 eq1}
    \sum_{j=1}^T(V^+_{jT})^2=\frac{1}{Tb}\e_0^\top M_{1,b,+}^{-1}M_{2,b,+}M_{1,b,+}^{-1}\e_0.
\end{align}
By repeating the Stieltjes-sum approximation used in the proof of Theorem \ref{th concentration fixed}, we know, for sufficiently large $T$, the inverse $M_{1,b,+}^-$ exists, and we can also obtain $M_{1,b,+}\to \Gamma_{1,S,+}$ and $M_{2,b,+}\to \Gamma_{2,S,+}$. 
Therefore, using \eqref{eq proof of th F1 eq1} asserts $\sum_{j=1}^T(V_{jT}^+)^2=\frac{v_{S,+}}{Tb}(1+o(1)).$\\

\noindent\textbf{Step 2}($\sum_{j=1}^T(A_{jT}^+)^2$) A key observation is that $A_{jT}^+=\sum_{t=1}^Tl_{th}^+(0)w_{jb}(x_{tT};0,1)$ can be nonzero only if simultaneously $0<x_{tT}\leq h$ and $|x_{jT}-x_{tT}|\leq b$. According to Assumption \ref{as fixed-design}, since the $\Psi$ function is bi-Lipschitz, for each fixed $j$, there are at most $O(2Tb)$-many indices $t$ involved. Lemma \ref{lemma F1} thus gives 
\begin{align*}
    |A_{jT}^+|\lesssim Tb \frac{1}{Th}\frac{1}{Tb}=O(\frac{1}{Th}).
\end{align*}
Furthermore, $A_{jT}^+\neq 0$ requires $x_{jT}\in [0,h+b]$. The number of non-zero $A_{jT}^+$'s is of order $O(Th)$. Consequently, 
\begin{align*}
    \sum_{j=1}^T(A_{jT}^+)^2\leq O(Th)O(\frac{1}{T^2h^2})=O(\frac{1}{Th}).
\end{align*}
\noindent\textbf{Step 3} By the Cauchy--Schwarz inequality and the results in Steps 1 and 2, we immediately have
\begin{align*}
    \Big|\sum_{j=1}^TA_{jT}^+V_{jT}^+\Big|\leq \sqrt{\sum_{j=1}^T(A_{jT}^+)^2}\sqrt{\sum_{j=1}^T(V_{jT}^+)^2}=O(\frac{1}{T\sqrt{hb}})=O(\frac{1}{Tb}),
\end{align*}
where the last equality is due to the assumption $b/h\to 0$. We finish the proof of Theorem \ref{th F1}.\\

\noindent\textbf{Proof of Corollary \ref{corollary F1}} The proof of the corollary is a direct consequence of Theorem \ref{th F1}. We thus only sketch it here. First, note that, by repeating the proof of Theorem \ref{th F1}, we can show
\begin{align*}
    \sum_{j=1}^T(q_{jT}^-)^2=\frac{v_{S,-}}{Tb}(1+o(1)).
\end{align*}
Using the Cauchy--Schwarz inequality, we can further show the cross-term $2\sum_{j=1}^Tq^+_{jT}q^-_{jT}=o(\frac{1}{Tb})$. Thus, we finish the proof.\\

\noindent\textbf{Proof of Corollary \ref{corollary F2}} Since $\E[\varepsilon_{tT}^2]=\sigma^2$ (see \eqref{model fixed-design}), some simple algebra shows
\begin{align*}
    \text{Var}(\hat B_h)&=\text{Var}(\hat B_h-\E[\hat B_h])=\text{Var}(\sum_{j=1}^Tq_{jT}\varepsilon_{jT})\\
    &=\sigma^2\sum_{j=1}^Tq^2_{jT}+\sum_{j=1}^T\sum_{k=1\atop k\neq j}^{T}q_{jT}q_{kT}\text{Cov}(\varepsilon_{jT},\varepsilon_{kT})\\
    &=\frac{\sigma^2(v_{S,+}+v_{S,-})}{Tb}(1+o(1))+\sum_{j=1}^T\sum_{k=1\atop k\neq j}^{T}q_{jT}q_{kT}\text{Cov}(\varepsilon_{jT},\varepsilon_{kT}),
\end{align*}
where the last equality is due to Theorem \ref{th F1}. Thus, we only need to control the second term. Using $2|xy|\leq x^2+y^2$ and symmetry of covariance yields
\begin{align*}
    &\Big|\sum_{j=1}^T\sum_{k=1\atop k\neq j}^{T}q_{jT}q_{kT}\text{Cov}(\varepsilon_{jT},\varepsilon_{kT})\Big|\leq \sum_{j=1}^T\sum_{k=1\atop k\neq j}^{T}|q_{jT}q_{kT}|\ |\text{Cov}(\varepsilon_{jT},\varepsilon_{kT})|\\
    \leq &\frac{1}{2}\sum_{j=1}^T\sum_{k=1\atop k\neq j}^{T}(q^2_{jT}+q^2_{kT}) |\text{Cov}(\varepsilon_{jT},\varepsilon_{kT})|=\sum_{k=1}^Tq_{kT}^2\sum_{j=1\atop j\neq k}^T|\mathrm{Cov}(\varepsilon_{kT},\varepsilon_{jT})|\\
    \leq &\left(\sup_{k\leq T;T\geq 1}\sum_{j=1}^T|\mathrm{Cov}(\varepsilon_{kT},\varepsilon_{jT})|\right)\sum_{k=1}^Tq_{kT}^2.
\end{align*}
Then, together with D2 of Assumption \ref{as misspecification 1} and Theorem \ref{th F1}, we finish the proof of $ \mathrm{Var}(\hat B_h)=O(\frac{1}{Tb})$.

Now we start to calculate the uniform bias, and the proof can be divided into four smaller steps.

\noindent\textbf{Step 1}(local counts and $l_1$ bounds for the primitive weights)
Let $I\subset[-C_\psi,C_\psi]$ be any interval of length $|I|$. Since $\Psi$ is bi-Lipschitz from above, for every $\eta>0$ and endpoints $u<v$ of $I$,
\begin{align*}
    \frac{1}{T}\#\{j:x_{jT}\in[u,v]\}\leq \Psi_T(v)-\Psi_T(u-\eta)
\leq\bar c_\psi(v-u+\eta)+2D_T.
\end{align*}
Since $b=T^{-1/(2S+1)}$ and $D_T=o(\log T/\sqrt{T})$, letting $\eta\downarrow0$ gives
\begin{align}
\label{1}
    \#\{j:x_{jT}\in I\}\leq T\{\bar c_\psi|I|+2D_T\},\ 
\frac{D_T}{b}=o\!\left(\log T\,T^{-1/2+1/(2S+1)}\right)=o(1).
\end{align}
Since $b/h=o(1)$, (2) also implies $D_T/h=o(1)$. Hence every interval of length $O(b)$ contains $O(Tb)$ design points and every interval of length $O(h)$ contains $O(Th)$ design points.

For the order-$S$ weights of the pilot estimator, the lower Lipschitz bound in Assumption \ref{as fixed-design} implies that the corresponding population moment matrix has its smallest eigenvalue bounded away from zero uniformly over one-sided evaluation points. The Stieltjes approximation used in Theorem \ref{th concentration fixed} gives an empirical-population matrix difference of order $D_T/b=o(1)$. Weyl's inequality therefore yields a uniform bound on the inverse empirical moment matrix. Since $K$ and $r_S$ are bounded on $[-1,1]$,
\begin{align}
    \label{2}
    \sup_{x\in[0,h]}\max_{1\leq j\leq T}|w_{jb}(x;0,1)|\lesssim\frac{1}{Tb}, \qquad \sup_{x\in[-h,0]}\max_{1\leq j\leq T}|w_{jb}(x;-1,0)|\lesssim\frac{1}{Tb}.
\end{align}
The same argument with the order-one local-linear moment matrices and bandwidth $h$ gives
\begin{align}
\label{3}
    \max_t|l^+_{th}(0)|+\max_t|l^-_{th}(0)|\lesssim\frac{1}{Th}.
\end{align}
Combining \eqref{1}-\eqref{3} thus implies
\begin{align}
    \label{4}
    \sup_{x\in[0,h]}\sum_{j=1}^T|w_{jb}(x;0,1)|+\sup_{x\in[-h,0]}\sum_{j=1}^T|w_{jb}(x;-1,0)|\leq C\ \text{and}\  \sum_{t=1}^T|l^+_{th}(0)|+\sum_{t=1}^T|l^-_{th}(0)|\leq C
\end{align}
hold for some $C>0$ independent of $T,h,b$. Indeed, a pilot weight can be nonzero only when $|x_{jT}-x|\leq b$, so at most $O(Tb)$ terms occur in each sum in \eqref{4}. Similarly, $l^+_{th}(0)$ and $l^-_{th}(0)$ are supported on intervals of length $h$.

\noindent\textbf{Step 2}(polynomial reproduction and bias) For $x\in[0,h]$, the definition of $w_{jb}(x;0,1)$ gives the finite-sample identity
\begin{align*}
    &\sum_{j=1}^Tw_{jb}(x;0,1)r_S\!\left(\frac{x_{jT}-x}{b}\right)^\top=e_0^\top,\\
    &\sum_{j=1}^Tw_{jb}(x;0,1)(x_{jT}-x)^s=\begin{cases}1,&s=0,\\0,&1\leq s\leq S.\end{cases}
\end{align*}
The identical identities hold for $w_{jb}(x;-1,0)$ on the left side as well. Then, for all sufficiently large $T$, $h+b<\epsilon_0$. If $x\in[0,h]$ and $w_{jb}(x;0,1)\neq0$, then $|x_{jT}-x|\leq b$. Assumption \ref{as smoothness} therefore yields
\begin{align*}
    f_+(x_{jT})=
\sum_{s=0}^S\frac{f_+^{(s)}(x)}{s!}(x_{jT}-x)^s+R_{jT}(x), \qquad |R_{jT} (x)|\leq M_Sb^S\rho_S(b).
\end{align*}
Together with the results above, using $E[Y_{jT}]=f_+(x_{jT})$, we obtain
\begin{align}
\label{6}
    \left|E[\widehat f_{+,b}(x)]-f_+(x)\right|&=
\left|\sum_{j=1}^Tw_{jb}(x;0,1)R_{jT}(x)\right|\notag\\
&\leq M_Sb^S\rho_S(b)\sum_{j=1}^T|w_{jb}(x;0,1)|\lesssim M_Sb^S\rho_S(b).
\end{align}
The same one-sided argument at the boundary point gives $\left|\E[\widehat f_{+,b}(0+)]-f_+(0+)\right|\lesssim M_Sb^S\rho_S(b)$.
Repeating this argument on the left side gives
\begin{align}
\label{7}
    \sup_{x\in[-h,0]}\left|E[\widehat f_{-,b}(x)]-f_-(x)\right|+\left|E[\widehat f_{-,b}(0-)]-f_-(0-)\right|\lesssim M_Sb^S\rho_S(b).
\end{align}

\noindent\textbf{Step 3} Note the following identity
\begin{align*}
    \E[\widehat B_h]-B_h=&\sum_{t=1}^Tl^+_{th}(0)\Big(\E[\widehat f_{+,b}(x_{tT})]-f_+(x_{tT})-\E[\widehat f_{+,b}(0+)]+f_+(0+)
\Big)\\
&-\sum_{t=1}^Tl^-_{th}(0)\Big(\E[\widehat f_{-,b}(x_{tT})]-f_-(x_{tT})-\E[\widehat f_{-,b}(0-)]+f_-(0-)\Big).
\end{align*}
Combining \eqref{6}, \eqref{7}, and the triangle inequality gives, uniformly over $f\in\mathcal F_S(\epsilon_0,M_S,\rho_S)$,
\begin{align*}
    \left|\E_f[\widehat B_h]-B_h\right|\lesssim M_Sb^S\rho_S(b).
\end{align*}
Finally, $\rho_S(b)\to0$ and $b=T^{-1/(2S+1)}$, so (14) implies. This proves the result.

%%%%%%%%%%%%%%%%%%%%%%%%%%%%%%%%%%%%%%%%%%%%%%%%%%%%%%%%%%%%%%%%%%%%%%%%%%%%%%%%%%%%%%%%%%%%%%%%%%%%%%%%%%%%%%%%%%%%%%%%%%%%%%%%%%%%%%%%%%%%%%%%%%%%%%%%%%%%%%%%%%%%
\subsection{Random-Design}
\label{app F Random}
Suppose $\{X_i,Y_i\}_{i\geq 1}$ admits regression model \eqref{model random-design} and Assumptions \ref{as random design}--\ref{as misspecificaiton 2} hold. In this section, we will investigate the asymptotic property of bias estimation 
\begin{align*}
\tilde B_h
:&=\sum_{i=1}^nL_{ih,+}(0)[\tilde f_{+,b}(X_i)-\tilde f_{+,b}(0)]-\sum_{i=1}^nL_{ih,-}(0)[\tilde f_{-,b}(X_i)-\tilde f_{-,b}(0)],\\
&\tilde f_{+,b}(x)=\sum_{j=1}^nW_{jb}(x;0,1)Y_j,\ \ \tilde f_{-,b}(x)=\sum_{j=1}^nW_{jb}(x;-1,0)Y_j;\notag
\end{align*}
where the weight function $W_{jb}(x;a,b)$ is the weight function of $S$-order local polynomial with truncated inputs, and its definition is introduced in Appendix \ref{app concentration: random}.

We first review the definition of the $\beta$-mixing coefficient of a stochastic process. Suppose $(\Omega,\mathcal{F},\P)$ is a probability space and $\{Z_{i}\}_{i\geq 1}$ is a $\mathbf{Z}$-valued process defined on it. Its $\beta$-mixing coefficient is defined as follows,
\begin{align}
    \label{eq beta-mixing}
    \beta(r)=\sup_{n\geq 1}\sup_{1\leq t\leq n-r}||\mathcal{L}(Z_1,...,Z_t,Z_{t+r},...,Z_n)-\mathcal{L}(Z_1,...,Z_t)\otimes\mathcal{L}(Z_{t+r},...,Z_{n})||_{\text{TV}},
\end{align}
where $\mathcal{L}$ denotes the law of a random vector and $||P-Q||_{\text{TV}}$ is the total variation distance between measures $P$ and $Q$. For a more detailed discussion and introduction of $\beta$-mixing condition, please refer to textbooks like \citeA{doukhan2012mixing} and \citeA{j2007weak}.

Based on the $\varepsilon_0$ defined in $\F$ from Assumption \ref{as smoothness}, which equals $1$ in Section \ref{sec 4}, let
\begin{align*}
    \triangle_+:=&\sup_{x\in (0,\varepsilon_0]}|\tilde f_{+,b}(x)-f(x)|,\ \triangle_-:=\sup_{x\in [-\varepsilon_0,0]}|\tilde f_{-,b}(x)-f(x)|,  \\
    \tilde f_{+,b}(x)&=\sum_{j=1}^nW_{jb}(x;0,1)Y_j,\ \ \tilde f_{-,b}(x)=\sum_{j=1}^nW_{jb}(x;-1,0)Y_j.
\end{align*}
The following theorem investigates the convergence rate of $ \triangle_+$ and $ \triangle_-$.
\begin{theorem}
    \label{th F3}
    Under Assumptions \ref{as random design}--\ref{as misspecificaiton 2}, when $b=n^{-\frac{1}{2S+1}}$, we have 
    \begin{align*}
       \lim_{n\to\infty}n^{\frac{2S}{2S+1}}\P( \triangle_+\lor \triangle_- > \log n /n^{\frac{S}{2S+1}})=0.
    \end{align*}
\end{theorem}

\begin{corollary}
    \label{corollary F4}
     Under Assumptions \ref{as random design}--\ref{as misspecificaiton 2}, when $b=n^{-\frac{1}{2S+1}}$ and $b=o(h)$, we have 
     \begin{align*}
         \lim_{n\to\infty}n^{\frac{2S}{2S+1}}  \P(|\tilde B_h-B_h|>\log n /n^{\frac{S}{2S+1}})=0
     \end{align*}
\end{corollary}

\noindent\textbf{Proof of Theorem \ref{th F3}} %It suffices to consider $\P(\triangle_+>\log n/(2n^{\frac{S}{2S+1}}))$ and $\P(\triangle_->\log n/(2n^{\frac{S}{2S+1}}))$ respectively. We only demonstrate the proof for obtaining the tail probability of the former one since the argument can be directly applied to the  
It suffices to prove $P\left(\Delta_+>\frac{\log n}{n^{S/(2S+1)}}\right)
=o\left(n^{-2S/(2S+1)}\right)$, since the argument for $\Delta_-$ is identical.

\noindent\textbf{Step 1}
Recall that $D_n=\sup_{x\in[-1,1]}|\Psi_n(x)-\Psi(x)|.$ Recall that indicator class $\{\mathbf{1}(X_i\le x):x\in[-1,1]\}$ has VC-dimension equal to one. Since $\sum_{k\ge1}\beta(k)<\infty$, the covariance inequality for bounded $\beta$-mixing variables implies that the variance of every consecutive block of centered indicator variables is bounded by a constant times its length. Applying Lemma \ref{lemma 3} with $l=\lfloor\sqrt n\rfloor$, together with VC bracketing, yields
\begin{align*}
P(D_n>d_n)\lesssim d_n^{-1}
\exp\left(-\frac{cnd_n^2}{1+\sqrt n\,d_n}\right)
+d_n^{-1}\sqrt n\,\beta(\lfloor\sqrt n\rfloor).
\end{align*}
Take $d_n=b^2=n^{-2/(2S+1)}$. Then $S\ge3$ implies $\frac{nd_n^2}{1+\sqrt n\,d_n}\asymp n^{1/2-2/(2S+1)}\longrightarrow\infty$. Meanwhile, Assumption \ref{as misspecificaiton 2} D5 implies
\begin{align*}
    d_n^{-1}\sqrt n\,\beta(\lfloor\sqrt n\rfloor)\lesssim n^{2/(2S+1)+(1-\gamma)/2}=o\left(n^{-2S/(2S+1)}\right).
\end{align*}
Consequently, $P(D_n>d_n)=o\left(n^{-2S/(2S+1)}\right)$.

Moreover, on the event $\{D_n\le d_n\}$, repeating Step 1 of the proof of Theorem \ref{th concentration fixed} gives
\begin{align*}
\sup_{x\in[0,\varepsilon_0]}\left\|M_{1b}(x;0,1)-M_{1b,\Psi}(x;0,1)\right\|_{\mathrm{op}}
\nonumber\lesssim \frac{D_n}{b}\leq \frac{d_n}{b}=O(b)=o(1).
\end{align*}
By Lemma \ref{lemma 2} and Weyl's inequality, for all sufficiently large $n$, $\inf_{x\in[0,\varepsilon_0]}\lambda_{\min}(M_{1b}(x;0,1))
\ge\frac{c_\psi c_{1,S,\mu}}{2}$ on $\{D_n\le d_n\}$. Hence, 
\begin{align*}
    \sup_{x\in[0,\varepsilon_0]}\|M_{1b}(x;0,1)^{-1}\|_{\mathrm{op}}\le C\ \  \text{and}\  \sup_{x\in [0,\varepsilon_0]}\frac{1}{nb}\sum_{i=1}^nK(U_{ib,x})\mathbf{1}\{X_i\in(0,1]\}\leq C
\end{align*}
hold for some $C>0$ independent of $n,h,b$.

\noindent\textbf{Step 2} By the definition of the order-$S$ local-polynomial estimator, polynomial reproduction, and Assumption 3,
\begin{align*}
|\tilde f_{+,b}(x)-f(x)|
\le C\|M_{1b}(x;0,1)^{-1}\|_{\mathrm{op}}\Big[M_Sb^S\rho_S(b)\frac{1}{nb}\sum_{i=1}^nK(U_{ib,x})\mathbf{1}\{X_i\in(0,1]\}+\max_{0\le j\le S}\Big|\frac{1}{nb}\sum_{i=1}^n g_{x,j,+}(Z_i)\Big|\Big].
\end{align*}
On condition that $D_n\leq d_n$, together with the results in Step 1, we obtain
\begin{align*}
    \Delta_+\lesssim b^S\rho_S(b)+\max_{0\le j\le S}\sup_{x\in[0,\varepsilon_0]}\left|\frac{1}{nb}\sum_{i=1}^n g_{x,j,+}(Z_i)\right|.
\end{align*}
Fix any $\eta\in(0.5,1)$. Lemma \ref{lemma 5} gives
\begin{align*}
    P\left(\max_{0\le j\le S}\sup_{x\in[0,\varepsilon_0]}\left|\frac{1}{nb}\sum_{i=1}^n g_{x,j,+}(Z_i)\right|>\frac{(\log n)^\eta}{\sqrt{nb}}\right)=o\left(n^{-2S/(2S+1)}\right).
\end{align*}
Together with the setting $b=n^{-\frac{1}{2S+1}}$, we complete the proof.

\noindent\textbf{Proof of Corollary \ref{corollary F4}} This proof is a direct consequence of Theorem \ref{th F3}. We thus omit its proof here.

%%%%%%%%%%%%%%%%%%%%%%%%%%%%%%%%%%%%%%%%%%%%%%%%%%%%%%%%%%%%%%%%%%%%%%%%%%%%%%%%%%%%%%%%%%%%%%%%%%%%%%%%%%%%%%%%%%%%%%%%%%%%%%%%%%%%%%%%%%%%%%%%%%%%%%%%%%%%%%%%%%%%%%%%%%%%%%%%%%%%%%%%%%%%%%%%%%%%%%%%%%%%%%%%%%%%%%%%%%%%%%%%%%%%%%%%%%%%%%%%%%%%%%%%%%%%%%%%%%%%%%%%%%%%%%%%%%%%%%%%%%%%%%%%%%%%%%%
\newpage
\section{Technical Lemmas}
\def\theequation{G.\arabic{equation}}
\setcounter{equation}{0}
\def\thelemma{G.\arabic{lemma}}
\setcounter{lemma}{0}
\def\thetheorem{G.\arabic{theorem}}
\setcounter{theorem}{0}
\label{app lemma}
\begin{lemma}
    \label{lemma 1}
    Provided that $Z$ is a standard normal random variable, for every fixed $\beta>1$ and $\alpha\in (0,0.5)$, there exists an $\eta_{\alpha,\beta}>0$ depending only on $\beta$ and $\alpha$ such that
    \begin{align*}
       \sup_{t\geq 0} \P(|\beta Z+t|\leq \text{cv}_{1-\alpha}(t))<1-\alpha-\eta_{\alpha,\beta},
    \end{align*}
    where $\text{cv}_{1-\alpha}(t)$ is the upper $\alpha$-quantile of the folded normal distribution $|N(t,1)|$.
\end{lemma}

\begin{lemma}
\label{lemma 2}
    Let $K(u)$ be a density with compact support $[-1,1]$, and assume $\mu(\{u: K(u)>0\})>0$, where $\mu$ is the Lebesgue measure. For some fixed $[a,b]\subset \mathbb{R}$, by denoting $\bar K:=\sup_{u\in [-1,1]}K(u)$, we have
    \begin{align*}
        & \inf_{x\in [a,b], 0<h\leq \frac{b-a}{2}}\lambda_{\min}(\Gamma_{k,q,\mu}(x,h))\geq c_{k,q,\mu}>0,\qquad \sup_{x\in [a,b], 0<h\leq \frac{b-a}{2}} \lambda_{\max}(\Gamma_{2,q,\mu}(x,h))\leq \bar{K} (q+1),
    \end{align*}
    where $\lambda_{\min}(\A)$ and $\lambda_{\max}(\A)$ are the smallest and largest eigenvalues of $\A\in \R^{d\times d}$, and
    \begin{align*}
    &\Gamma_{k,q,\mu}^{+}=\int_{0}^1K^k(u)\r_q(u)\r^\top_q(u)d\mu,\ \Gamma_{k,q}^{-}=\int_{-1}^0K^k(u)\r_q(u)\r^\top_q(u)d\mu,\\ 
&I_{x,h}=[\frac{a-x}{h},\frac{b-x}{h}]\cap[-1,1],\ \Gamma_{k,q}(x,h)=\int_{I_{x,h}}K^k(u)\r_q(u)\r^\top_q(u)d\mu,\\
&\qquad c_{k,q,\mu}=\lambda_{\min}(\Gamma_{k,q,\mu}^{+})\land \lambda_{\min}(\Gamma_{k,q,\mu}^{-}),\qquad \ k=1,2.\notag
    \end{align*}
\end{lemma}

\begin{lemma}
\label{lemma 3}
    Let $\{\Xi_i\}_{i\geq 1}$ be a $\R^{d_1\times d_2}$-valued stochastic process satisfying $\beta$-mixing condition with coefficient $\beta(k)\leq Ck^{-\gamma}$. Given an integer $1\leq l\leq [n/2]$, let $J=[n/2l]$ and partition $\{1,...,n\}$ into consecutive blocks $I_{j}=\{(j-1)l+1,...,\min\{jl\}\}$, $1\leq j\leq J$. By defining $B_j=\sum_{i\in I_j}\Xi_{i}$, $\mathcal{J}_{0}=\{j\leq J:\ \text{$j$ is odd}\}$ and $\mathcal{J}_{e}=\{j\leq J:\ \text{$j$ is even}\}$, assume 
    \begin{align*}
        \sigma_{l}^2:= \max_{r\in \{o,e\}}\max\left\{\left\|\sum_{j\in \mathcal{J}_r}\E[B_{j}B_{j}^{\top}]\right\|,\left\|\sum_{j\in \mathcal{J}_r}\E[B_{j}^{\top}B_{j}]\right\|\right\}<\infty.
    \end{align*}
    Then, provided that $\E[\Xi_i]=0$ and $||\Xi_i||_{\text{op}}\leq R_n$, for arbitrary given $t>0$, we have
    \begin{align*}
        \P\left(\Big\|\sum_{i=1}^n\Xi_i\Big\|_{\text{op}}\geq t\right)\leq 2(d_1+d_2)\exp\left(-\frac{t^2}{8\sigma_l^2+\frac{4}{3}lRt}\right)+J\beta(l).
    \end{align*}
\end{lemma}

\begin{lemma}
    \label{lemma 4}
    For each given $S\geq 3$, define $Z_i=(X_i,\varepsilon_i)$, $g_{x,j,+}(Z_i)=K(U_{ib,x})\1[X_i\in (0,1]]U_{ib,x}^j\varepsilon_i$ and $g_{x,j,-}(Z_i)=K(U_{ib,x})\1[X_i\in [-1,0]]U_{ib,x}^j\varepsilon_i$ where $U_{ib,x}=\frac{X_i-x}{b}$. Under Assumptions \ref{as random design}-\ref{as misspecificaiton 2}, for every consecutive set $I\subset \{1,...,n\}$, we have
    \begin{align*}
    \max_{j\leq S} \sup_{x\in [0,\varepsilon_0]}   \mathrm{Var}(\sum_{i\in I}g_{x,j,+}(Z_i))\lesssim |I|b,\qquad \max_{j\leq S} \sup_{x\in [-\varepsilon_0,0]}   \mathrm{Var}(\sum_{i\in I}g_{x,j,-}(Z_i))\lesssim |I|b,
    \end{align*}
    where $|I|$ denotes the cardinality of the set $I$.
\end{lemma}

\begin{lemma}
    \label{lemma 5} Based on the $Z_i$ and $g_{x,j,\pm}(x)$ defined in Lemma \ref{lemma 4}, for each given $0\leq j\leq S$, under Assumptions \ref{as random design}--\ref{as misspecificaiton 2}, the following inequality holds for each $ s\in \{+,-\}$ and $0.5<\eta<1$,
    \begin{align*}
      \lim_{n\to\infty} n^{\frac{2S}{2S+1}}\P\Big( \max_{0\leq j\leq S} \sup_{x\in \mathcal I_s}\Big|\frac{1}{nb}\sum_{i=1}^ng_{x,j,s}(Z_i)\Big|>\frac{(\log n)^\eta}{\sqrt{nb}}\Big)=0,
    \end{align*}
    where $\mathcal I_+:=[0,\varepsilon_0]$ and $\mathcal I_-:=[-\varepsilon_0,0]$.
\end{lemma}

%%%%%%%%%%%%%%%%%%%%%%%%%%%%%%%%%%%%%%%%%%%%%%%%%%%%%%%%%%%%%%%%%%%%%%%%%%%%%%%%%%%%%%%%%%%%%%%%%%%%%%%%%%%% Proof of Lemma 1
%%%%%%%%%%%%%%%%%%%%%%%%%%%%%%%%%%%%%%%%%%%%%%%%%%%%%%%%%%%%%%%%%%%%%%%%%%%%%%%%%%%%%%%%%%%%%%%%%%%%%%%%%%%%%%%%%%%%%%%%%%%

\noindent\textbf{Proof of Lemma \ref{lemma 1}} Define $G_{\alpha}(t,\beta):=\P(|\beta Z+t|\leq \text{cv}_{1-\alpha}(t))$. The proof is divided into two steps. In Step 1, we prove that, for each fixed $\beta>1$ and $\alpha\in (0,0.5)$, $G_{\alpha}(t,\beta)< G_{\alpha}(t,1)=1-\alpha$ holds pointwise on $t\geq 0$. In Step 2, we further prove that, for each fixed $\alpha\in (0,0.5)$ and $\beta>1$, $\lim_{t\to\infty}G_\alpha(t,\beta)<1-\alpha$ as well. We then finish the proof by using the continuity of the mapping $t\to G_{\alpha}(t,\beta)$ on $[0,\infty]$.

\noindent\textbf{Step 1} First, note that, when $\beta=1$, the definition of $\text{cv}_{1-\alpha}(t)$ yields the identity
\begin{align*}
    \P(|Z+t|\leq \text{cv}_{1-\alpha}(t))=\Phi(\text{cv}_{1-\alpha}(t)-t)-\Phi(-\text{cv}_{1-\alpha}(t)-t)=1-\alpha,\ \forall\ t\geq 0,
\end{align*}
where $\Phi$ is the standard Gaussian distribution function. Since $\alpha<0.5$ implies $1-\alpha>0.5$, the fact that $\Phi(0)=0.5$ yields 
\begin{align}
    \label{eq proof of lemma 1 eq1}
    \text{cv}_{1-\alpha}(t)>t,\ \forall\ t\geq 0.
\end{align}
Then, for any fixed $t\geq 0$, using \eqref{eq proof of lemma 1 eq1} and some simple algebra asserts
\begin{align*}
      \partial_{\beta} G(t,\beta)=-\frac{\text{cv}_{1-\alpha}(t)-t}{\beta^2}\phi(\frac{\text{cv}_{1-\alpha}(t)-t}{\beta})-\frac{\text{cv}_{1-\alpha}(t)+t}{\beta^2}\phi(\frac{-\text{cv}_{1-\alpha}(t)-t}{\beta})<0,
\end{align*}
which implies that, for each fixed $\beta>1$ and $\alpha\in (0,0.5)$,
\begin{align}
    \label{eq proof of lemma 1 eq2}
    G_{\alpha}(t,\beta)< G_{\alpha}(t,1)=1-\alpha,\ \forall\ t\geq 0.
\end{align}

\noindent\textbf{Step 2} Note the facts that $\lim_{t\to 0}\text{cv}_{1-\alpha}(t)=z_{1-\alpha/2}$ and $\lim_{t\to \infty}\text{cv}_{1-\alpha}(t)-t=z_{1-\alpha}$ hold for every fixed $\alpha\in (0,1)$, where $z_{1-\alpha}$ is the $\alpha$-level upper quantile of the standard Gaussian distribution. Then, for any fixed $\beta>1$, since the mapping $t\to G_{\alpha}(t,\beta)$ is continuous on $[0,\infty]$, the following results can be obtained immediately for any fixed $\alpha\in (0,0.5)$ and $\beta>1$.
\begin{align}
    \label{eq proof of lemma 1 eq3}
    &\qquad\qquad \lim_{t\to 0^+}G_{\alpha}(t,\beta)=2\Phi(\frac{z_{1-\alpha/2}}{\beta})-1<1-\alpha,\\
        \label{eq proof of lemma 1 eq4}
   & \lim_{t\to+\infty}G_{\alpha}(t,\beta)=\lim_{t\to\infty}\Phi\Big(\frac{\text{cv}_{1-\alpha}(t)-t}{\beta}\Big)-\Phi\Big(\frac{-\text{cv}_{1-\alpha}(t)-t}{\beta}\Big)=\Phi(\frac{z_{1-\alpha}}{\beta})<1-\alpha.
\end{align}
\eqref{eq proof of lemma 1 eq3} and \eqref{eq proof of lemma 1 eq4} indicate that, for every fixed $\alpha\in (0,0.5)$, there exist $0<T_1<T_2<\infty$ such that $G_{\alpha}(t,\beta)<1-\alpha$ holds uniformly over all $t\in [0,T_1]\cup [T_2,\infty]$. Meanwhile, for any fixed $\alpha,\beta$, since $t\to G_{\alpha}(t,\beta)$ is continuous on closed interval $[T_1,T_2]$, \eqref{eq proof of lemma 1 eq2} implies that $\sup_{t\in [T_1,T_2]}|G_{\alpha}(t,\beta)|<1-\alpha$. Altogether, we finish the proof of Lemma \ref{lemma 1}.\\

%%%%%%%%%%%%%%%%%%%%%%%%%%%%%%%%%%%%%%%%%%%%%%%%%%%%%%%%%%%%%%%%%%%%%%%%%%%%%%%%%%%%%%%%%%%%%%%%%%%%%%%%%%%% Proof of Lemma 2
%%%%%%%%%%%%%%%%%%%%%%%%%%%%%%%%%%%%%%%%%%%%%%%%%%%%%%%%%%%%%%%%%%%%%%%%%%%%%%%%%%%%%%%%%%%%%%%%%%%%%%%%%%%%%%%%%%%%%%%%%%%
\noindent\textbf{Proof of Lemma \ref{lemma 2}} We first prove the case where $k=1$. The proof is divided into two steps. 

\noindent\textbf{Step 1} We first focus on proving that the following two matrices are positive definite,
\begin{align*}
    \Gamma_{1,q}^+=\int_{0}^1 K(u)\r_q(u)\r^\top_q(u)d\mu,\qquad \Gamma_{1,q}^-=\int_{-1}^0 K(u)\r_q(u)\r^\top_q(u)d\mu.
\end{align*}
For $\Gamma_{1,q}^+$, note that, for any given $\v=(v_0,...,v_{q})\in \mathbb{R}^{q+1}$ such that $\v\neq 0$, we have 
\begin{align*}
    \v^\top \Gamma_{1,q}^+\v=\int_{0}^1 K(u)(\v^\top \r_q(u))^2d\mu=:\int_{0}^1K(u)p_\v^2(u)d\mu.
\end{align*}
Since $p_\v(u)=\sum_{j=0}^qv_ju^j$ is a non-zero polynomial, there are at most finitely many solutions of the equation $p_\v(u)=0$. This implies $\mu(\{p^2_\v(u)=0\})=0$. Then, condition $\mu(\{K(u)>0\})>0$ implies $\Gamma_{1,q}^+\succ 0$. The same argument works for the proof of  $\Gamma_{1,q}^-\succ 0$. By denoting $c_{1,q}=\lambda_{\min}(\Gamma_{1,q}^+)\land\lambda_{\min}(\Gamma_{1,q}^-)$, we have $c_{1,q}>0$.

\noindent\textbf{Step 2} Note that, for any $x\in [a,b]$ and $0<h\leq \frac{b-a}{2}$,  $x\geq a+h$ or $x\leq b-h$ holds. Otherwise, we have $b-a=(b-x)+(x-a)<2h$, which violates the definition of $h$. When $x\in [a+h,b]$, $[-1,0]\subset I_{x,h}$ holds and, since $K(u)\r_q(u)\r^\top_q(u)\in \R^{(q+1)^2}$ is non-negative definite for each $u\in \R$, we have 
\begin{align*}
    \Gamma_{1,q}(x,h)-\Gamma_{1,q}^-=\int_{I_{x,h}\backslash [-1,0]}K(u)\r_q(u)\r^\top_q(u)d\mu\succeq 0,
\end{align*}
which implies $\lambda_{\min}(\Gamma_{1,q}(x,h))\geq \lambda_{\min}(\Gamma_{1,q}^-)\geq c_{1,q,\mu}$. Similarly, we can prove that the same result holds for all $x\in [a,b-h]$. The argument above also works for the proof of case $k=2$, since $K^2(u)\r_q(u)\r_q^\top(u)$ is non-negative definite for every given $u\in [-1,1]$. We thus omit it here. For the second inequality, we only need to notice that $\Gamma_{2,q}(x,h)\leq \Gamma_{1,q}^++\Gamma_{1,q}^-$ and $K^2(u)\r_q(u)\r_q^\top(u)\leq q+1$.
Then, we finish the proof of Lemma \ref{lemma 2}.\\

%%%%%%%%%%%%%%%%%%%%%%%%%%%%%%%%%%%%%%%%%%%%%%%%%%%%%%%%%%%%%%%%%%%%%%%%%%%%%%%%%%%%%%%%%%%%%%%%%%%%%%%%%%%% Proof of Lemma 3
%%%%%%%%%%%%%%%%%%%%%%%%%%%%%%%%%%%%%%%%%%%%%%%%%%%%%%%%%%%%%%%%%%%%%%%%%%%%%%%%%%%%%%%%%%%%%%%%%%%%%%%%%%%%%%%%%%%%%%%%%%%
\noindent\textbf{Proof of Lemma \ref{lemma 3}} This proof is nearly the same as that of Theorem 4.2 in \citeA{chen2015optimal}. We thus omit it here.

%%%%%%%%%%%%%%%%%%%%%%%%%%%%%%%%%%%%%%%%%%%%%%%%%%%%%%%%%%%%%%%%%%%%%%%%%%%%%%%%%%%%%%%%%%%%%%%%%%%%%%%%%%%% Proof of Lemma 4
%%%%%%%%%%%%%%%%%%%%%%%%%%%%%%%%%%%%%%%%%%%%%%%%%%%%%%%%%%%%%%%%%%%%%%%%%%%%%%%%%%%%%%%%%%%%%%%%%%%%%%%%%%%%%%%%%%%%%%%%%%%
\noindent\textbf{Proof of Lemma \ref{lemma 4}} The proof is based on Theorem 1 in \citeA{hansen2008uniform}. We first focus on proving the variance upper bound for $g_{x,j,+}(Z_i)$. The proof holds for arbitrary fixed $j\leq S$.

\noindent\textbf{Step 1} (Moments) On the support of the kernel, $|U_{ib,x}|\le1$, and $g_{x,j,+}(Z_i)\neq0$ only if
$0<X_i\le 1=\varepsilon_0$. For $1\le q\le\varsigma$,
conditional H\"older's inequality and Assumption \ref{as moment} M1 yield, for
$u\in[0,\varepsilon_0]$,
\begin{align*}
&\E\bigl(|\varepsilon_i|^q\mid X_i=u\bigr)f_X(u)
\le \bigl[\E\bigl(|\varepsilon_i|^{\varsigma}\mid X_i=u\bigr)\bigr]^{q/\varsigma}f_X(u)\le B_1^{q/\varsigma}f_X(u)^{1-q/\varsigma}=:C
\end{align*}
Consequently, using boundedness and compact support of $K$,
\begin{align*}
\E|g_{x,j,+}(Z_i)|^q
&\le C\int_0^{\varepsilon_0}\mathbf{1}\{|u-x|\le b\}
\E\bigl(|\varepsilon_i|^q\mid X_i=u\bigr)f_X(u)\,du
\notag\\
&\le C\int_0^{\varepsilon_0}\mathbf{1}\{|u-x|\le b\}\,du
\le Cb.
\end{align*}
Exogeneity gives $\E [g_{x,j,+}(Z_i)]=0$, because the kernel multiplier depends only on $X_i$. In particular,
\begin{align}
\label{eq proof of lemma 4 eq1}
    \text{Var}(g_{x,j,+}(Z_i))\le Cb,
\qquad \|g_{x,j,+}(Z_i)\|_{L^{\varsigma}}\le Cb^{1/\varsigma}.
\end{align}

\medskip
\noindent\textbf{Step 2. Short-lag covariances ($1\le k<k_0$).}
Write, as before,
$C_{i,k}(x)=\operatorname{Cov}(g_{x,j,+}(Z_i),g_{x,j,+}(Z_{i+k}))$
whenever $i+k\le T$. By Cauchy--Schwarz and \eqref{eq proof of lemma 4 eq1},
\begin{align}
\label{eq proof of lemma 4 eq2}
&|C_{i,k}(x)|
\le\sqrt{\operatorname{Var}(g_{x,j,+}(Z_i))
\operatorname{Var}(g_{x,j,+}(Z_{i+k}))}\le Cb,
\notag\\
&\sum_{k=1}^{k_0-1}|C_{i,k}(x)|
\le C(k_0-1)b\le Cb.
\end{align}
The integer $k_0$ is fixed and does not depend on $b$ or $T$.

\medskip
\noindent\textbf{Step 3. Intermediate-lag covariances
($k_0\le k\le[ b^{-1}]$).}
Take $b$ sufficiently small that $[ b^{-1}]\ge k_0$. For $x\in[0,\varepsilon_0]$, let $A_x:=[x-b,x+b]\cap(0,1]$, an interval contained in $[-1,1]=[-\varepsilon_0,\varepsilon_0]$ with $|A_x|\le 2b$. Since $g_{x,j,+}(Z_i)=0$ unless $X_i\in A_x$, and $|K(U_{ib,x})U_{ib,x}^j|\le\bar K$ on the support of the kernel, $\E g_{x,j,+}(Z_i)=0$ and Assumption \ref{as moment} M2 give
\begin{align*}
|C_{i,k}(x)|
&\le \E|g_{x,j,+}(Z_i)g_{x,j,+}(Z_{i+k})|
\le \bar K^2\,\E\left[|\varepsilon_i\varepsilon_{i+k}|\,\mathbf{1}\{X_i\in A_x,\ X_{i+k}\in A_x\}\right]
\notag\\
&\le \bar K^2\left(c_k|A_x|+B_2|A_x|^2\right)\le C\left(c_kb+b^2\right),
\end{align*}
which implies
\begin{align}
    \label{eq proof of lemma 4 eq3}
    \sum_{k=k_0}^{[ b^{-1}]}|C_{i,k}(x)|
\le Cb\sum_{k\ge k_0}c_k+Cb^2[ b^{-1}]\le Cb.
\end{align}
The first term accounts for pairs $(X_i,X_{i+k})$ that coincide or nearly coincide, which M2 permits. Under the joint-density condition discussed after Assumption \ref{as misspecificaiton 2}, $c_k=0$ and only the second term remains.

\medskip
\noindent\textbf{Step 4. Long-lag covariances
($k>[ b^{-1}]$).} Davydov's covariance inequality, joint absolute regularity, and \eqref{eq proof of lemma 4 eq1} give
\begin{align}
    \label{eq proof of lemma 4 eq4}
|C_{i,k}(x)|
&\le C\beta_k^{1-2/\varsigma}
\|g_{x,j,+}(Z_i)\|_{L^{\varsigma}}
\|g_{x,j,+}(Z_{i+k})\|_{L^{\varsigma}}
\notag\\
&\le Cb^{2/\varsigma}k^{-\gamma(1-2/\varsigma)}.
\end{align}
Assumption \ref{as misspecificaiton 2} D5 ensures $\gamma\left(1-\frac{2}{\varsigma}\right)
>2-\frac{2}{\varsigma}>1$ and $\frac{2}{\varsigma}
+\gamma\left(1-\frac{2}{\varsigma}\right)-1>1.$
Because $[ b^{-1}]\ge b^{-1}/2$ for sufficiently small
$b$, integral comparison and these two inequalities give
\begin{align}
   \label{eq proof of lemma 4 eq5}
\sum_{k>[ b^{-1}]}|C_{i,k}(x)|
&\le Cb^{2/\varsigma}
\sum_{k>[ b^{-1}]}k^{-\gamma(1-2/\varsigma)}\le Cb^{2/\varsigma}
\int_{[ b^{-1}]}^{\infty}
 t^{-\gamma(1-2/\varsigma)}\,dt
\notag\\
&=\frac{Cb^{2/\varsigma}}{\gamma(1-2/\varsigma)-1}
[ b^{-1}]^{1-\gamma(1-2/\varsigma)}\le Cb^{2/\varsigma+\gamma(1-2/\varsigma)-1}
\le Cb.
\end{align}

\medskip
\noindent\textbf{Step 5. Covariance summability and the block variance.}
Combining the diagonal bound \eqref{eq proof of lemma 4 eq1}-\eqref{eq proof of lemma 4 eq5}, uniformly in $i$, we have
\begin{align*}
    \text{Var}(g_{x,j,+}(Z_i))+\sum_{\substack{k\ge1\\i+k\le T}}|C_{i,k}(x)|\le Cb.
\end{align*}
The same bounds apply to the backward covariances: for $k\ge1$ with
$i-k\ge1$, use $C_{i-k,k}(x)$ and the uniformity of Steps 2--4 in the
initial index $i-k$. Thus
\begin{align*}
    \sup_{1\le i\le T}\sum_{k=1}^T\big|\text{Cov}(g_{x,j,+}(Z_i),g_{x,j,+}(Z_k))\bigr|\le Cb.
\end{align*}
For any consecutive $I\subseteq\{1,\ldots,T\}$, we immediately obtain
\begin{align}
\text{Var}\left(\sum_{i\in I}g_{x,j,+}(Z_i)\right)
&=\sum_{i\in I}\sum_{k\in I}
\operatorname{Cov}(g_{x,j,+}(Z_i),g_{x,j,+}(Z_k))
\notag\\
&\le\sum_{i\in I}\sum_{k=1}^T
\bigl|\operatorname{Cov}(g_{x,j,+}(Z_i),g_{x,j,+}(Z_k))\bigr|
\notag\\
&\le C|I|b.
\tag{12}
\end{align}
Every bound above is uniform over $x\in[0,\varepsilon_0]$ and $0\le j\le S$.\\

%%%%%%%%%%%%%%%%%%%%%%%%%%%%%%%%%%%%%%%%%%%%%%%%%%%%%%%%%%%%%%%%%%%%%%%%%%%%%%%%%%%%%%%%%%%%%%%%%%%%%%%%%%%% Proof of Lemma 5
%%%%%%%%%%%%%%%%%%%%%%%%%%%%%%%%%%%%%%%%%%%%%%%%%%%%%%%%%%%%%%%%%%%%%%%%%%%%%%%%%%%%%%%%%%%%%%%%%%%%%%%%%%%%%%%%%%%%%%%%%%%
\noindent\textbf{Proof of Lemma \ref{lemma 5}} The proof is divided into multiple steps. We only demonstrate the proof for $g_{x,j,+}(x)$  since the argument can be directly applied to $g_{x,j,-}$.

Throughout, let $a:=\frac{S}{2S+1}$, so that $nb=n^{2a}$, and let $t:=(\log n)^{\eta}/\sqrt{nb}=(\log n)^{\eta}n^{-a}$. All constants below are independent of $n$, $x$, $j$ and $f$.

\noindent\textbf{Step 1} (Truncation level and block length) Since $1+2a=\frac{4S+1}{2S+1}$ and $\frac{a}{1+2a}=\frac{S}{4S+1}$, the joint condition in Assumption \ref{as misspecificaiton 2} D5 is equivalent to $\frac{1+2a}{\varsigma}+\frac{1+2a}{\gamma+1}<a$. Hence we can fix constants $\lambda,\rho>0$ such that
\begin{align*}
    \lambda>\frac{1+2a}{\varsigma},\qquad \rho>\frac{1+2a}{\gamma+1},\qquad \lambda+\rho<a.
\end{align*}
Let $L_n:=n^{\lambda}$ and $r:=\lceil n^{\rho}\rceil$. The three conditions control the truncation, the coupling and the range of the blocks, respectively.

\noindent\textbf{Step 2} (Truncation) Define $\varepsilon_{in}=\varepsilon_i\1[|\varepsilon_i|\leq L_n]$ and $\varepsilon_{in}^-=\varepsilon_i-\varepsilon_{in}$. For fixed $j$ and $x\in[0,\varepsilon_0]$,
\begin{align*}
\frac{1}{nb}\sum_{i=1}^ng_{x,j,+}(Z_i)=&\frac{1}{nb}\sum_{i=1}^nQ_{ig}(x)+\frac{1}{nb}\sum_{i=1}^n\E\big[K(U_{ib,x})\1[X_i\in (0,1]]U_{ib,x}^j\varepsilon_{in}\big]\\
&+\frac{1}{nb}\sum_{i=1}^nK(U_{ib,x})\1[X_i\in (0,1]]U_{ib,x}^j\varepsilon^-_{in},
\end{align*}
where $Q_{ig}(x)=K(U_{ib,x})\1[X_i\in (0,1]]U_{ib,x}^j\varepsilon_{in}-\E[K(U_{ib,x})\1[X_i\in (0,1]]U_{ib,x}^j\varepsilon_{in}]$. By Assumption \ref{as moment} M1, $\max_i\E|\varepsilon_i|^{\varsigma}=\max_i\int_{-1}^1\E[|\varepsilon_i|^{\varsigma}\mid X_i=u]f_X(u)du\leq 2B_1$, and, for every $x$, $\E[|\varepsilon_i|^{\varsigma}\1\{|X_i-x|\leq b\}]\leq 2bB_1$.

The third term vanishes on $\bigcap_{i\leq n}\{|\varepsilon_i|\leq L_n\}$, and
\begin{align*}
    \P\Big(\bigcup_{i=1}^n\{|\varepsilon_{i}|>L_n\}\Big)\leq nL_n^{-\varsigma}\max_{i}\E|\varepsilon_{i}|^{\varsigma}\leq 2B_1n^{1-\lambda\varsigma}=o(n^{-2a}),
\end{align*}
because $\lambda\varsigma>1+2a$. The second term is deterministic. Since $\E[\varepsilon_i\mid X_i]=0$, we have $\E[K(U_{ib,x})\1[X_i\in (0,1]]U_{ib,x}^j\varepsilon_{in}]=-\E[K(U_{ib,x})\1[X_i\in (0,1]]U_{ib,x}^j\varepsilon_i\1\{|\varepsilon_i|>L_n\}]$, and $|\varepsilon_i|\1\{|\varepsilon_i|>L_n\}\leq L_n^{1-\varsigma}|\varepsilon_i|^{\varsigma}$. Hence
\begin{align*}
    \sup_{x\in[0,\varepsilon_0]}\Big|\frac{1}{nb}\sum_{i=1}^n\E\big[K(U_{ib,x})\1[X_i\in (0,1]]U_{ib,x}^j\varepsilon_{in}\big]\Big|\leq \frac{\bar K}{b}\,L_n^{1-\varsigma}\,2bB_1=2\bar KB_1n^{-\lambda(\varsigma-1)}.
\end{align*}
Since $\lambda(\varsigma-1)=\lambda\varsigma-\lambda>1+2a-a>a$, this is $o(t)$, and so it is smaller than $t/3$ for all sufficiently large $n$. Therefore, for all sufficiently large $n$,
\begin{align*}
 \P\left(\max_{0\leq j\leq S} \sup_{x\in [0,\varepsilon_0]}\Big|\frac{1}{nb}\sum_{i=1}^ng_{x,j,+}(Z_i)\Big|>t\right)
 \leq \sum_{j=0}^S\P\left(\sup_{x\in [0,\varepsilon_0]}\Big|\frac{1}{nb}\sum_{i=1}^nQ_{ig}(x)\Big|>\frac{t}{3}\right)+o(n^{-2a}).
\end{align*}

\noindent\textbf{Step 3} (Blocking and coupling) Let $K_n:=\lceil n/r\rceil$, $I_k=\{(k-1)r+1,\ldots,\min\{kr,n\}\}$ for $k=1,\ldots,K_n$, and $B_k(x):=\sum_{i\in I_k}Q_{ig}(x)$. Then
\begin{align*}
   \P\left(\sup_{x\in [0,\varepsilon_0]}\Big|\frac{1}{nb}\sum_{i=1}^nQ_{ig}(x)\Big|>\frac{t}{3}\right)\leq& \P\Big(\sup_{x\in [0,\varepsilon_0]}\Big|\frac{1}{nb}\sum_{k\ \text{odd}}B_{k}(x)\Big|>\frac{t}{6}\Big)\\
   &+\P\Big(\sup_{x\in[0,\varepsilon_0]}\Big|\frac{1}{nb}\sum_{k\ \text{even}}B_{k}(x)\Big|>\frac{t}{6}\Big)=:\P_{o}+\P_{e}.
\end{align*}
Consecutive odd blocks are separated by a gap of length $r$. By Berbee's lemma \cite{Berbee1979}, there exist mutually independent blocks $\{\tilde Z_i\}_{i\in I_k}$, $k$ odd, such that each $\{\tilde Z_i\}_{i\in I_k}$ has the same law as $\{Z_i\}_{i\in I_k}$ and $\P(\exists\,k\ \text{odd}:\{\tilde Z_i\}_{i\in I_k}\neq\{Z_i\}_{i\in I_k})\leq K_n\beta(r)$. Let $\tilde B_k(x)$ be $B_k(x)$ computed from $\{\tilde Z_i\}_{i\in I_k}$. Then
\begin{align*}
    \P_{o}\leq \P\Big(\sup_{x\in [0,\varepsilon_0]}\Big|\frac{1}{nb}\sum_{k\ \text{odd}}\tilde B_{k}(x)\Big|>\frac{t}{6}\Big)+K_n\beta(r),\qquad K_n\beta(r)\lesssim n^{1-\rho}n^{-\rho\gamma}=n^{1-\rho(1+\gamma)}=o(n^{-2a}),
\end{align*}
because $\rho(1+\gamma)>1+2a$. The same holds for $\P_e$.

\noindent\textbf{Step 4} (Bernstein inequality at a fixed $x$) Fix $x\in [0,\varepsilon_0]$. The variables $\tilde B_k(x)$, $k$ odd, are independent and have mean zero. The proof of Lemma \ref{lemma 4} applies verbatim to the truncated and centered summands $Q_{ig}(x)$: $|\varepsilon_{in}|\leq|\varepsilon_i|$, and, by Step 2, centering adds at most $(2\bar KB_1bL_n^{1-\varsigma})^2\leq Cb^2$ to each covariance in its Step 3. Hence $\mathrm{Var}(\tilde B_k(x))=\mathrm{Var}(B_k(x))\leq C|I_k|b$, so that $\sum_{k\ \text{odd}}\mathrm{Var}(\tilde B_k(x))\leq Cnb$. Moreover, $|Q_{ig}(x)|\leq 2\bar KL_n$, so $|\tilde B_k(x)|\leq 2\bar KrL_n$. Lemma \ref{lemma 3} with $l=1$, which is Bernstein's inequality for the independent variables $\tilde B_k(x)$, gives
\begin{align*}
    \P\Big(\Big|\frac{1}{nb}\sum_{k\ \text{odd}}\tilde B_{k}(x)\Big|>\frac{t}{12}\Big)
    \leq 4\exp \Big(-\frac{(nbt/12)^2}{8Cnb+\frac{4}{3}\cdot 2\bar KrL_n\cdot nbt/12}\Big).
\end{align*}
Here $(nbt)^2=nb(\log n)^{2\eta}$ and $rL_n\cdot nbt\lesssim n^{\rho+\lambda}\,nb\,n^{-a}(\log n)^{\eta}=o(nb)$, because $\lambda+\rho<a$. Hence the right-hand side is at most $4\exp(-c(\log n)^{2\eta})$ for all sufficiently large $n$, with $c>0$ independent of $x$ and $f$. The same bound holds for the even blocks.

\noindent\textbf{Step 5} (Uniformity in $x$) The map $u\mapsto K(u)u^j$ is Lipschitz on $[-1,1]$, $|U_{ib,x}-U_{ib,y}|=|x-y|/b$, and every truncated error is bounded by $L_n$. Hence, pathwise for both the original and the coupled blocks, for any $x,y\in [0,\varepsilon_0]$,
\begin{align*}
 \Big| \frac{1}{nb}\sum_{k\ \text{odd}}\big(\tilde B_{k}(x)-\tilde B_{k}(y)\big)\Big|\leq \frac{C_1L_n}{b^2}|x-y|.
\end{align*}
Let $\delta=\frac{b^2t}{12C_1L_n}$ and choose a deterministic grid $\mathcal{X}_{\delta}\subset [0,\varepsilon_0]$ containing both endpoints with maximal mesh $\delta$. Then $N_{\delta}:=\textbf{Card}(\mathcal{X}_{\delta})\leq 2+\frac{12C_1\varepsilon_0L_n}{b^2t}$, which grows polynomially in $n$, and every $x\in[0,\varepsilon_0]$ has a grid point within distance $\delta$. Consequently,
\begin{align*}
    \P\Big(\sup_{x\in [0,\varepsilon_0]}\Big|\frac{1}{nb}\sum_{k\ \text{odd}}\tilde B_{k}(x)\Big|>\frac{t}{6}\Big)\leq \P\Big(\max_{y\in\mathcal X_\delta}\Big|\frac{1}{nb}\sum_{k\ \text{odd}}\tilde B_{k}(y)\Big|>\frac{t}{12}\Big)\leq 4N_{\delta}\exp(-c(\log n)^{2\eta})=o(n^{-2a}),
\end{align*}
since $2\eta>1$. The same holds for the even blocks. Combining Steps 2--5 over $0\leq j\leq S$ finishes the proof.

%%%%%%%%%%%%%%%%%%%%%%%%%%%%%%%%%%%%%%%%%%%%%%

\end{document}